\documentclass[sigconf,nonacm]{acmart}

\usepackage{algorithm}
\usepackage{algpseudocode}
\usepackage{booktabs}
\usepackage{subfig}

\AtBeginDocument{%
  \providecommand\BibTeX{{%
    \normalfont B\kern-0.5em{\scshape i\kern-0.25em b}\kern-0.8em\TeX}}}

\copyrightyear{2020}
\acmYear{2020}
\setcopyright{licensedothergov}\acmConference[SIGSPATIAL '20]{28th International Conference on Advances in Geographic Information Systems}{November 3--6, 2020}{Seattle, WA, USA}
\acmBooktitle{28th International Conference on Advances in Geographic Information Systems (SIGSPATIAL '20), November 3--6, 2020, Seattle, WA, USA}
\acmPrice{15.00}
\acmDOI{10.1145/3397536.3422210}
\acmISBN{978-1-4503-8019-5/20/11}

\begin{document}

\title{Succinct Trit-array Trie for Scalable Trajectory Similarity Search}
 
\author{Shunsuke Kanda}
\affiliation{%
  \institution{RIKEN AIP}
}
\email{shunsuke.kanda@riken.jp}

\author{Koh Takeuchi}
\affiliation{%
  \institution{Kyoto University and RIKEN AIP}
}
\email{takeuchi@i.kyoto-u.ac.jp}

\author{Keisuke Fujii}
\affiliation{%
  \institution{Nagoya University and RIKEN AIP}
}
\email{fujii@i.nagoya-u.ac.jp}

\author{Yasuo Tabei}
\affiliation{%
  \institution{RIKEN AIP}
}
\email{yasuo.tabei@riken.jp}

\begin{abstract}
Massive datasets of spatial trajectories representing the mobility of a diversity of moving objects are ubiquitous in research and industry. Similarity search of a large collection of trajectories is indispensable for turning these datasets into knowledge. Locality sensitive hashing (LSH) is a powerful technique for fast similarity searches.
Recent methods employ LSH and attempt to realize an efficient similarity search of trajectories; however, those methods are inefficient in terms of search time and memory when applied to massive datasets. To address this problem, we present the \emph{trajectory-indexing succinct trit-array trie (tSTAT)}, which is a scalable method leveraging LSH for trajectory similarity searches. tSTAT quickly performs the search on a tree data structure called \emph{trie}. We also present two novel techniques that enable to dramatically enhance the memory efficiency of tSTAT. One is a node reduction technique that substantially omits redundant trie nodes while maintaining the time performance. The other is a space-efficient representation that leverages the idea behind \emph{succinct data structures} (i.e., a compressed data structure supporting fast data operations). 
We experimentally test tSTAT on its ability to retrieve similar trajectories for a query from large collections of trajectories and show that tSTAT performs superiorly in comparison to state-of-the-art similarity search methods.
\end{abstract}



\maketitle

\newcommand{\Frechet}{Fr\'echet}
\newcommand{\DFD}[1]{\mathrm{Frec}(#1)}
\newcommand{\Ham}[1]{\mathrm{Ham}(#1)}
\newcommand{\Popcnt}[1]{\mathrm{Popcnt}(#1)}
\newcommand{\Database}{\mathcal{P}}
\newcommand{\IdSet}{\mathcal{I}}
\newcommand{\SolutionTP}{{\mathcal{R}}_\mathrm{TP}}
\newcommand{\SolutionT}{\mathcal{R}_\mathrm{T}}
\newcommand{\SolutionP}{\mathcal{R}_\mathrm{P}}
\newcommand{\Nodes}{N}
\newcommand{\NodesTR}{N_{\mathrm{tra}}}
\newcommand{\NodesIN}{N_{\mathrm{in}}}
\newcommand{\NodesOUT}{N_{\mathrm{out}}}
\newcommand{\Dict}{\mathcal{S}}
\newcommand{\Index}{\mathcal{X}}
\newcommand{\Cand}{\mathcal{C}}
\newcommand{\BigO}[1]{\mathrm{O}(#1)}
\newcommand{\BigOdisp}[1]{\mathrm{O}\left(#1\right)}
\newcommand{\SmallO}[1]{\mathrm{o}(#1)}
\newcommand{\SmallOdisp}[1]{\mathrm{o}\left(#1\right)}
\newcommand{\Ceil}[1]{\lceil{#1}\rceil}
\newcommand{\Floor}[1]{\lfloor{#1}\rfloor}
\newcommand{\Tuple}[1]{({#1})}
\newcommand{\Rank}{\mathrm{Rank}}
\newcommand{\Select}{\mathrm{Select}}
\newcommand{\Child}{\mathrm{Child}}
\newcommand{\Children}{\mathrm{Children}}
\newcommand{\Hex}[1]{\textrm{0x#1}}
\newcommand{\Times}[1]{#1$\times$}

\renewcommand{\algorithmicrequire}{\textbf{Input:}}
\renewcommand{\algorithmicensure}{\textbf{Output:}}

\newcommand{\qref}[1]{(\ref{#1})}
\newcommand{\fref}[1]{Figure \ref{#1}}
\newcommand{\ffref}[2]{Figures \ref{#1} and \ref{#2}}
\newcommand{\ftfref}[2]{Figures \ref{#1}--\ref{#2}}
\newcommand{\tref}[1]{Table \ref{#1}}
\newcommand{\ttref}[2]{Tables \ref{#1} and \ref{#2}}
\newcommand{\sref}[1]{Section \ref{#1}}
\newcommand{\srefs}[2]{Sections \ref{#1} and \ref{#2}}
\newcommand{\gref}[1]{Algorithm \ref{#1}}
\newcommand{\pref}[1]{Property \ref{#1}}
\newcommand{\aref}[1]{Appendix \ref{#1}}

\newcommand{\argmin}{\mathrm{arg\,min}}

\algnewcommand{\IfThen}[2]{%
  \State \algorithmicif\ #1\ \algorithmicthen\ #2}

\newcommand{\TrieSearch}{\textsc{TrieSearch}}
\newcommand{\TrieSearchRecur}{\textsc{TrieSearchRecur}}

\newcommand{\LB}{\mathrm{LB}}
\newcommand{\SB}{\mathrm{SB}}
\newcommand{\LUT}{\mathrm{LUT}}

\section{Introduction}

With advances in location-acquisition technology and mobile computing, 
spatial trajectory data representing the mobility of a diversity of moving objects such as people, vehicles, and animals are ubiquitous in research and industry \cite{zheng2015trajectory}. 
For example, a city monitoring system with the global positioning system (GPS) enables us to record a huge number of complex trajectories from vehicles  \cite{datasets:PortoTaxi}.
A camera-based tracking system such as SportVU also enables us to collect a large number of fast and dynamic movements of sports players precisely~\cite{datasets:NBA}.
There is, therefore, a strong and growing demand for developing new powerful methods to make the best use of huge collections of trajectories toward data mining applications such as prediction of a taxi demand in a city in the near future and development of an effective game strategy in sports.
 
Searching for similar trajectories in a large collection for a given query is an example of effective use of huge collections of trajectories, and it has a wide variety of applications from route recommendation \cite{shang2014personalized,luo2013finding} to sports play retrieval and clustering \cite{10.1145/3054132,sha2016chalkboarding}. 
Several distance measures for trajectories such as \Frechet{} distance, Hausdorff distance, dynamic time warping (DTW), and longest common subsequence (LCSS) \cite{toohey2015trajectory} have been proposed thus far.
\Frechet{} distance is the \emph{defacto} standard measure \cite{alt1995computing} 
to evaluate the similarity of trajectories. 
\Frechet{} distance can be intuitively explained as the length of the shortest leash enabling 
a person to walk a dog such that they move along each of their own trajectories at their own speed without going back.
\Frechet{} distance has several advantages compared to other distance measures.
For example, \Frechet{} distance captures the directions of trajectories in contrast to Hausdorff distance, and it satisfies the distance conditions in contrast to DTW and LCSS.
Owing to those advantages, \Frechet{} distance has been successfully applied in various applications including detection of commuting patterns \cite{buchin2011detecting}, handwriting recognition \cite{sriraghavendra2007frechet}, protein structure alignment \cite{wylie2013protein}, and many other applications \cite{campbell2015clustering,konzack2017visual,zhu2010mining}.
In addition, an algorithm competition focusing on similarity searches of trajectories using \Frechet{} distance was held in ACM SIGSPATIAL Cup 2017 \cite{werner2018acm}.
While these applications of \Frechet{} distance show its practicality, 
similarity search of trajectories using \Frechet{} distance is computationally demanding because the computation time of the \Frechet{} distance between a pair of trajectories is quadratic to their length \cite{bringmann2014walking},
which limits large-scale applications using \Frechet{} distance in practice.

\emph{Locality sensitive hashing (LSH) for \Frechet{} distance} has been proposed for scalable similarity searches of trajectories using \Frechet{} distance. 
The first LSH method proposed by Indyk \cite{indyk2002approximate} approximates \Frechet{} distance via a product metric.
Recently, more efficient LSH methods with tighter approximation bounds have been proposed \cite{driemel2017locality,astefanoaei2018multi,ceccarello2019fresh}, which map trajectories into non-negative integer vectors (called \emph{sketches}) such that \Frechet{} distance is preserved as the Hamming distance among sketches.
LSH for \Frechet{} distance has been successfully applied to trajectory clustering \cite{sanchez2016fast,rayatidamavandi2017comparison}.

\emph{FRESH} \cite{ceccarello2019fresh} is the state-of-the-art method applying LSH for \Frechet{} distance to similarity searches of trajectories in practice.
FRESH uses an inverted index implemented by the hash table data structure, which stores values associated with each key. 
For a collection of sketches of fixed length $L$, FRESH builds $L$ hash tables whose key is an integer at each position of sketches and value is the set of sketch identifiers with the integer.
Given a query sketch of the same length $L$, FRESH retrieves the hash table with the integer at each position of the query and computes a set of sketch identifiers with the integer at each position, resulting in $L$ sets of sketch identifiers in total for $L$ positions of the query. 
The Hamming distances between each sketch in the set and the query are computed by sequentially counting the number of appearances of each integer in these sets. 
However, this approach suffers from performance degradation caused by the sequential counting of integers 
if large sets of identifiers are computed. 
In addition, FRESH consumes a large amount of memory for indexing large collections of trajectories because of the memory inefficiency of hash tables, which limits large-scale applications in practice. 
Thus, an important open challenge is to develop a memory-efficient data structure for fast similarity searches of sketches for trajectories. 
 
\emph{Trie} \cite{fredkin1960trie} is an ordered labeled tree data structure for a set of strings and supports various string operations such as string search and prefix search with a wide variety of applications in string processing.
Examples are string dictionaries \cite{kanda2017compressed}, $n$-gram language models \cite{pibiri2017efficient,pibiri2019handling}, and range query filtering \cite{zhang2018surf}.
A typical pointer-based representation of trie consumes a large amount of memory. 
Thus, recent researches have focused on space-efficient representations of trie (e.g., \cite{kanda2017compressed,zhang2018surf,pibiri2017efficient,pibiri2019handling}).
To date, trie has been applied only to the limited application domains listed above.
However, as we will see, there remains great potential for a wide variety of applications related to trajectories.

\paragraph{Our Contribution}

We present a novel method called \emph{trajectory-indexing succinct trit-array trie (tSTAT)} that efficiently performs a similarity search for trajectories with LSH for \Frechet{} distance.
As in FRESH, multiple data structures according to the sketch length are built for efficient query retrievals. 
However, in contrast to FRESH, tSTAT enables faster similarity searches by effectively partitioning sketches into 
several blocks by the \emph{multi-index approach} \cite{greene1994multi} and building the trie data structure for each block. 
tSTAT's query retrieval is performed by an efficient trie traversal that bypasses the sequential counting of integers for computing  Hamming distance. 
While preserving fast similarity searches, tSTAT successfully reduces the memory usage by presenting 
two novel techniques of \emph{node reduction} and a space-efficient representation of trie by leveraging 
\emph{succinct data structures} (i.e., compressed representations of data structures while supporting various data operations in the compressed format) \cite{jacobson1989space, patrascu2008succincter}. 

We experimentally test tSTAT on its ability to retrieve similar trajectories for a query from large collections of real-world trajectories.
The performance comparison with state-of-the art similarity search methods for trajectories demonstrates that 
tSTAT performs superiorly with respect to time and memory efficiency. 

\begin{table*}[tb]
\centering
\caption{
Summary of similarity search methods, where $n$ is the number of trajectories (or sketches) in a collection,
$L$ is the fixed length of sketches,
$\sigma$ is the alphabet size,
$B$ is the number of blocks in the multi-index $(B \leq L)$,
$K$ is the Hamming distance threshold, and
$\NodesIN$ is the number of internal nodes in tSTAT.
$C_\mathrm{fresh}$, $C_\mathrm{hms}$ and $C_\mathrm{tstat}$ are the verification times for candidates obtained in FRESH, HmSearch and tSTAT, respectively.
The search times are obtained assuming that sketches are uniformly distributed.
}
\label{tab:complexity}
\begin{tabular}{|c||c|c|c|} \hline
{\bf Method} & {\bf Data structure} & {\bf Memory usage (bits)} & {\bf Search time} \\ \hline\hline
FRESH \cite{ceccarello2019fresh} & Hash table & $\BigO{Ln \log n + Ln \log \sigma}$ & $\BigO{L \cdot \max(1, n/\sigma)} + C_\mathrm{fresh}$ \\ \hline
HmSearch \cite{zhang2013hmsearch} & Hash table & $\BigO{Ln \log n + Ln \log \sigma}$ & $\BigO{L \cdot \max(1,Ln/\sigma^{L-{2L}/{K}+1})} + C_\mathrm{hms}$  \\ \hline
tSTAT (this study) & STAT & $\BigO{\sigma \NodesIN + Bn \log n + Ln \log \sigma}$ & $\BigO{B(L/B)^{K/B+2}} + C_\mathrm{tstat}$ \\ \hline
\end{tabular}
\end{table*}
\section{Problem Setting}
\label{sect:problem}

This section introduces discrete \Frechet{} distance \cite{eiter1994computing} as a similarity measure for trajectories and then presents the similarity search problem on the \Frechet{} distance. 
Although tSTAT is presented with discrete \Frechet{} distance, it is also applicable to continuous \Frechet{} distance \cite{alt1995computing}, as presented in \cite{ceccarello2019fresh}.
In the remainder of this paper \Frechet{} distance is used for discrete \Frechet{} distance.

\subsection{\Frechet{} Distance}

A trajectory $P$ of length $m$ is a sequence of $m$ coordinates $P = (p_1,p_2,\dots,p_{m})$, where $p_i \in \mathbb{R}^d$ is a coordinate with $d$ dimensions.
Let $P = (p_1,p_2,\dots,p_{m_1})$ and $Q = (q_1,q_2,\dots,q_{m_2})$ be two trajectories of lengths $m_1$ and $m_2$, respectively.
A \emph{traversal} $\tau$ is a sequence of coordinate pairs $(p_i,q_j) \in P \times Q$ such that (i) $\tau$ begins with $(p_1,q_1)$ and ends with $(p_{m_1},q_{m_2})$, and (ii) $(p_i,q_j) \in \tau$ can be followed by one of $(p_{i+1},q_j)$, $(p_{i},q_{j+1})$, or $(p_{i+1},q_{j+1})$.
For the set of all possible traversals $\mathcal{T}$ for $P$ and $Q$, the (discrete) \Frechet{} distance between $P$ and $Q$ is defined as 
\[
    \DFD{P,Q} = \min_{\tau \in \mathcal{T}} \max_{(p_i,q_j) \in \tau} \| p_i - q_j \|_2,
\]
where $\|\cdot\|_2$ is the $L_2$ norm. 
The computation time for $\DFD{P,Q}$ is $\BigO{m_1m_2}$ by the dynamic programming \cite{eiter1994computing}.

\subsection{Similarity Search for Trajectories}
\label{subsect:simsearch}

We formulate the similarity search problem for trajectories with LSH for \Frechet{} distance \cite{ceccarello2019fresh}.
Given a collection of $n$ trajectories $\Database = \{P_1,P_2,\dots,P_{n}\}$, 
LSH for \Frechet{} distance projects each $P_i$ in $\Database$ into sketch $S_i$ of fixed length $L$, i.e., 
$S_i=(s_1,s_2,\dots,s_L)$ for each $s_j \in \{0,1,\dots,\sigma-1\}$, where $\sigma$ is the alphabet size.
That is, it produces a collection of $n$ sketches $\Dict = \{S_1,S_2,\dots,S_n\}$.
Given a query sketch $T$ projected from a query trajectory $Q$ and of the same length $L$, the task of similarity search is to retrieve from the collection $\Dict$ all the identifiers of sketches $\IdSet =\{i_1,i_2,\dots,i_k\}$ such that the Hamming distance between $S_{i_j}$ and $T$ is no more than a threshold $K$, i.e., $\Ham{S_{i_j},T} \leq K$ for each $i_j \in \IdSet$,
where $\Ham{\cdot, \cdot}$ denotes the Hamming distance between two sketches (i.e., the number of positions at which the corresponding integers between two sketches are different). We shall call the problem for computing $\IdSet$ the \emph{Hamming distance problem}. 

$\IdSet$ is an approximation set of solutions that can include two types of errors consisting of false positive and false negative: (i) false positive is identifier  $i \in \IdSet$ such that the Hamming distance between $S_{i}$ and $T$ is no more than $K$ (i.e., $\Ham{S_{i}, T} \leq K$), but the \Frechet{} distance between $P_{i}$  and $Q$ is more than threshold $R$ (i.e., $\DFD{P_{i}, Q} > R$) and
(ii) false negative is identifier $i \not\in \IdSet$ such that the Hamming distance between 
$S_{i}$ and $T$ is more than $K$ (i.e., $\Ham{S_{i}, T} > K$), but 
the \Frechet{} distance between $P_{i}$ and $Q$ is no more than $R$ (i.e., $\DFD{P_{i}, Q} \leq R$).

False positives in $\IdSet$ can be filtered out by computing the \Frechet{} distance between $P_{i}$ for each $i \in \IdSet$ and $Q$, resulting in solution set $\IdSet^\prime \subseteq \IdSet$.
False negatives for $\IdSet$ can be reduced by setting $K$ to be large, but such a setting can increase the search time.

\section{Literature Review}
\label{sect:related}

Several efficient data structures have been proposed for similarity search over sketches in the research areas of trajectory data analysis and data mining. 
We briefly review the state-of-the-art methods, which are also summarized in Table~\ref{tab:complexity}.

\subsection{Related Work on Trajectory Data Analysis}
\label{sect:related:traj}

For car trajectories, a large number of similarity search methods have been proposed 
and most of them employ map matching that projects such trajectories into a sequence of roads~\cite{wang2018torch,song2014press,yuan2019distributed,krogh2016efficient}.
While the applicability of the map matching is limited to car trajectories, 
there are important similarity searches for several other types of trajectories.

There are several existing methods for trying to compute exact solutions for trajectory similarity searches. 
In ACM SIGSPATIAL Cup 2017 \cite{werner2018acm}, similarity search methods for \Frechet{} distance were developed \cite{baldus2017fast,buchin2017efficient,dutsch2017filter}.
Recently, Bringmann et al. \cite{bringmann2019walking} improved the winning algorithm \cite{baldus2017fast}.
In another line of research, several similarity search methods for distributed query processing were proposed \cite{xie2017distributed,shang2018dita}.
Those methods commonly build spatial indexes (e.g., R-trees or KD-trees) using spatial points and minimum bounding rectangles of trajectories and 
solve similarity searches with spatial indexes. 
However, owing to the computational demand of \Frechet{} distance, designing an efficient data structure for its fast similarity searches is difficult. 
Thus, recent researches have focused on approximate similarity searches using LSH for \Frechet{} distance because the LSH enables us to solve the trajectory similarity search problem as the Hamming distance problem for sketches.

The first data structure for approximate nearest neighbor searches for trajectories was proposed by Indyk \cite{indyk1998approximate}. 
While the search time is $\BigO{m^{\BigO{1}} \log n}$ for the number of trajectories $n$ and the maximum length of trajectory $m$, it consumes a large amount of memory in $\BigO{|\mathcal{A}|^{\sqrt{m}}(m^{\sqrt{m}} n)^2}$ words for the size of domain $|\mathcal{A}|$ on which the trajectories are defined, which limits practical applications of the data structure. 
Later, improved data structures \cite{driemel2017locality, astefanoaei2018multi} were presented, and they take $\BigO{n \log n + nm}$ memory words while performing a similarity search in $\BigO{m \log n}$ time. 

FRESH~\cite{ceccarello2019fresh} is a practical data structure for approximate similarity search of trajectories, and it uses $L$ inverted indexes with hash tables. 
The search time of FRESH is $\BigO{L \cdot \max(1, n/\sigma)} + C_\mathrm{fresh}$ assuming that the hash table has a uniform distribution, where $C_\mathrm{fresh}$ is the verification time for candidates.
The memory usage is $\BigO{Ln \log n + Ln \log \sigma}$ bits.

\subsection{Related Work on Data Mining}
\label{sect:related:dm}

In data mining, recent similarity search methods for sketches use the \emph{multi-index} approach~\cite{greene1994multi} as follows.  
The approach partitions sketches in $\Dict$ into several blocks of short sketches and builds multiple inverted indexes from the short sketches in each block.
The similarity search consists of two steps: filtering and verification.
The filtering step roughly obtains candidates of similar sketches by querying the inverted indexes.
The verification step removes false positives from these candidates by computing the Hamming distances.

Several data structures with the multi-index approach were presented specifically for similarity searches on binary sketches (i.e., $\sigma = 2$) \cite{norouzi2014fast,qin2019generalizing,gog2016fast}. 
These methods are not directly applicable to similarity searches on integer sketches (i.e., $\sigma > 2$).
HmSearch \cite{zhang2013hmsearch} is an efficient similarity search with the multi-index approach for integer sketches. 
It generates sketches similar to the ones in $\Dict$ and adds them to $\Dict$, which enables fast similarity searches with hash tables. 
The search time is $\BigO{L \cdot \max(1,Ln/\sigma^{L-{2L}/{K}+1})} + C_\mathrm{hms}$ assuming that the sketches in $\Dict$ are uniformly distributed, where $C_\mathrm{hms}$ is the verification time for candidates.
HmSearch needs a large memory consumption of $\BigO{Ln \log n + Ln \log \sigma}$ bits because of the amplification of the sketches in $\Dict{}$. Although several similarity search methods applicable to integer sketches have been proposed (e.g., \cite{manku2007detecting,li2008efficient,liu2011large}), HmSearch performs the best \cite{zhang2013hmsearch}. 

Despite the importance of a scalable similarity search for massive trajectories, no previous work 
achieving fast search and memory efficiency in trajectory data analysis and data mining exists.
As reviewed in this section, the current best possible methods are (i) to use $L$ inverted indexes with hash tables or (ii) to use the multi-index approach with hash tables.
However, those methods scale poorly due to the large computation cost of the search algorithm and memory inefficiency of the data structure.

\section{Trajectory-indexing succinct trit-array trie}
\label{sect:mstat}

tSTAT proposes to use an efficient approach called multi-index \cite{greene1994multi} for solving the similarity searches in the Hamming space and leverages a trie \cite{fredkin1960trie} for implementing the {multi-index} approach. 
tSTAT efficiently computes Hamming distance by traversing tries. 
A straightforward implementation of tries using pointers is not memory efficient for indexing a large collection of sketches. Instead, we present a novel node reduction technique and memory-efficient representation of trie called \emph{succinct trit-array trie (STAT)}.  

\subsection{Multi-Index Approach}
\label{sect:mstat:mi}

\begin{figure}[tb]
\centering
\includegraphics[scale=0.25]{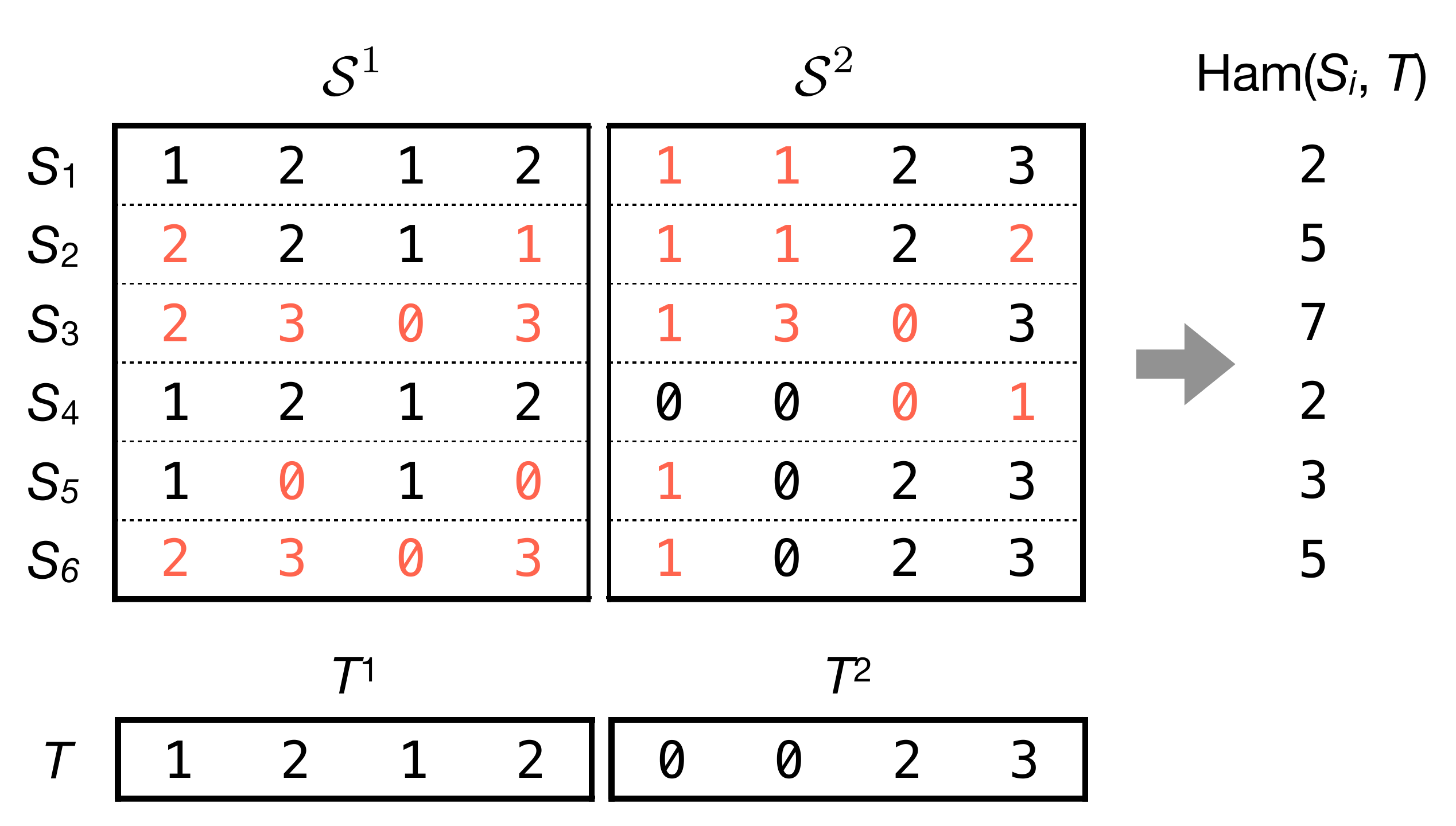}
\caption{
Collection of six sketches $\Dict$ (respectively, query $T$) of $L = 8$ and $\sigma = 4$, partitioned into two blocks $S^1$ and $S^2$ (respectively, $T^1$ and $T^2$). Each integer in $\Dict$ mismatched to the corresponding integer in $T$ are colored in red.
}
\label{fig:collection}
\end{figure}

The {multi-index approach} builds multiple indexes from collection $\Dict$ for solving the similarity search problem of sketches on the Hamming space.
Each sketch $S_i \in \Dict$ is partitioned into $B$ disjoint blocks $S^1_i, S^2_i, \dots, S^B_i$ of fixed length $L/B$.
For convenience, we assume that $L$ is divisible by $B$.
The $j$-th block of $\Dict$ for $j=1,2,\dots,B$ is denoted by $\Dict^j = \{S^j_1, S^j_2, \dots, S^j_n\}$.
\fref{fig:collection} shows an example collection of six sketches $\Dict$, and 
$\Dict$ is partitioned into two blocks $\Dict^1$ and $\Dict^2$.

For the similarity search for query $T$, we first partition $T$ into $B$ blocks $T^1, T^2, \dots, T^B$ in the same manner. 
We then assign smaller Hamming distance thresholds $K^1,K^2,\dots,K^B$ to each block, which is detailed later.
We obtain the candidate sets of sketch ids $\Cand^j = \{i \mid \Ham{S^j_i,T^j} \leq K^j\}$ for each block $j = 1,2,\dots,B$ and compute the union of the candidate sets as $\Cand = \Cand^1 \cup \Cand^2 \cup \dots \cup \Cand^B$.
Finally, we remove false positives  by computing $\Ham{S_i,T}$ for each $i \in \Cand$, resulting in solution set $\IdSet$.

The threshold assignment requires $K^j$ such that each solution in $\IdSet$ is included in $\Cand^j$ for any $j=1,2,\dots,B$.  Such assignment is satisfied by setting $K^j$ such that $K^1 + K^2 + \dots + K^B = K - B + 1$, which is ensured by the general pigeonhole principle \cite{qin2019generalizing}.

For the similarity search with $K=3$ in \fref{fig:collection}, 
we assign $K^1=1$ and $K^2=1$ such that $K^1+K^2=K-B+1 = 2$ based on the general pigeonhole principle.
Then, $\Cand^1 = \{ 1,4 \}$ and $\Cand^2 = \{ 5,6 \}$ are obtained.
Since $\IdSet = \{1,4,5\}$ is a subset of $\Cand = \{ 1,4,5,6 \}$, we can obtain $\IdSet$ by verifying each element in $\Cand$.

\subsection{Trie Implementing {Multi-Index}}
\label{sect:mstat:trie}

\begin{figure}[tb]
\centering
\subfloat[Trie index $\Index^1$]{
\includegraphics[scale=0.22]{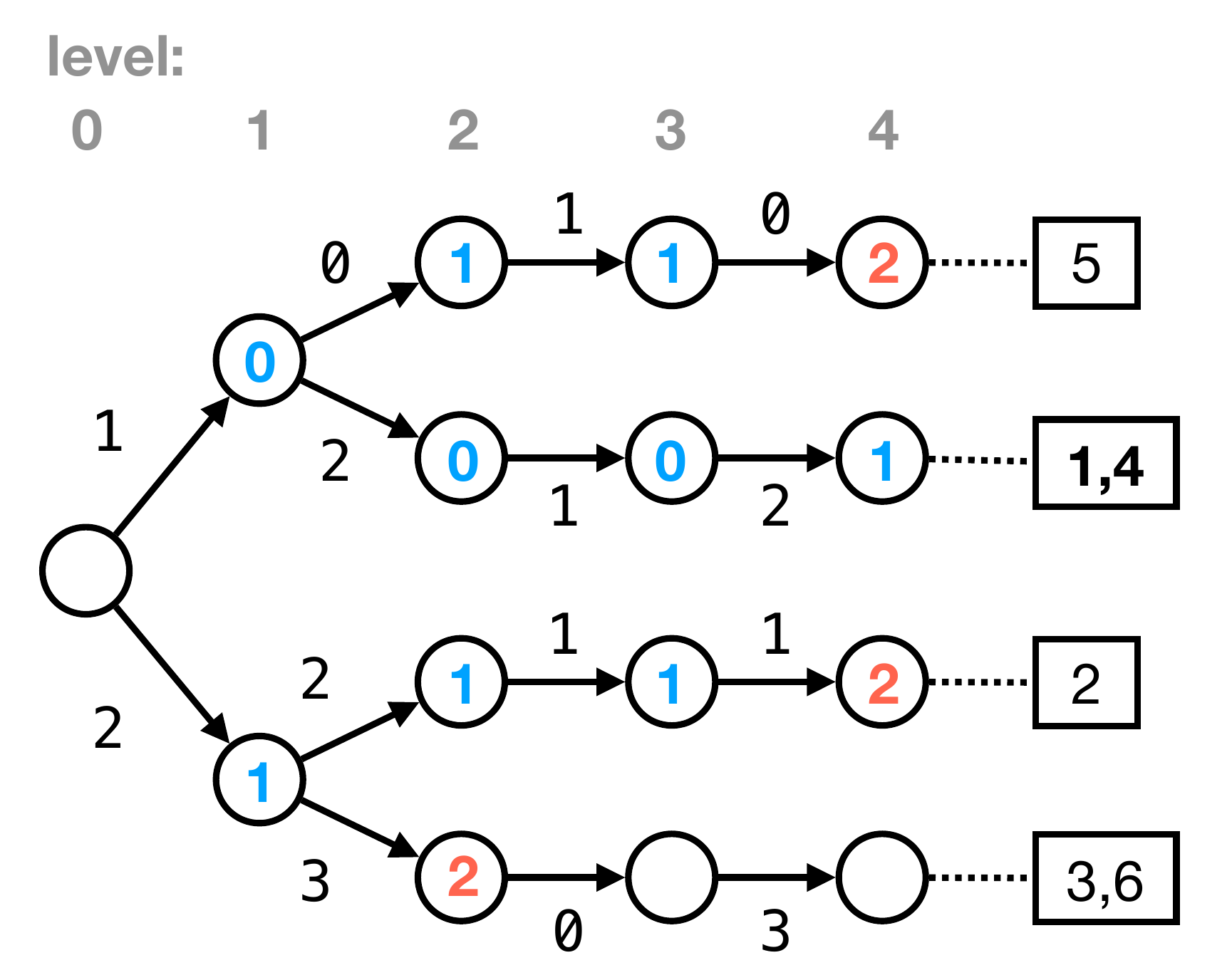}
\label{fig:trie:1}
}
\subfloat[Trie index $\Index^2$]{
\includegraphics[scale=0.22]{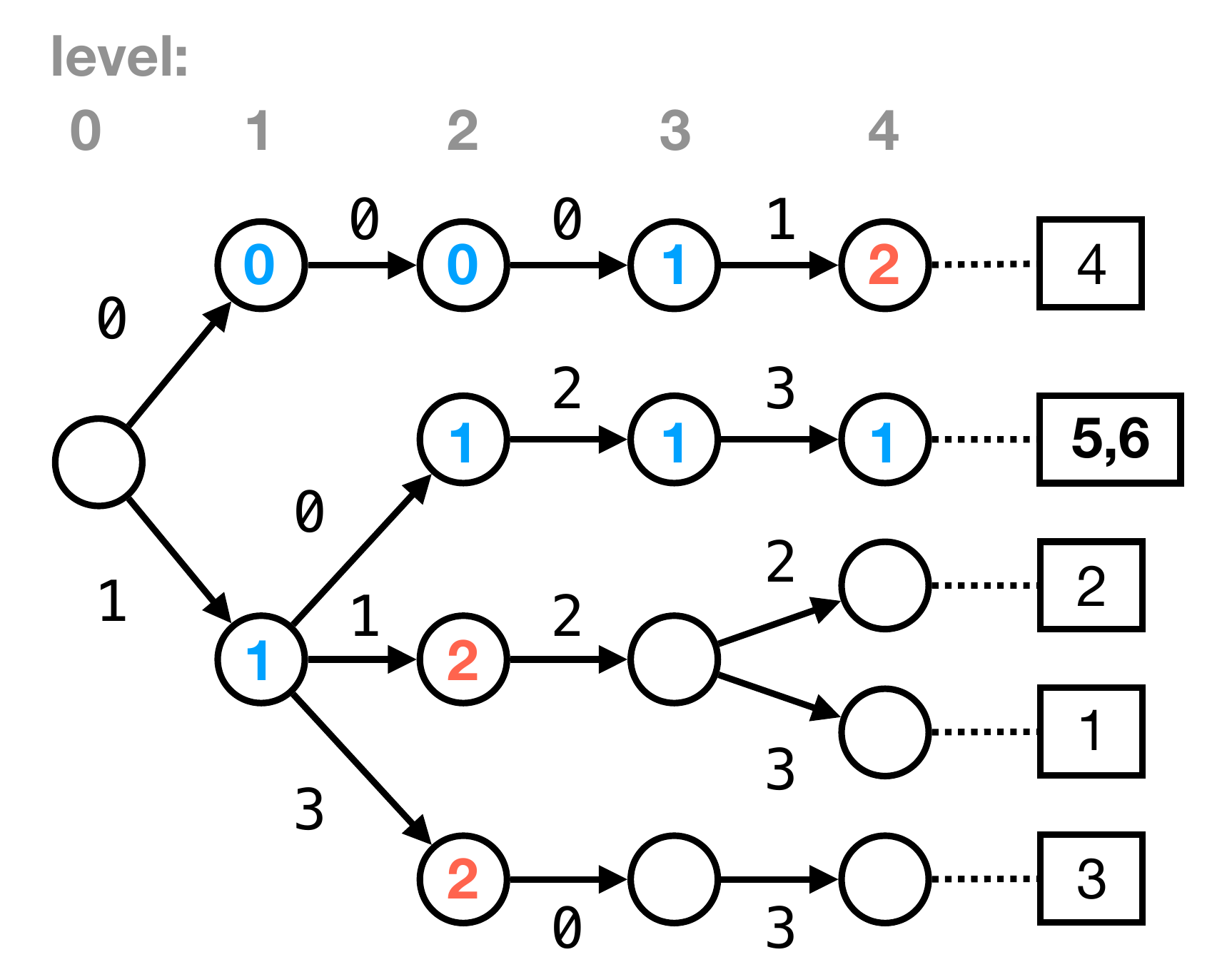}
\label{fig:trie:2}
}
\caption{
Tries $\Index^1$ and $\Index^2$ built from $\Dict^1$ and $\Dict^2$ of \fref{fig:collection}, respectively.
In each traversed node $u$, $dist_u$ to $T$ of \fref{fig:collection} is denoted by red/blue numbers when $K^1 = 1$ and $K^2 = 1$.
}
\label{fig:trie}
\end{figure}

Trie $\Index^j$ is an edge-labeled tree indexing sketches in $\Dict^j$ (see \fref{fig:trie}).
Each node in $\Index^j$ is associated with the common prefix of a subset of sketches in $\Dict^j$, and the root (i.e., node without a parent) is not associated with any prefix. 
Each leaf (i.e., node without any children) is associated with input sketches of the same integers and has the list of their sketch ids.
All outgoing edges of a node are labeled with distinct integers.
For a node $u$ in $\Index^j$, the number of edges from the root to $u$ is the {level} of $u$.
\fref{fig:trie} shows an example of $\Index^1$ and $\Index^2$ built from $\Dict^1$ and $\Dict^2$ in \fref{fig:collection}, respectively.

Searching similar sketches for query sketch $T^j$ and threshold $K^j$ is performed 
by traversing trie $\Index^j$ in a depth-first manner. 
We start to traverse nodes from the root at level 0.
At each level $\ell$, we compare the $\ell$-th integer of $T^j$ with labels associated with outgoing edges from nodes at level $\ell$.
For each node $u$ at level $\ell$, we compute the Hamming distance $dist_u$ between the sequence of edge labels from the root to $u$ and the subsequence from the first position to the $\ell$-th position in $T^j$.
If $dist_u$ becomes more than $K^j$ at each reached node $u$, 
we can safely stop traversing down to all the descendants under node $u$.
The candidate set $\Cand^j$ can be obtained by accessing the list of sketch ids at each reached leaf $v$, which means $dist_v \leq K^j$.

In \fref{fig:trie}, $dist_u$ for each node $u$ and query $T^1$ or $T^2$ is represented in each node. 
When $K^1=1$ and $K^2=1$, we stop traversing at each node including a red number. 
We obtain candidates $\Cand^1 = \{ 1,4 \}$ and $\Cand^2 = \{ 5,6 \}$, associated with the reached leaves.

The algorithm can prune unnecessary portions of the search space depending on $K^j$.
The number of traversed nodes in $\Index^j$ is bounded by $\BigO{(L/B)^{K^j + 2}}$ when assuming the complete $\sigma$-ary trie \cite{arslan2002dictionary}.
When we assign $K^j$ in a round robin manner, it holds $K^j \leq K / B$;
thus, the algorithm takes $\BigO{(L/B)^{K/B + 2} + |\Cand^j|}$ time for $\Index^j$.

When we represent trie $\Index^j$ by using pointers straightforwardly, the scalability becomes a critical issue since the memory usage is $\BigO{\Nodes^j \log (\sigma\Nodes^j)}$ bits.
In the remaining subsections, we present two techniques of node reduction and STAT data structure for compactly representing  $\Index^j$.

\subsection{Node Reduction}
\label{sect:mstat:node}
A crucial observation for candidate set $\Cand$ is that it allows false positives for solution set $\IdSet$ for the Hamming distance problem.  
This observation enables us to make tries more memory efficient by removing their redundant nodes. 

The \emph{weight} of node $u$ is defined as the total number of sketch ids associated with leaves of the substree with root $u$ and 
satisfies an \emph{antimonotonicity}, i.e., the weight of node $u$ is no less than the one of $u$'s child. 
The weight of the root in trie $\Index^j$ is $n$. 

Our node reduction is a depth-first traversal leveraging the antimonotonicity. 
The algorithm starts from the root and visits each node. 
If the weight of node $u$ is no more than hyper-parameter $\lambda$, the algorithm eliminates the subtree with root $u$ from the tree. 
Then, $u$ becomes a new leaf of the tree and has the sketch ids of previous leaves associated with the subtree with root $u$.  

\begin{figure}[tb]
\centering
\subfloat[$\Index^2$ with $\lambda = 1$]{
    \includegraphics[scale=0.22]{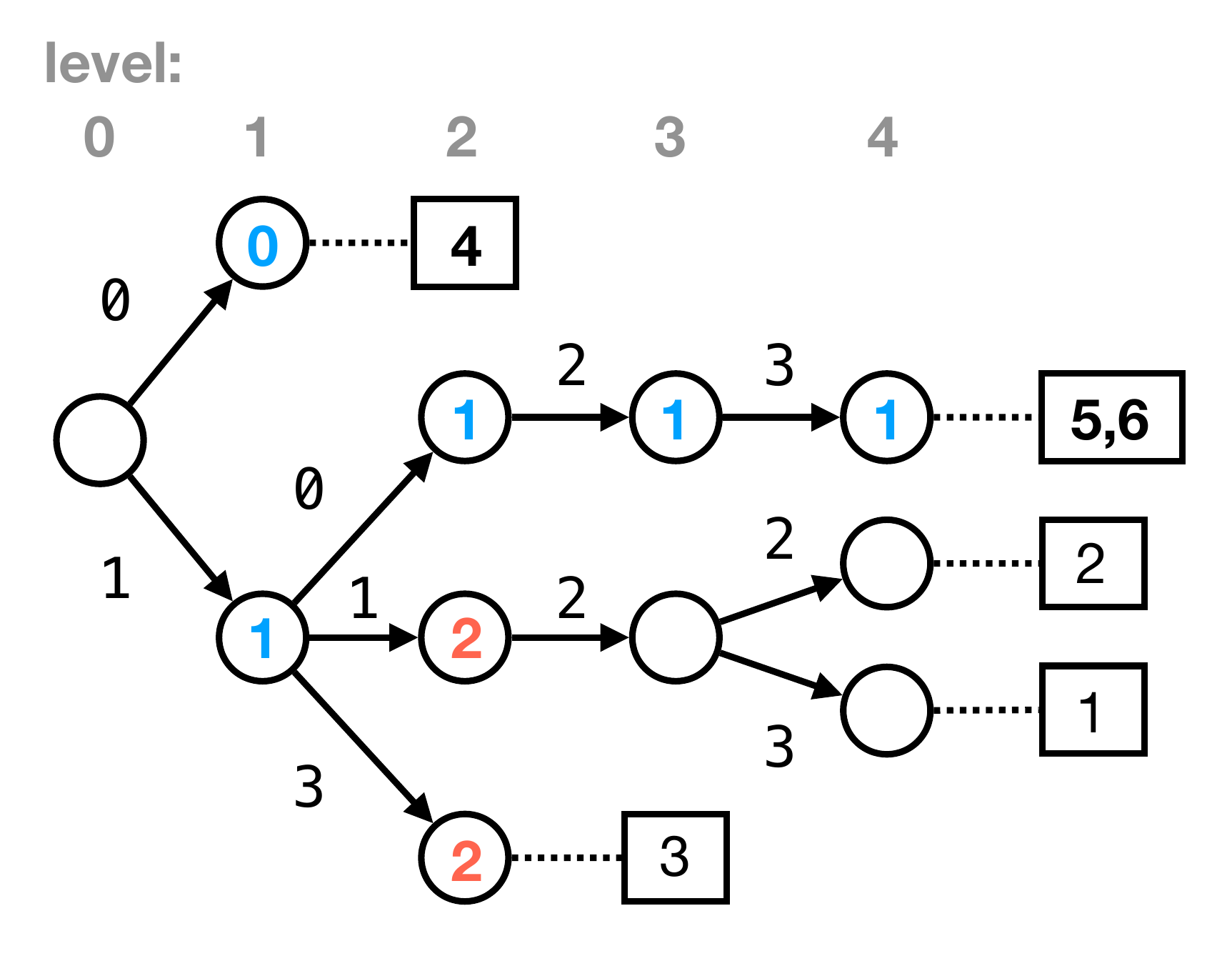}
    \label{fig:reduce:1}
}
\subfloat[$\Index^2$ with $\lambda = 2$]{
    \includegraphics[scale=0.22]{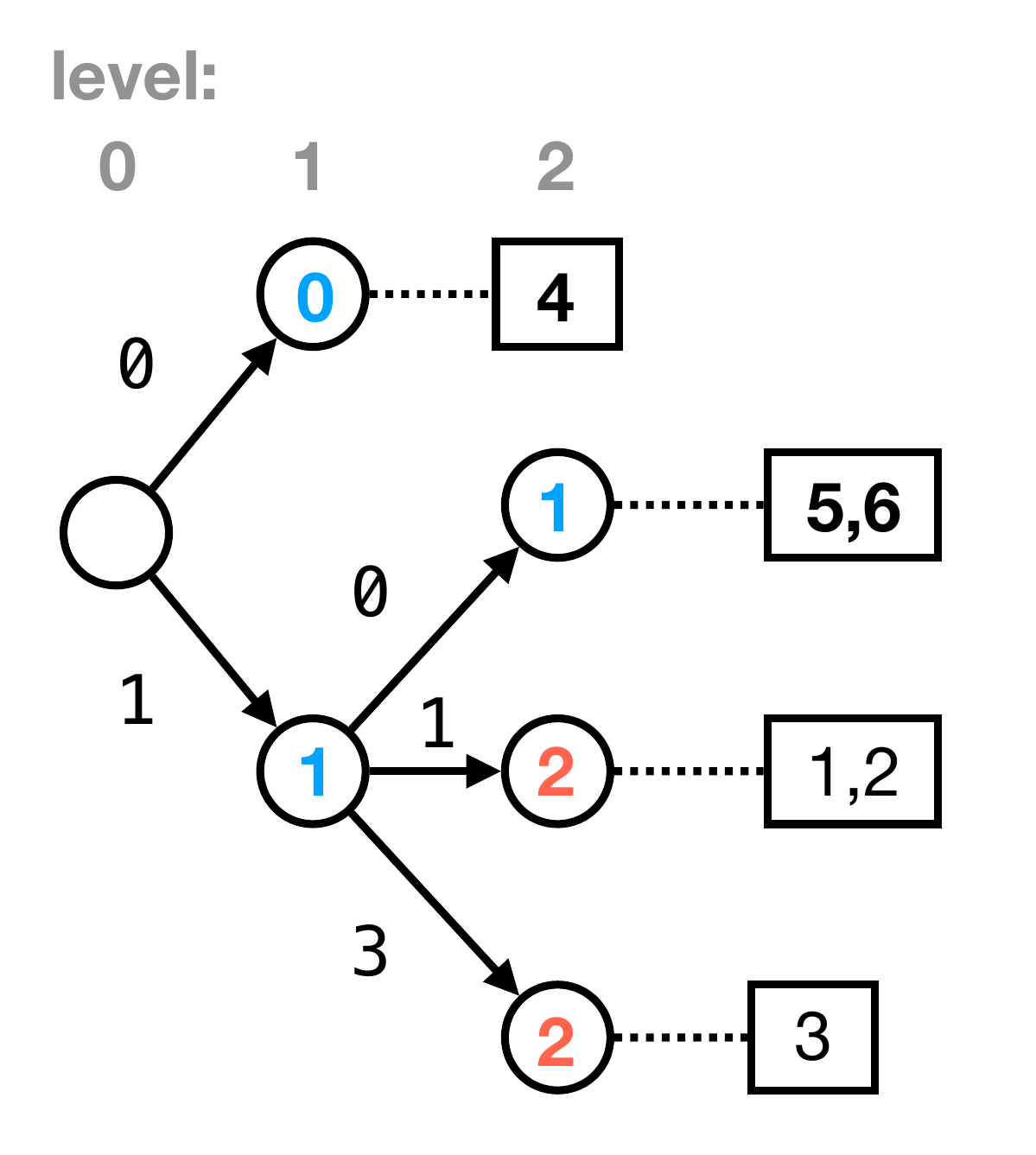}
    \label{fig:reduce:2}
}
\caption{
Trie $\Index^2$ after node reduction in \fref{fig:trie}.
}
\label{fig:reduce}
\end{figure}

\fref{fig:reduce} shows an example of node reductions with $\lambda=1$ and $\lambda=2$ for trie $\Index^2$ in \fref{fig:trie}. 
The trie with 16 nodes in \fref{fig:trie:2} is reduced to the tries with 11 nodes in  \fref{fig:reduce:1} and 6 nodes in \fref{fig:reduce:2} by the node reduction algorithm with $\lambda=1$ and $\lambda=2$, respectively. 

$\lambda$ is a trade-off parameter that can control the balance between the tree traversal time and the verification time for removing false positives. 
The larger $\lambda$ becomes, the smaller the number of nodes becomes and the larger the number of false positives becomes, resulting in a smaller tree traversal time and 
a larger verification time. 
This trade-off is investigated with various $\lambda$ in \sref{sect:ex}.

\subsection{Succinct Rank and Select Data Structures}

STAT is an integer array representation of trie leveraging succinct Rank and Select data structures \cite{jacobson1989space}.
Given an integer array $A$ of length $M$, the data structures support the following operations in compressed space:
\begin{itemize}
    \item $\Rank_c(A,i)$ returns the number of occurrences of integer $c$ between positions $0$ and $i-1$ on $A$.
    \item $\Select_c(A,i)$ returns the position in $A$ of the $i$-th occurrence of integer $c$; if $i$ exceeds the number of $c$'s in $A$, it always returns $M$.
\end{itemize}
For $A = (0,1,0,2,1,1,2,0)$, $\Rank_0(A,3)=2$ and $\Select_2(A,1)=6$.

For a bit array $A$ consisting of integers from $\{0, 1\}$, $A$ can be stored in $M + \SmallO{M}$ bits of space while supporting the operations in $\BigO{1}$ time \cite{jacobson1989space,vigna2008broadword}.
For a trit array $A$ consisting of integers from $\{0, 1, 2\}$, $A$ can be stored in $M \log_2 3 + \SmallO{M}$ bits of space while supporting the operations in $\BigO{1}$ time \cite{patrascu2008succincter,fischer2016glouds}.

\subsection{STAT Data Structure}
\label{sect:mstat:stat}

A basic idea behind the STAT data structure is to represent a trie using three arrays each of which represents children of a node, the number of sketch ids associated with leaves, and the sketch ids at each level. 
While the array representing children of a node consists of integers from $\{0,1,2\}$ and is indexed by the succinct rank and select data structure for trits, the array representing the number of sketch ids associated with leaves consists of integers from $\{0,1\}$ and 
is indexed by the succinct rank and select data structure for bit arrays.
Detailed explanation of the STAT data structure is subsequently presented.

\begin{figure}[tb]
\centering
\includegraphics[scale=0.28]{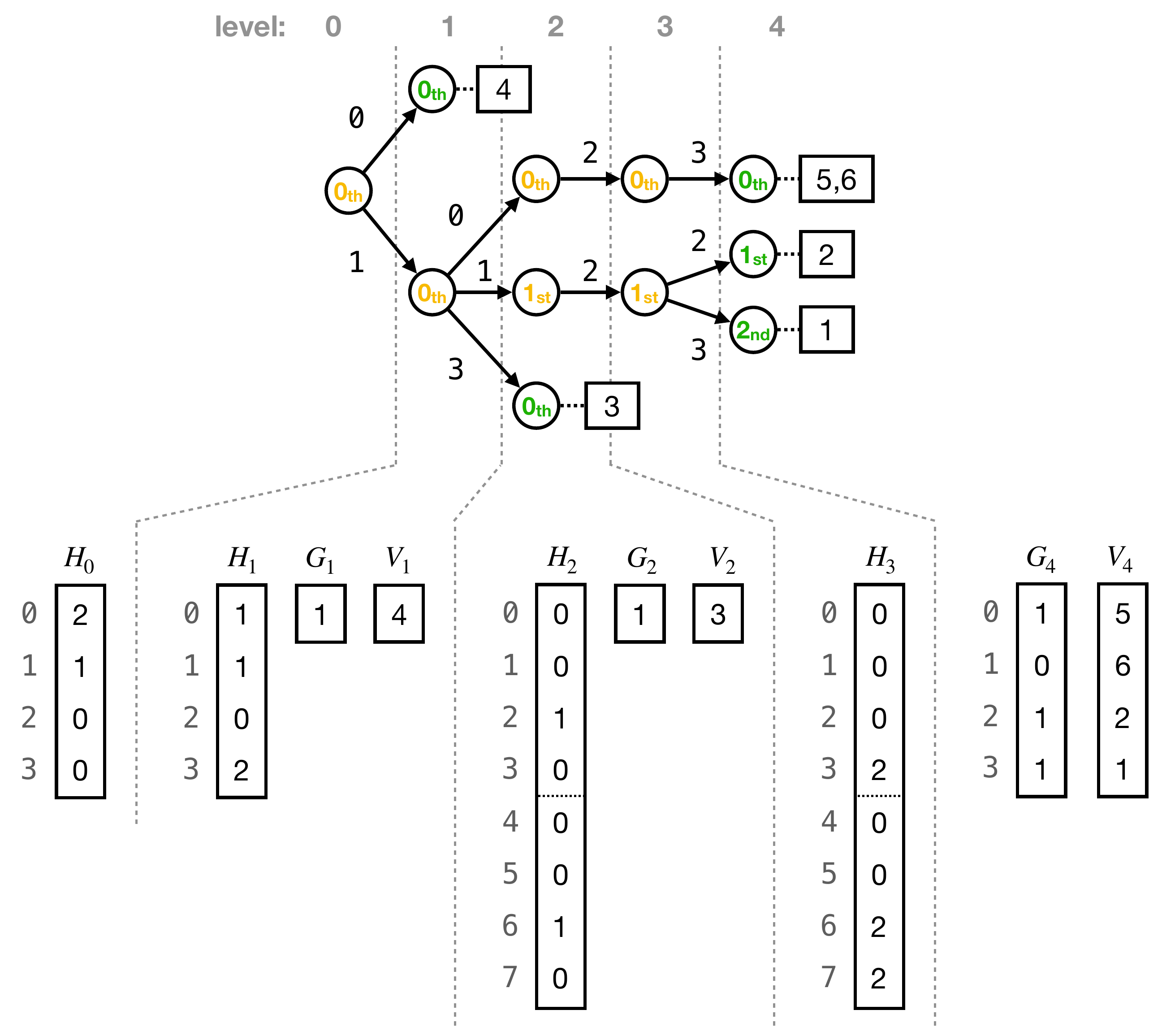}
\caption{
STAT for trie  $\Index^2$ in \fref{fig:reduce:1}.
The orders of internal nodes and leaves at each level are denoted by yellow and green numbers, respectively.
}
\label{fig:stat}
\end{figure}

STAT consists of three arrays $H_\ell$, $G_\ell$, and $V_\ell$ for compactly representing nodes at level $\ell$ of trie $\Index^j$ (see \fref{fig:stat}).
To define the arrays, we introduce the order of internal nodes and leaves to each level of the trie as follows.
We traverse $\Index^j$ in a breadth-first order.
The $i$-th internal node at a level is the $i$-th encountered node for internal nodes at the level.
Similarly, the $i$-th leaf at a level is the $i$-th encountered node for leaves at the level.
Ids starting at zero are assigned to nodes.
For example, level 2 of $\Index^2$ in \fref{fig:stat} consists of the zeroth internal node, the first internal node, and the zeroth leaf from top to bottom.

$H_\ell$ is a trit (i.e., ternary digit) array and represents children of internal nodes at level $\ell$.
The children of an internal node is represented in a trit array of length $\sigma$, and $H_\ell$ is constructed by concatenating such trit arrays in the order of internal nodes.
That is, the subarray of $H_\ell$ between positions $i \sigma$ and $(i+1)  \sigma - 1$ corresponds to the $i$-th internal node at level $\ell$.
The data structure is as follows:
\begin{itemize}
\item $H_\ell[i \sigma + c] = 0$ if the $i$-th internal node at level $\ell$ does not have a child indicated with edge label $c$ in the trie,
\item $H_\ell[i \sigma + c] = 1$ if the $i$-th internal node at level $\ell$ has a child indicated with edge label $c$ as an internal node, and
\item $H_\ell[i \sigma + c] = 2$ if the $i$-th internal node at level $\ell$ has a child indicated with edge label $c$ as a leaf.
\end{itemize}
$H_\ell$ is implemented by the Rank data structure for trits.
Let us consider the zeroth internal node at level 1 in \fref{fig:stat}.
$H_1[0]=1$ and $H_1[1]=1$ because it has two children as internal nodes indicated with edge labels 0 and 1;
$H_1[2]=0$ because it does not have a child indicated with edge label 2; and
$H_1[3]=2$ because it has a child as a leaf indicated with edge label 3.

$G_\ell$ is a bit array representing the numbers of sketch ids associated with leaves at level $\ell$ in unary encoding.
That is, for $g$ sketch ids associated with a leaf, the number $g$ is encoded into $(g-1)$ $0$s following $1$.
$G_\ell$ is constructed by concatenating the unary codes from the zeroth leaf at level $\ell$ and 
is implemented by the Select data structure for bits \cite{jacobson1989space}.
In \fref{fig:stat}, $G_4=(1,0,1,1)$ because the zeroth, first, and second leaves at level 4 have two, one, and one sketch ids, respectively.

$V_\ell$ is an array of sketch ids associated with leaves at level $\ell$ and is constructed by concatenating the sketch ids from the zeroth leaf at level $\ell$.
In \fref{fig:stat}, $V_4=(5,6,2,1)$ because the zeroth, first, and second leaves at level 4 have sketch ids $(5,6)$, $(2)$, and $(1)$, respectively.

We present operations for computing children for a given internal node and sketch ids associated with a given leaf. 
Those operations are necessary for computing Hamming distances for a query on trie $\Index^j$.
Given the $i$-th internal node at level $\ell$ and integer $c$, there is not a child indicated with edge label $c$ if $H_\ell[i \sigma + c] = 0$;
If $H_\ell[i \sigma + c] = 1$, the child is the $i'$-th internal node at level $\ell + 1$, where $i' = \Rank_1(H_\ell, i \sigma + c)$;
If $H_\ell[i \sigma + c] = 2$, the child is the $i'$-th leaf node at level $\ell + 1$, where $i' = \Rank_2(H_\ell, i \sigma + c)$.
Given the $i$-th leaf node at level $\ell$, the associated sketch ids are the elements between $s$ and $e$ in $V_\ell$, where $s=\Select_1(G_\ell,i)$ and $e=\Select_1(G_\ell,i+1)-1$.

In \fref{fig:stat}, for the zeroth internal node at level 1, the child indicated with edge label $1$ is the first internal node at level 2 because $H_1[1] = 1$ and $\Rank_1(H_1, 1)=1$; the child indicated with edge label $3$ is the zeroth leaf at level 2 because $H_1[3] = 2$ and $\Rank_2(H_1, 3)=0$.
For the zeroth leaf at level 4, the associated sketch ids are elements $V_4[0]=5$ and $V_4[1]=6$ because $\Select_1(G_4,1)=0$ and $\Select_1(G_4,1)-1=1$.

\paragraph{Analysis for Memory Efficiency}

Let $\NodesIN^j$ denote the number of internal nodes in trie $\Index^j$.
The total length of $H_\ell$ is $\sigma \NodesIN^j$, and the total memory usage of $H_\ell$ is $\sigma \NodesIN^j \log_2{3} + \SmallO{\sigma \NodesIN^j} = \BigO{\sigma \NodesIN^j}$ bits.
The total length of $G_\ell$ (or $V_\ell$) is $n$.
The total memory usage of $G_\ell$ and $V_\ell$ is $n + \SmallO{n} + n \log_2 n = \BigO{n \log n}$ bits.

The \emph{information-theoretic lower bound (ITLB)} is the defacto standard criteria for investigating the memory efficiency of data structures and is defined as the minimum memory usage for representing tries.  
We analyze the memory efficiency of STAT by comparing the memory usage of STAT with that of ITLB for tries. 

We ignore  $G_\ell$ and $V_\ell$ in the comparison because the arrays are independent from trie data structures.
The memory usage of STAT for a trie with $\NodesIN^j$ internal nodes is $\sigma \NodesIN^j \log_2{3} + \SmallO{\sigma \NodesIN^j}$ bits.
ITLB for a trie with $N^j$ nodes is $N^j (\sigma \log_2 \sigma - (\sigma - 1) \log_2 (\sigma - 1))$ bits \cite{benoit2005representing}.
Roughly, STAT becomes smaller than ITLB if $$\frac{\sigma \log_2 3}{\sigma \log_2 \sigma - (\sigma - 1) \log_2 (\sigma - 1)} < \frac{\Nodes^j}{\NodesIN^j}.$$
For example, when $\sigma = 256$, STAT becomes smaller if $43 < {\Nodes^j}/{\NodesIN^j}$.
The comparison shows that STAT is efficient for representing tries with large $\Nodes^j/\NodesIN^j$.

$\Nodes^j / \NodesIN^j$ is related to the average number of children for each internal node, i.e., the \emph{load factor} of $H_\ell$.
Since nodes at a deeper level in a trie have fewer children, STAT is more effective for representing shallow tries.
Thus, applying node reduction to a trie for eliminating trie nodes at deep levels 
enhances the space efficiency of STAT, as demonstrated in \sref{sect:ex}.

\subsection{Complexity Analysis}
\label{sect:mstat:complexity}

The space complexity of tSTAT is derived as follows.
Let $\NodesIN = \sum^{B}_{j=1} \NodesIN^j$.
$B$ STATs are represented in $\BigO{\sigma \NodesIN + B n \log n}$ bits of space, where the left term is for $H_\ell$ and the right term is for $G_\ell$ and $V_\ell$.
In addition, we need to store collection $\Dict$ in $\BigO{Ln\log \sigma}$ bits of space for verifying $\Cand$.

All the methods in \tref{tab:complexity} require $\BigO{Ln \log \sigma}$ bits of space to store $\Dict$.
In addition to the space, tSTAT uses $\BigO{\sigma \NodesIN + B n \log n}$ bits of space although FRESH and HmSearch use $\BigO{Ln\log n}$ bits of space.
The factor of $\BigO{n \log n}$ is obviously large for a massive collection with large $n$.
tSTAT can relax the large factor to $B/L$.
Although tSTAT needs $\BigO{\sigma \NodesIN}$ bits of space, $\NodesIN$ can be reduced by the node reduction approach.

The time complexity of tSTAT is derived by simply assuming $\lambda = 0$.
The algorithm consists of traversing nodes and verifying candidates of $\Cand$.
The traversal time is $\BigO{B(L/B)^{K/B+2}}$ as presented in \sref{sect:mstat:trie}.
The verification time $C_{\mathrm{tstat}}$ contains the times of removing duplicated candidates in $\Cand^1,\Cand^2,\dots,\Cand^B$ and verifying candidates in $\Cand$.

\section{Experiments}
\label{sect:ex}

\newcommand{\ChartWidth}{41mm}

\begin{table}[tb]
\centering
\caption{Statistics of datasets. Min, Max, Mean, and Median regarding trajectory length are presented.}
\label{tab:dataset}
\begin{tabular}{|l||r|r|r|r|r|} \hline
{\bf Dataset} & {\bf Number} & {\bf Min} & {\bf Max} & {\bf Mean} & {\bf Median} \\ \hline\hline
Taxi & 1,704,769 & 1 & 3,881 & 48.9 & 41 \\ \hline
NBA & 3,288,264 & 1 & 898 & 85.3 & 73 \\ \hline
OSM & 19,113,525 & 1 & 2,000 & 13.3 & 7 \\ \hline
\end{tabular}
\end{table}

We evaluated the performance of tSTAT through real-world trajectory datasets.   
We used three datasets with $d=2$, as shown in \tref{tab:dataset}.
\emph{Taxi} is 1.7 million trajectories of 442 taxis driving in the city of Porto for one year \cite{datasets:PortoTaxi}. 
The GPS locations (i.e., latitude and longitude) were recorded every 15 seconds with mobile data terminals installed in the taxis. 
Each trajectory represents a taxi trip taken by a passenger.
\emph{NBA} is 3.3 million trajectories of 636 games in the 2015/16 NBA seasons \cite{datasets:NBA}. 
Ten player locations were captured every 40 milliseconds by SportVU player tracking system. 
Each trajectory is segmented by stationary points (i.e., small moving distances).
\emph{OSM} is 19 million trajectories of various moving objects including cars, bikes and humans  traced by GPS in Western United States, and 
it is created from OpenStreetMap project \cite{datasets:OSM}. 
OSM is the largest dataset of 19 million trajectories among three datasets and enables us to evaluate  the performance of similarity search methods on a huge dataset. 

For each dataset, we sampled 1,000 trajectories as queries and excluded them from the collection.
We selected \Frechet{} distance thresholds $R$ such that 1, 10, and 100 solutions are found on average per query, respectively, which resulted in $R = 567, 2720, 7263$ for Taxi, $R = 0.15, 0.26, 0.45$ for NBA, and $R = 899, 2506, 7824$ for OSM.
LSH for \Frechet{} distance has two parameters of hash functions $\delta$ and $k$ \cite{ceccarello2019fresh}. 
Following the original paper~\cite{ceccarello2019fresh}, those parameters were set to $\delta = 8dR$ and $k=1$. 

We conducted all experiments on one core of quad-core Intel Xeon CPU E5--2680 v2 clocked at 2.8 GHz in a machine with 256 GB of RAM, running the 64-bit version of CentOS 6.10 based on Linux 2.6.
We implemented all data structures in {C++17} and compiled the source code with g++ (version 7.3.0) in optimization mode -O3.
We implemented succinct Rank and Select data structures on bits using the \emph{succinct data structure library} \cite{gog2014theory}.
As far as we know, there is no any available implementation on trits; therefore, we developed a practical implementation of the data structures for trits.
For a fast computation of Hamming distance on integers, we used an efficient computation algorithm exploiting a bit-parallelism technique \cite{zhang2013hmsearch}.
The implementation details are presented in Appendix \ref{appx:impl}.
The source code used in our experiments is available at \url{https://github.com/kampersanda/frechet_simsearch}.

\subsection{Performance of LSH}
\label{sect:ex:lsh}

We analyzed the performance of LSH for \Frechet{} distance while varying $L$ and $\sigma$.
All the combinations of $L = 32, 64, 128$ and $\sigma = 2^{1}, 2^{8}, 2^{32}$ were tested.
Setting $\sigma$ to $2^{1}$, $2^{8}$ and $2^{32}$ means that the resulting sketches are binary, byte, and four-byte strings, respectively.
For each combination of $L$ and $\sigma$, we computed recalls and precisions by varying Hamming distance threshold $K$ from $0$ to $L/4$ with step width $L/32$.

\begin{figure}[tb]
    \centering
    \setlength{\tabcolsep}{0mm}
    \subfloat[Taxi]{
        \begin{tabular}{c}
        \includegraphics[width=\ChartWidth]{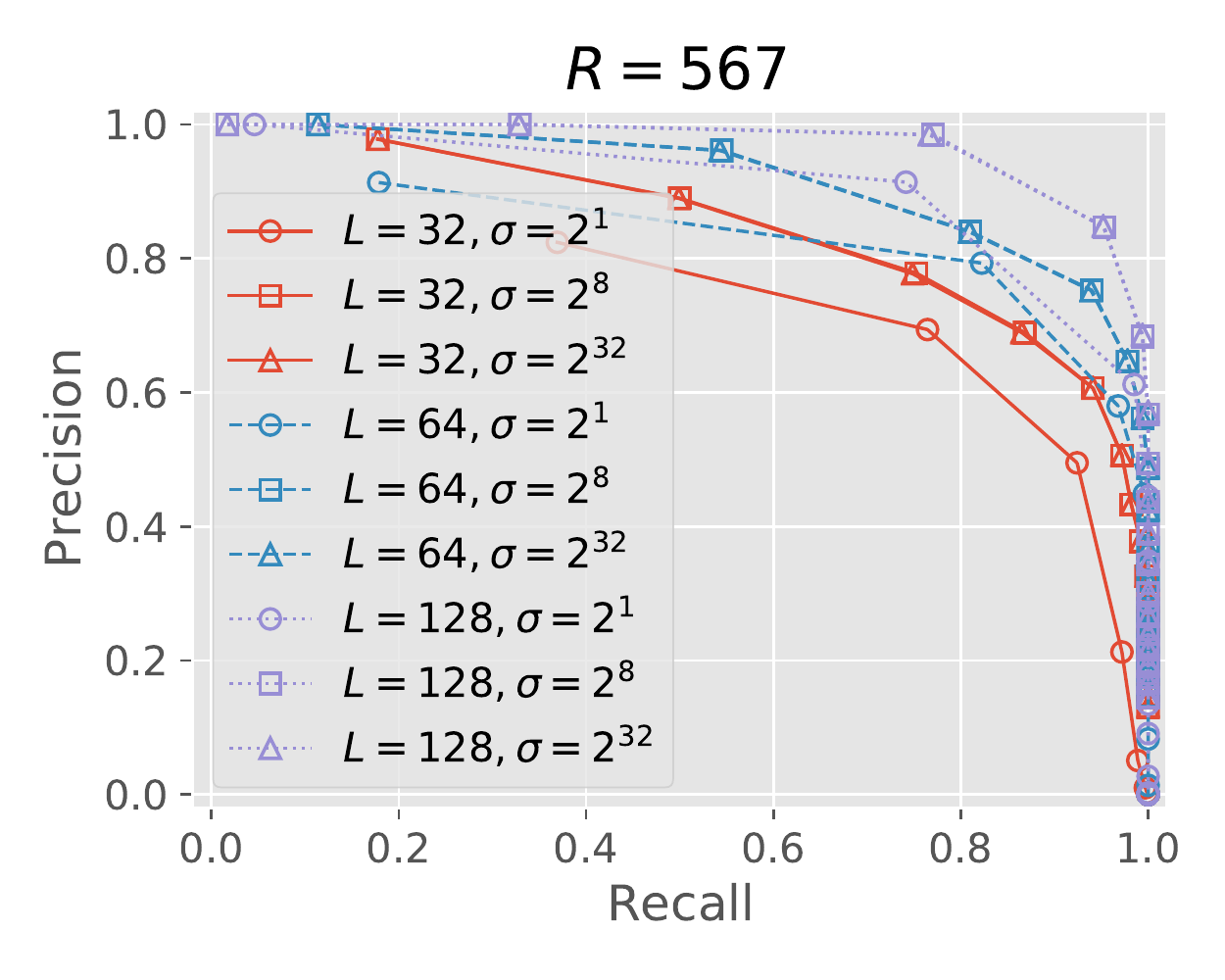}\\
        \includegraphics[width=\ChartWidth]{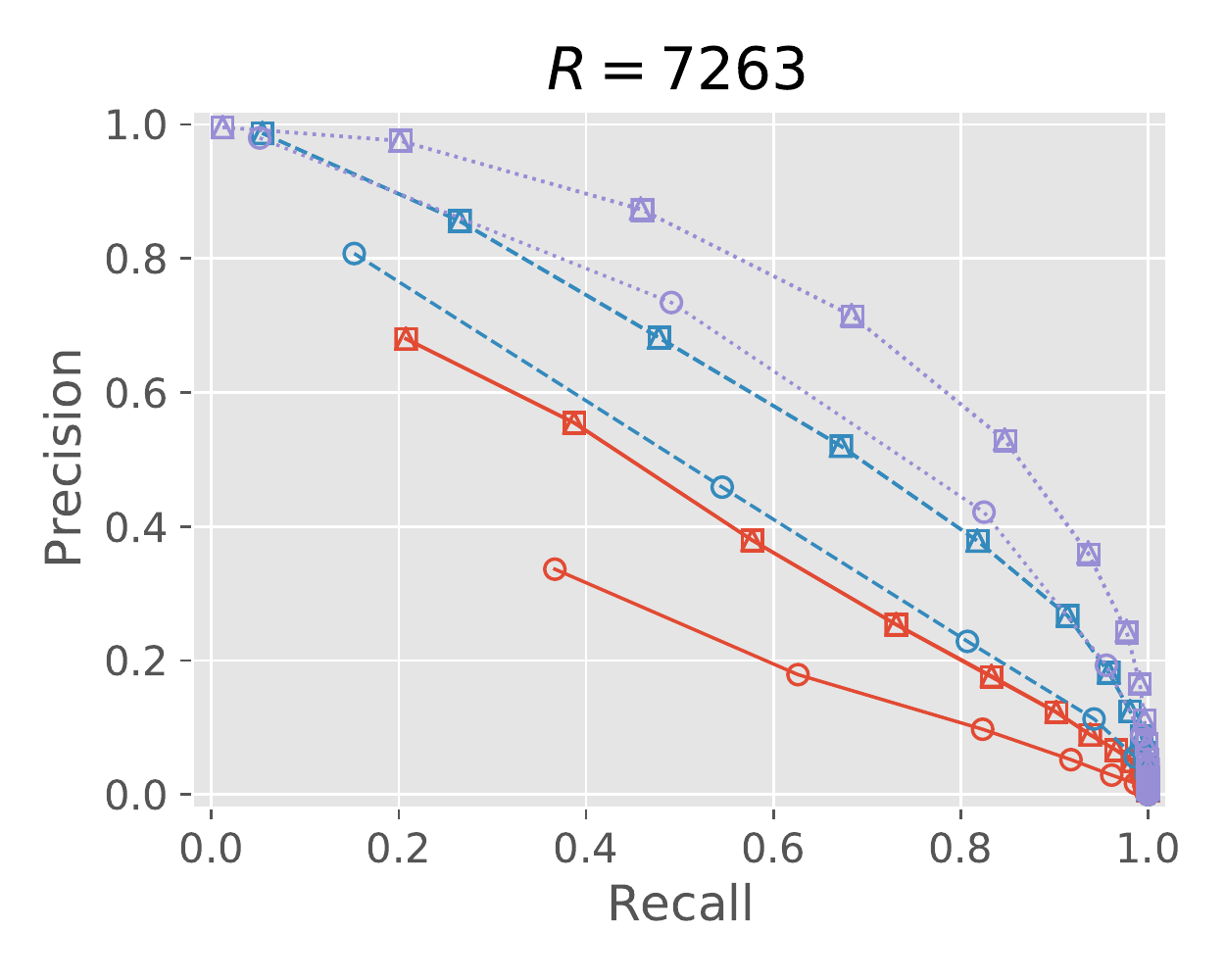}
        \end{tabular}
        \label{charts:LSH:Taxi}
    }
    \subfloat[NBA]{
        \begin{tabular}{c}
        \includegraphics[width=\ChartWidth]{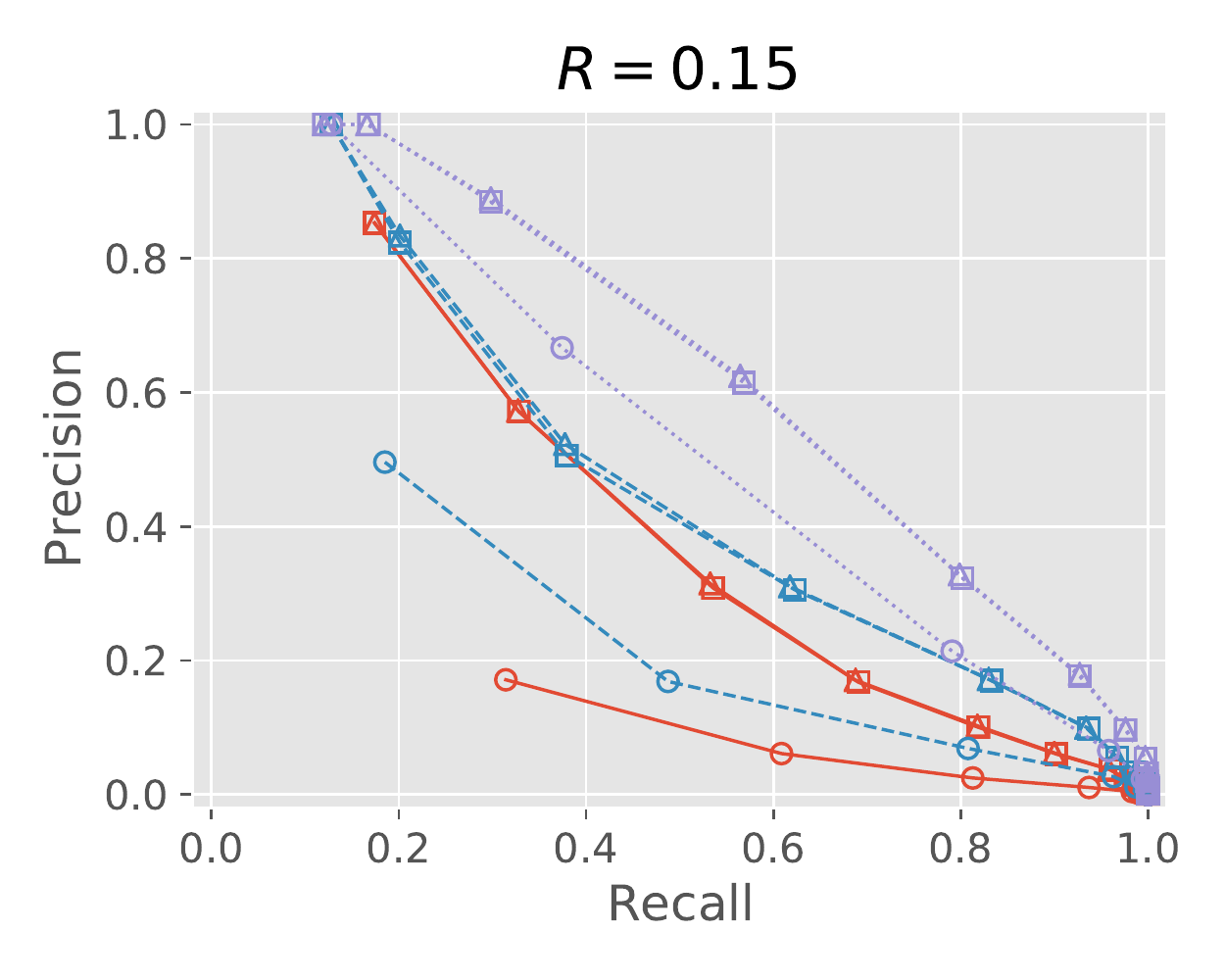}\\
        \includegraphics[width=\ChartWidth]{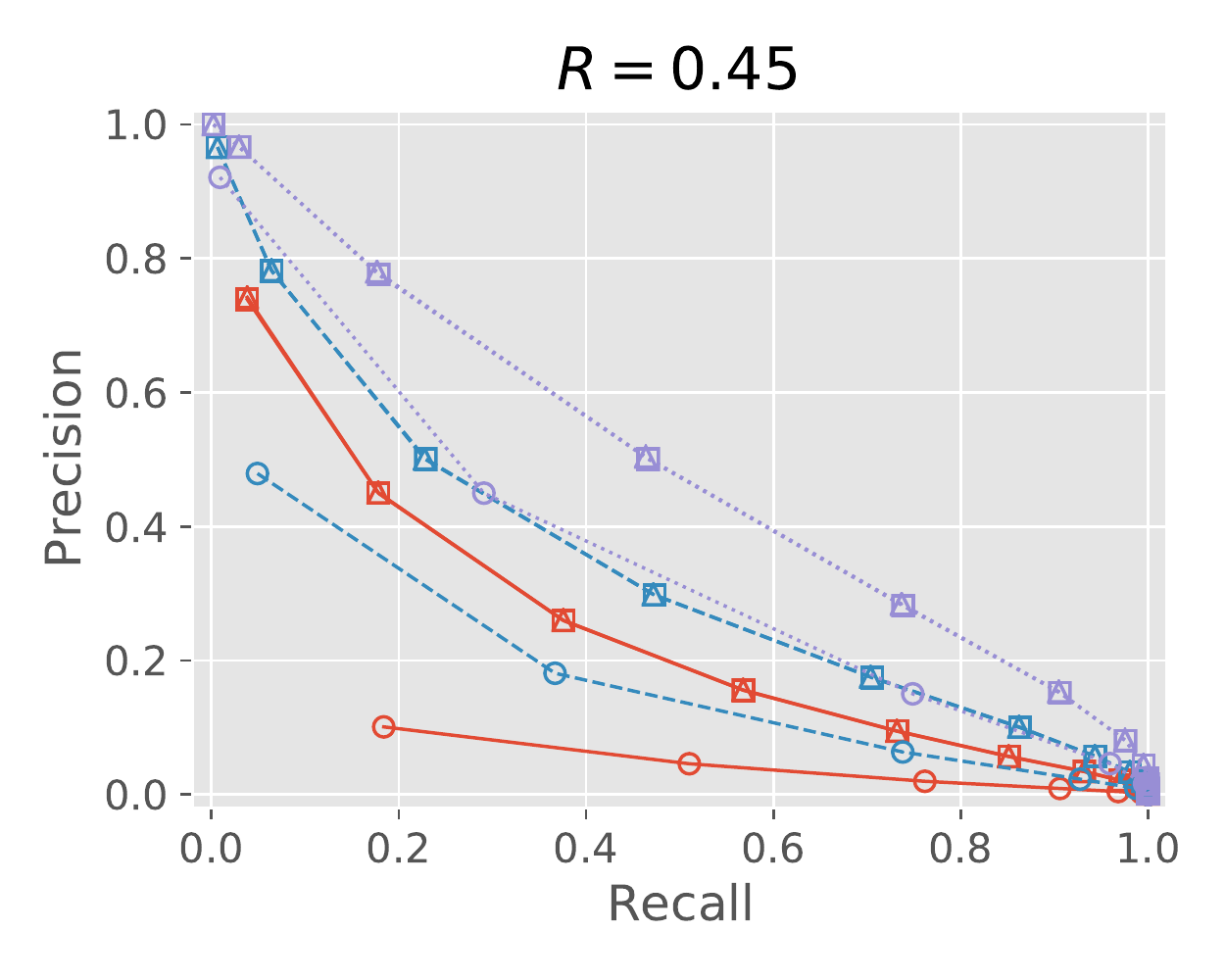}
        \end{tabular}
        \label{charts:LSH:NBA}
    }
    \caption{Precision-recall curve for varying Hamming distance threshold $K$ under fixed $L$ and $\sigma$ on each dataset.}
    \label{charts:LSH}
\end{figure}

\fref{charts:LSH} shows recall-precision curves on Taxi and NBA.
Each recall value (or precision value) was averaged per query. 
Overall, the larger Hamming distance thresholds were used, the larger the recall values became and the smaller the precision values became on each dataset. 
Under each fixed $L$, the recall and precision values for $\sigma=2^8$ were better than those for $\sigma=2^1$ on each dataset while 
the recall and precision values for $\sigma=2^8$ were almost the same as those for $\sigma=2^{32}$, which showed $\sigma=2^8$ was reasonable for 
achieving high recall and precision values of similarity searches using LSH. 
Under each fixed $\sigma$, $L=64$ achieved reasonable recall and precision values for similarity searches using LSH. 
In addition, $L=64$ enables us to efficiently implement the fast algorithm for computing Hamming distance \cite{zhang2013hmsearch}, which is applicable to not only tSTAT but also HmSearch and linear search (see Appendix \ref{appx:impl:ham}).
Thus, the following experiments were performed using $\sigma=2^8$ and $L=64$.

\subsection{Efficiency of Node Reduction and STAT Data Structure}
\label{sect:ex:node}

\begin{figure}[tb]
    \centering
    \subfloat[Number of internal nodes]{
        \includegraphics[width=\ChartWidth]{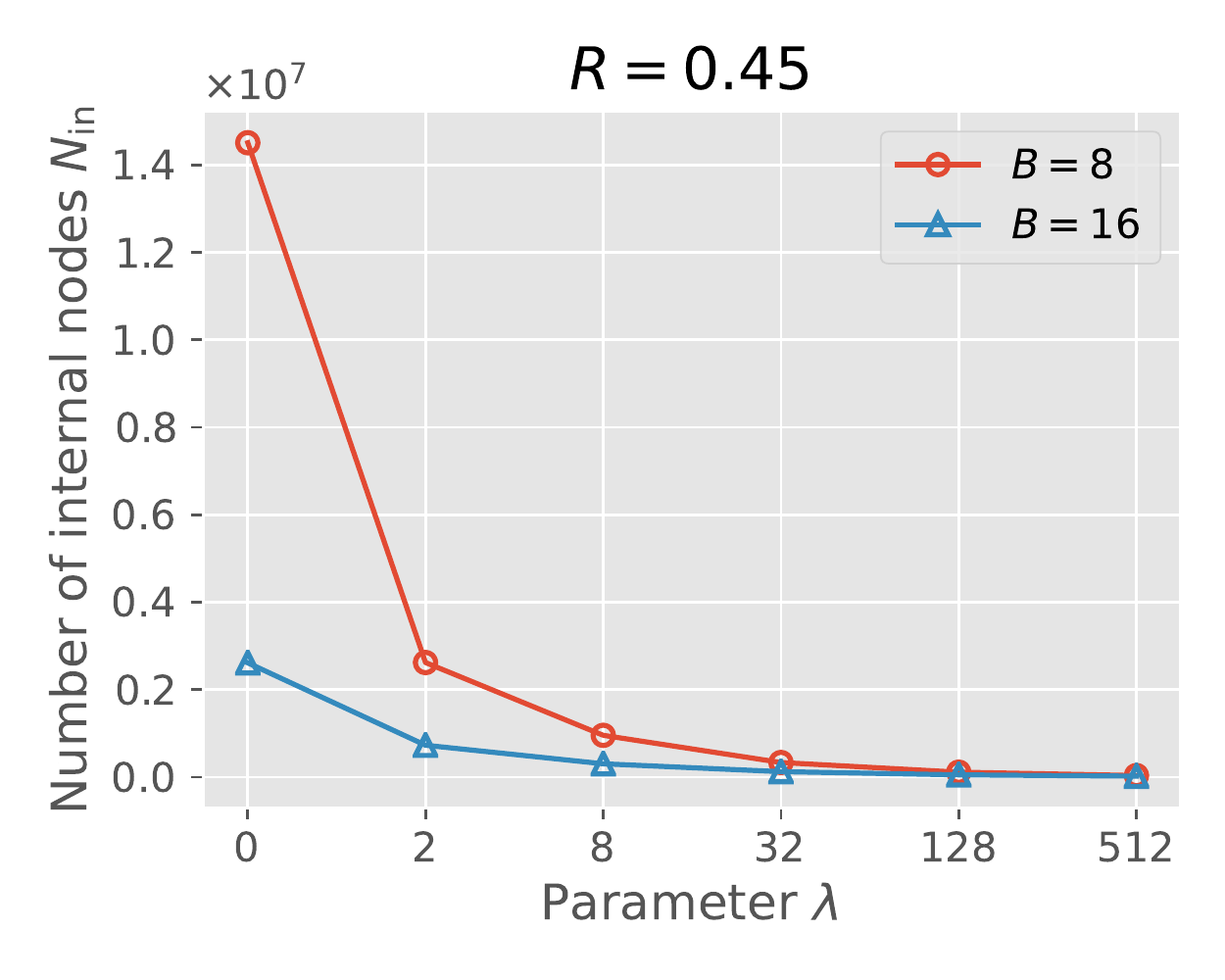}
        \label{charts:reduce_nodes}
    }
    \subfloat[Memory usage in GiB]{
        \includegraphics[width=\ChartWidth]{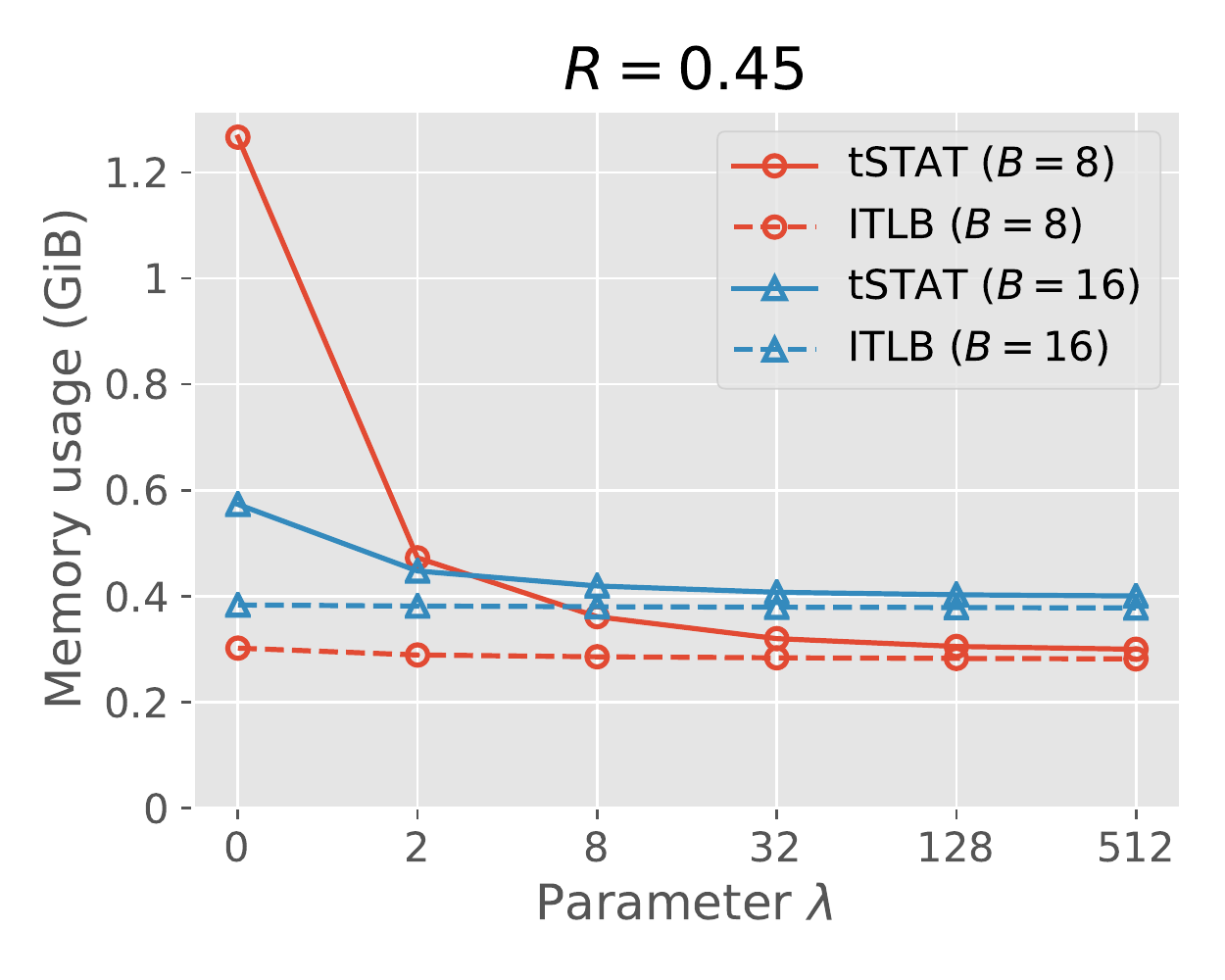}
        \label{charts:reduce_memory}
    }\\
    \subfloat[Number of candidates $|\Cand|$]{
        \includegraphics[width=\ChartWidth]{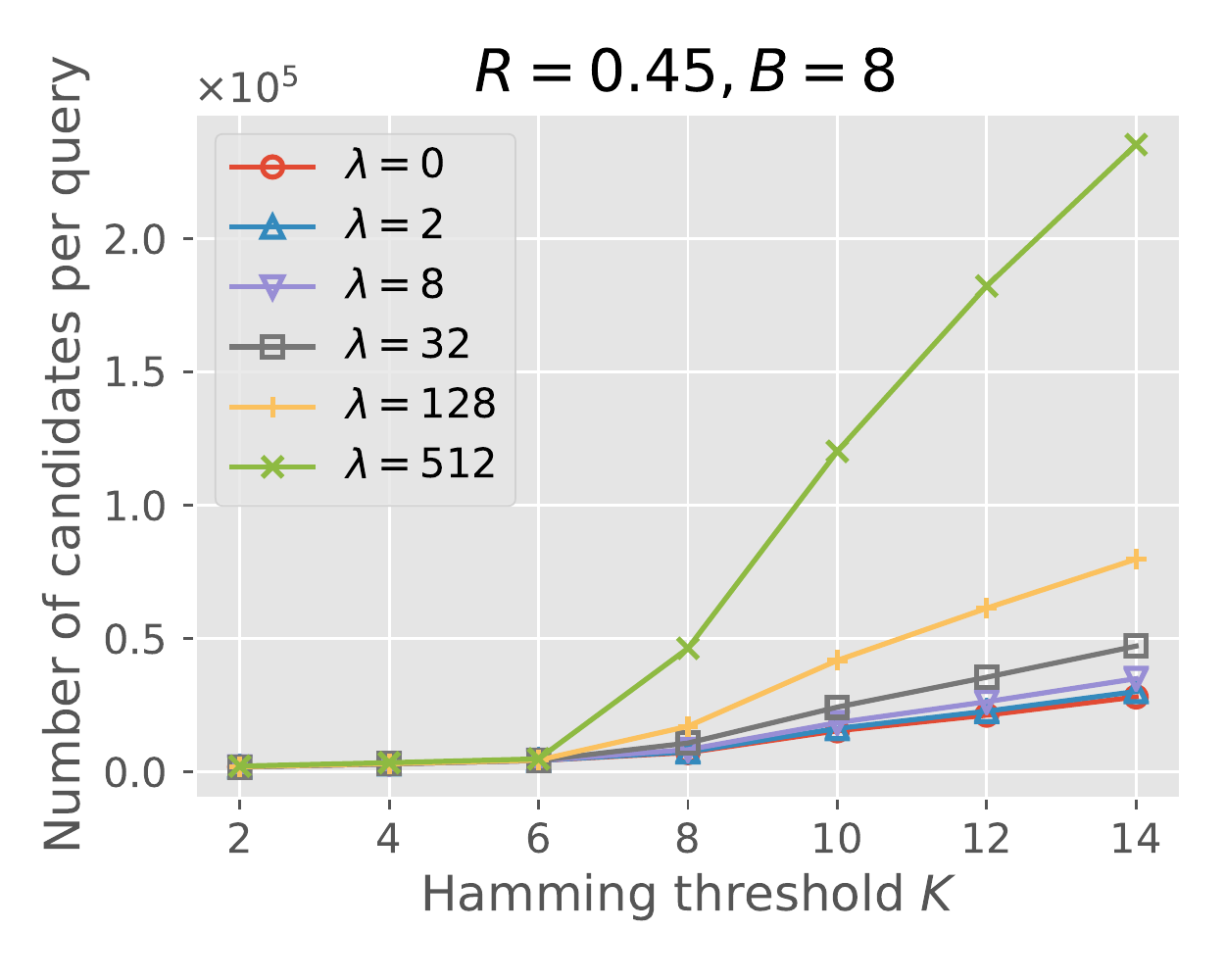}
        \includegraphics[width=\ChartWidth]{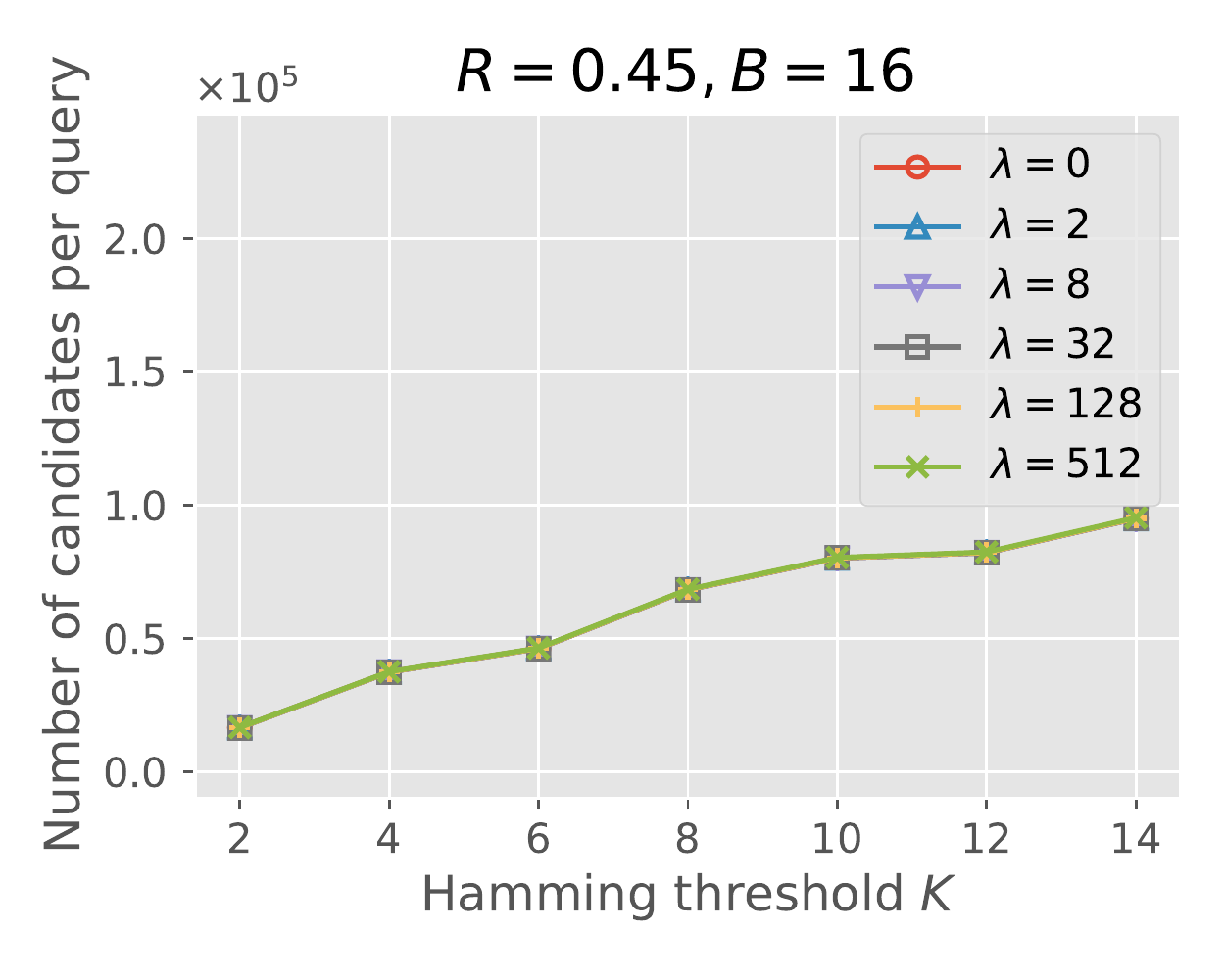}
        \label{charts:reduce_cands}
    }\\
    \subfloat[Search time in milliseconds (ms) per query]{
        \includegraphics[width=\ChartWidth]{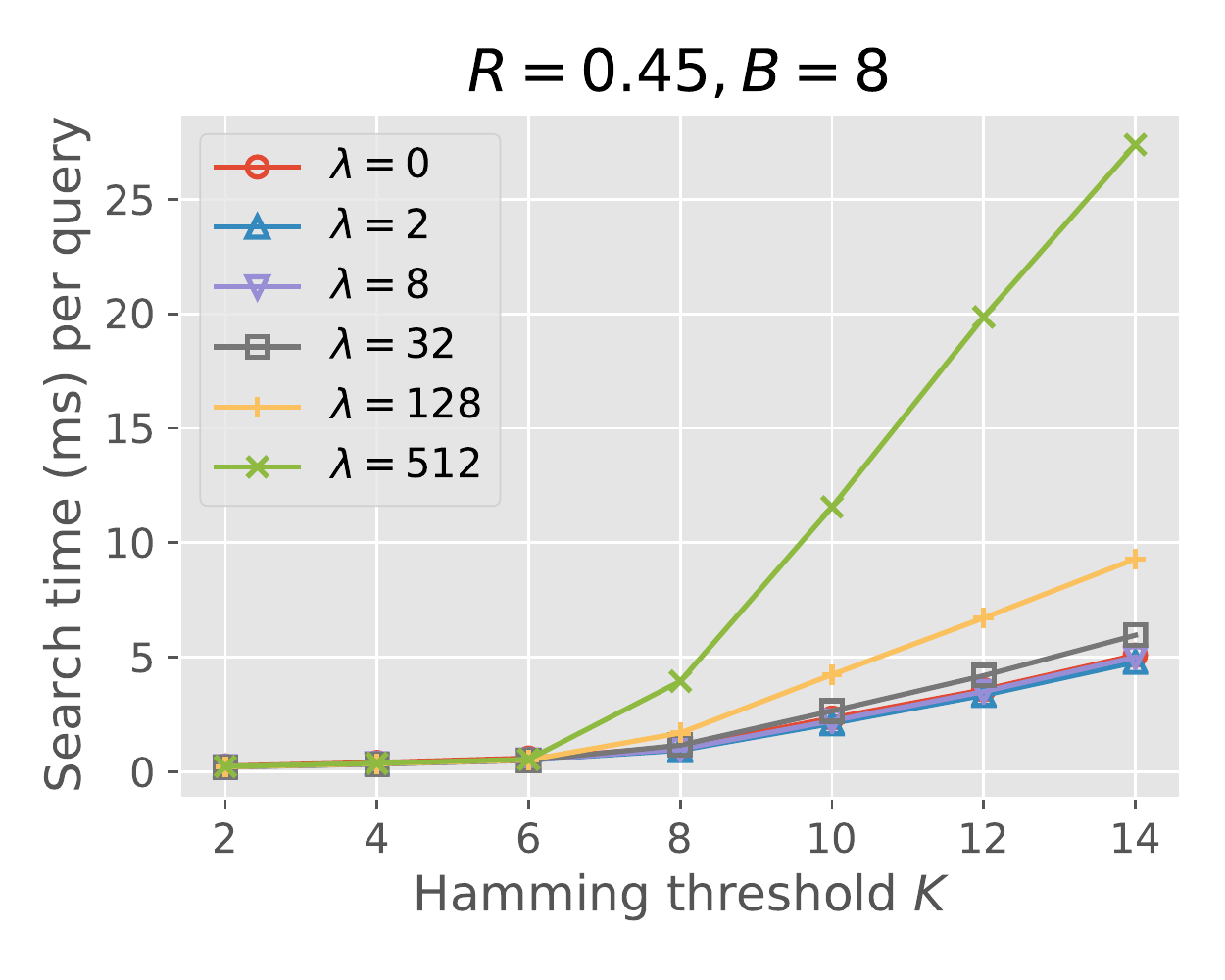}
        \includegraphics[width=\ChartWidth]{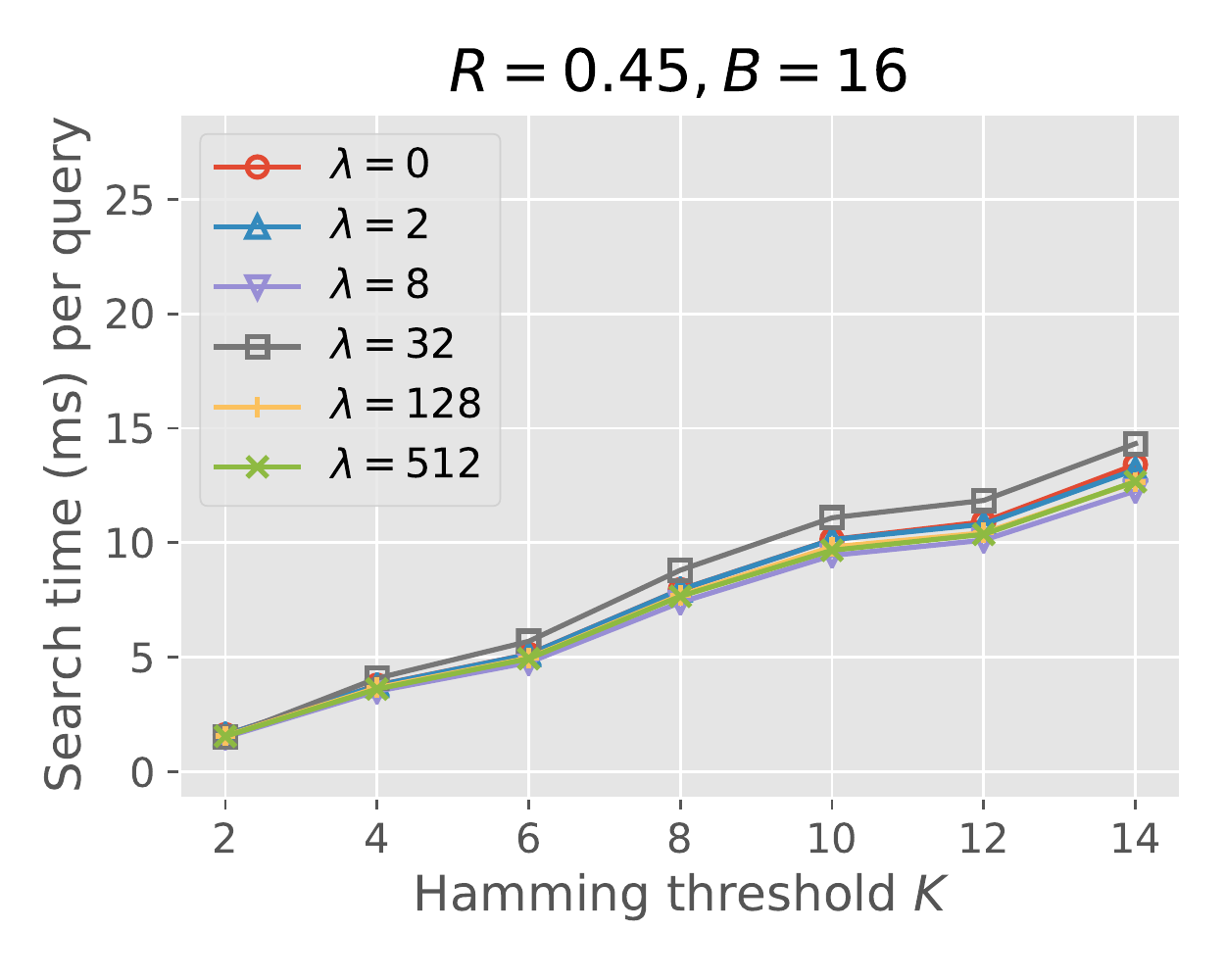}
        \label{charts:reduce_search}
    }
    \caption{Results of node reduction on NBA ($R=0.45$).}
    \label{charts:reduce}
\end{figure}

We evaluated the efficiency of node reduction in \sref{sect:mstat:node} and STAT data structure 
in \sref{sect:mstat:stat} in terms of the memory efficiency and search time of tSTAT for solving the Hamming distance problem. 
We measured (a) the number of internal nodes $\NodesIN$, (b) the memory usage, (c) the number of candidates $|\Cand|$, and (d) the average search time per query by testing parameters $\lambda = 0,2,8,32,128,512$.
Setting $\lambda = 0$ results in no node being eliminated by the node reduction, i.e., the original trie is represented by the STAT data structure. 
All the combinations of block numbers $B=8,16$ and Hamming distance thresholds $K=2,4,6,\dots,12,14$ were tested.  
\fref{charts:reduce} shows the experimental results on NBA for $R=0.45$. 
The results with other parameter settings are presented in \aref{appx:ex}.

\ffref{charts:reduce_nodes}{charts:reduce_memory} show the number of internal nodes $\NodesIN$ and the memory usage of tSTAT, respectively.
To observe the memory efficiency of tSTAT, \fref{charts:reduce_memory} also shows the ITLB estimated from the number of nodes $\Nodes = \sum^B_{j=1} \Nodes^j$.
As $\lambda$ grew, the number of internal nodes $\NodesIN$ and the memory usage were dramatically reduced.
As shown in \sref{sect:mstat:stat}, since the value $\Nodes/\NodesIN$ increased, the memory usage of tSTAT approached the ITLB.
The reduction of the number of internal nodes and the memory usage converged at around $\lambda=32$. 

\ffref{charts:reduce_cands}{charts:reduce_search} show the number of candidates $|\Cand|$ and the search time, respectively.
Since the search time was affected by the number of candidates, both figures showed a similar tendency. 
For $B = 8$ and $K \geq 8$, as $\lambda$ grew, the number of candidates and search time also increased.
However, the effect of $\lambda$ for the number of candidates and search time also converged at around $\lambda=32$. 
Those results showed that setting $\lambda=8$ or $32$ was beneficial for improving the search and memory efficiencies of tSTAT. 
For $\lambda=8$, $R=0.45$, and $B=8$, tSTAT achieved an average search time of 5 milliseconds and a memory usage of 0.36 GiB. 
The next subsection shows the efficiency of tSTAT with $\lambda=8$ and tSTAT can achieve a memory efficiency and fast search time.  

\subsection{Efficiency of tSTAT}
\label{sect:ex:comp}

\begin{figure*}[tb]
    \centering
    \setlength{\tabcolsep}{0mm}
    \subfloat[Search time on Taxi]{
        \begin{tabular}{c}
        \includegraphics[width=\ChartWidth]{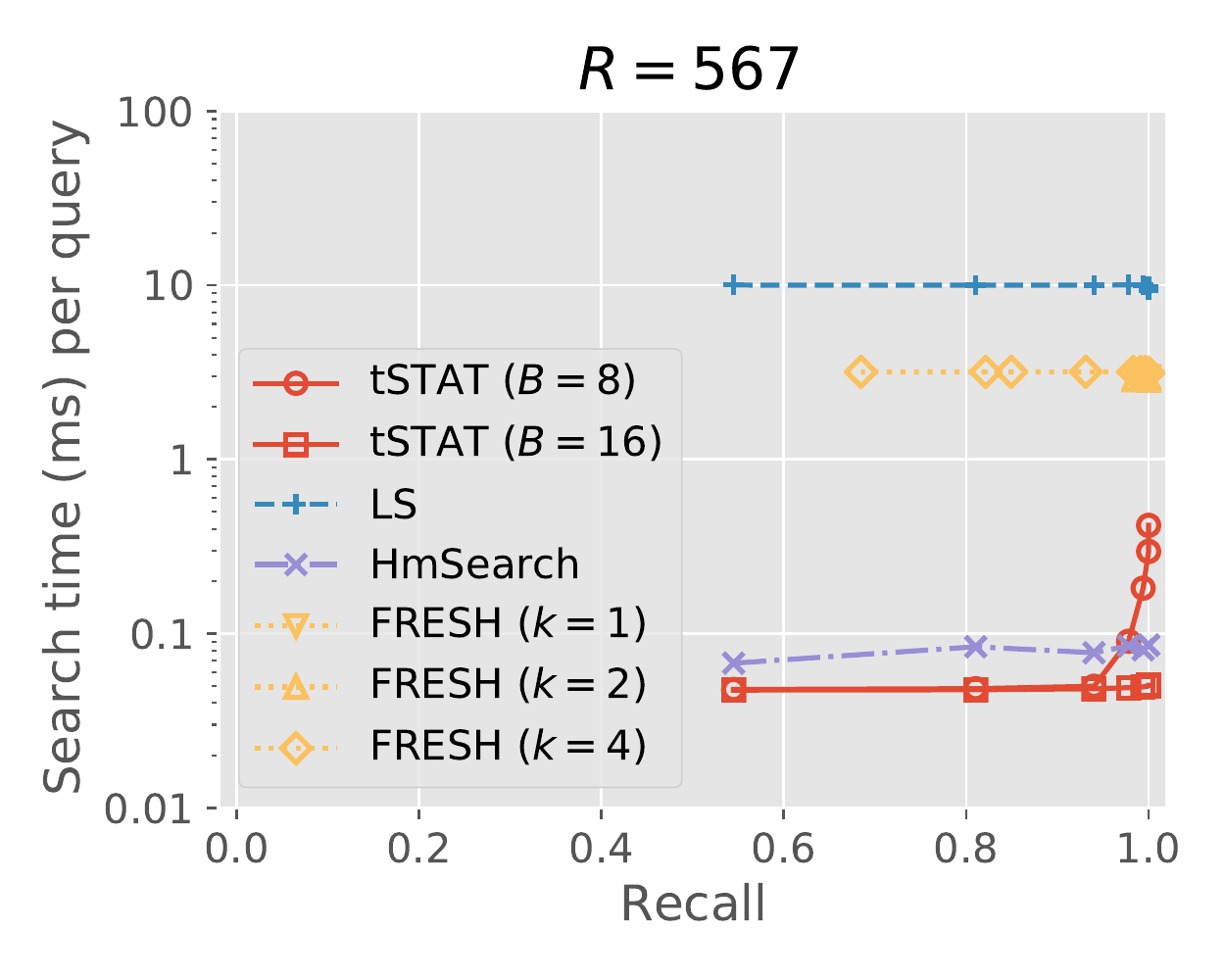}\\
        \includegraphics[width=\ChartWidth]{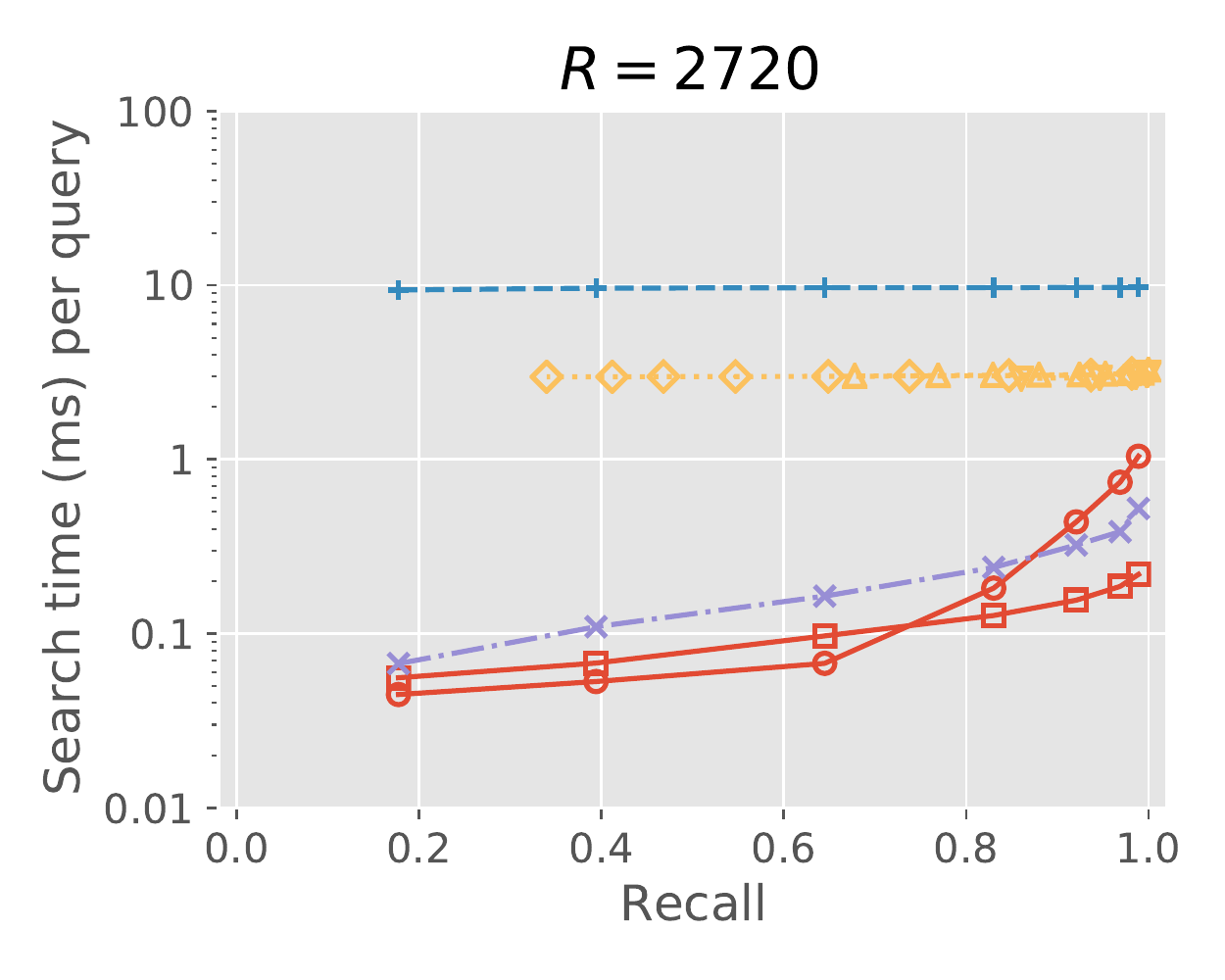}\\
        \includegraphics[width=\ChartWidth]{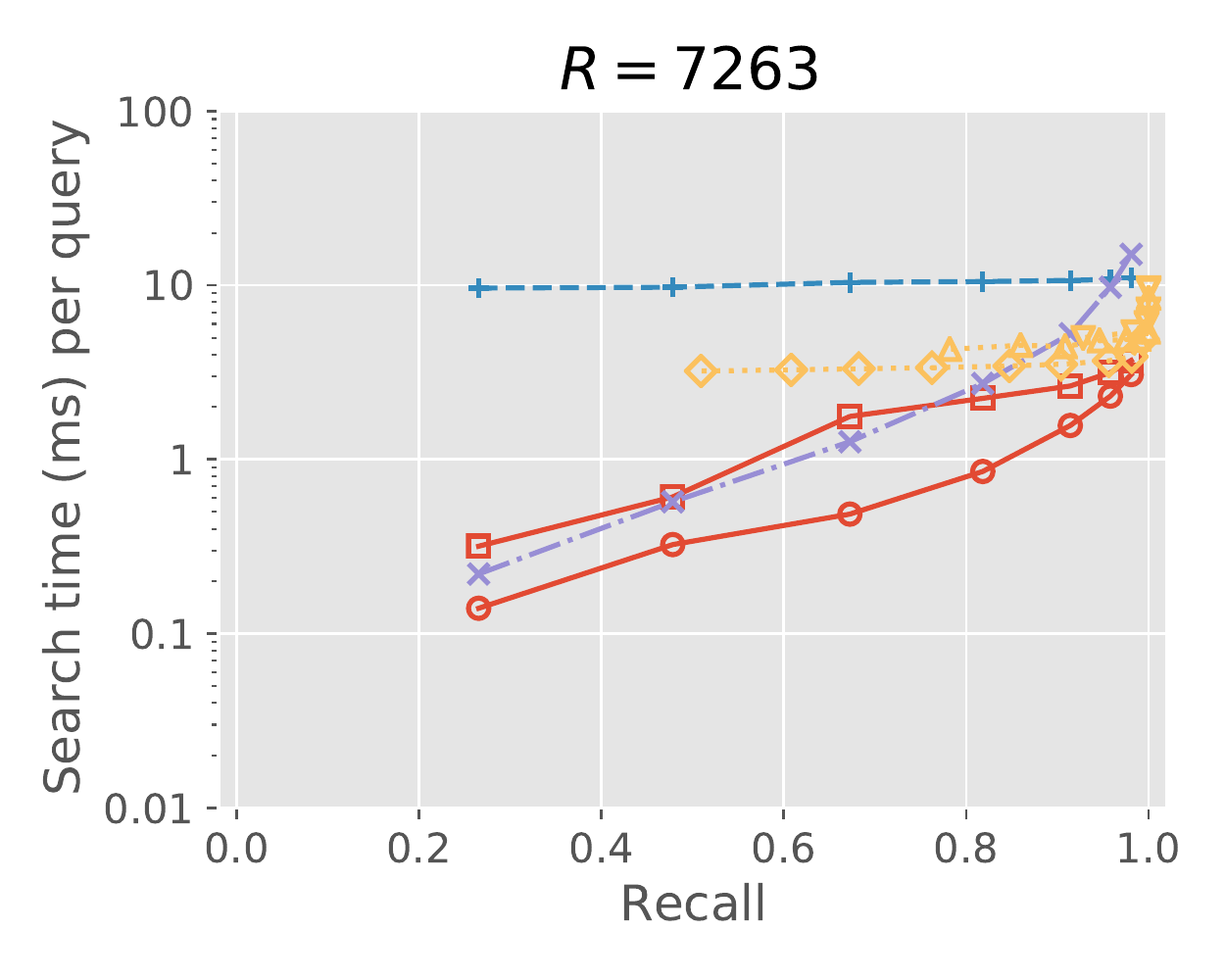}
        \end{tabular}
        \label{charts:search:Taxi}
    }
    \subfloat[Search time on NBA]{
        \begin{tabular}{c}
        \includegraphics[width=\ChartWidth]{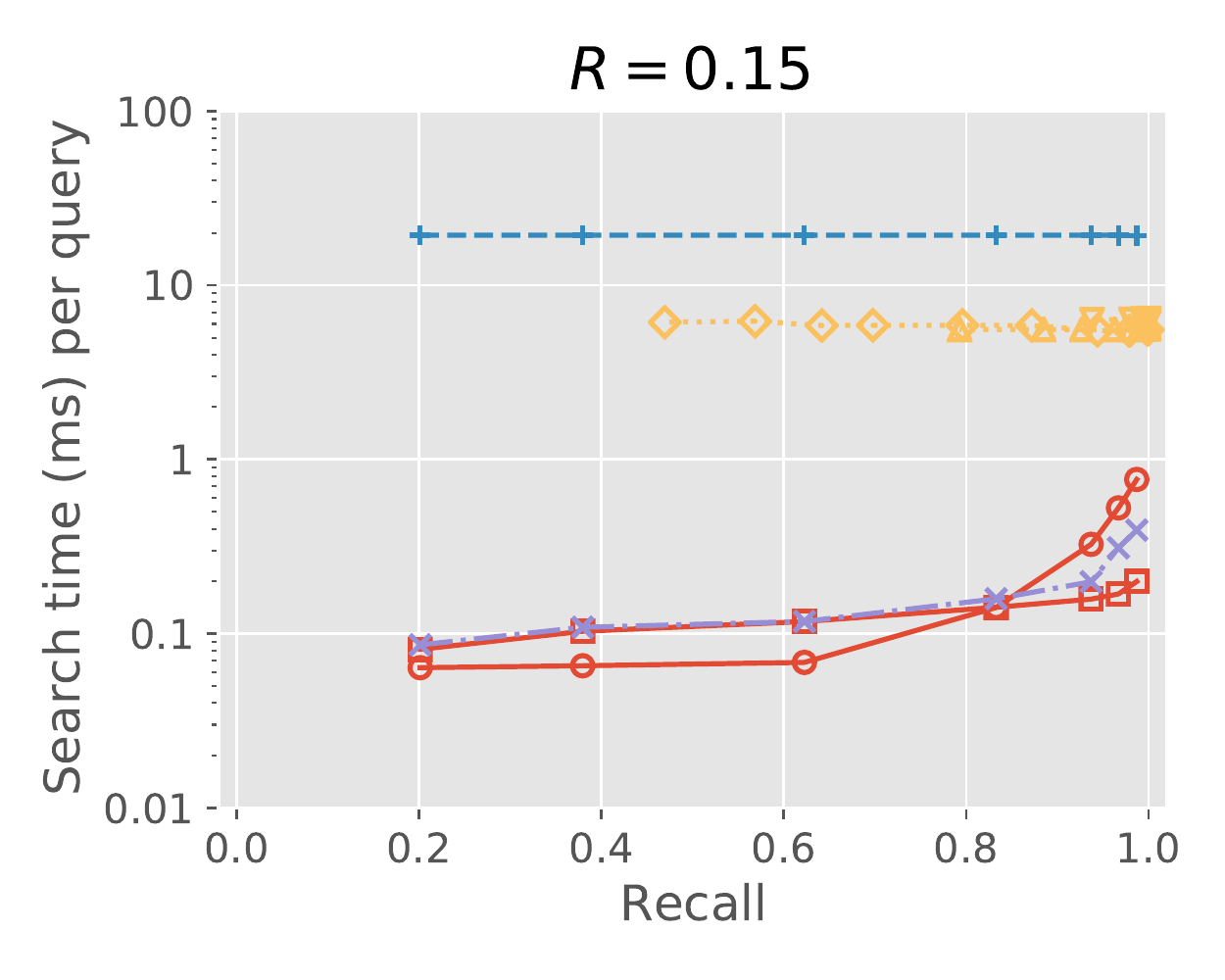}\\
        \includegraphics[width=\ChartWidth]{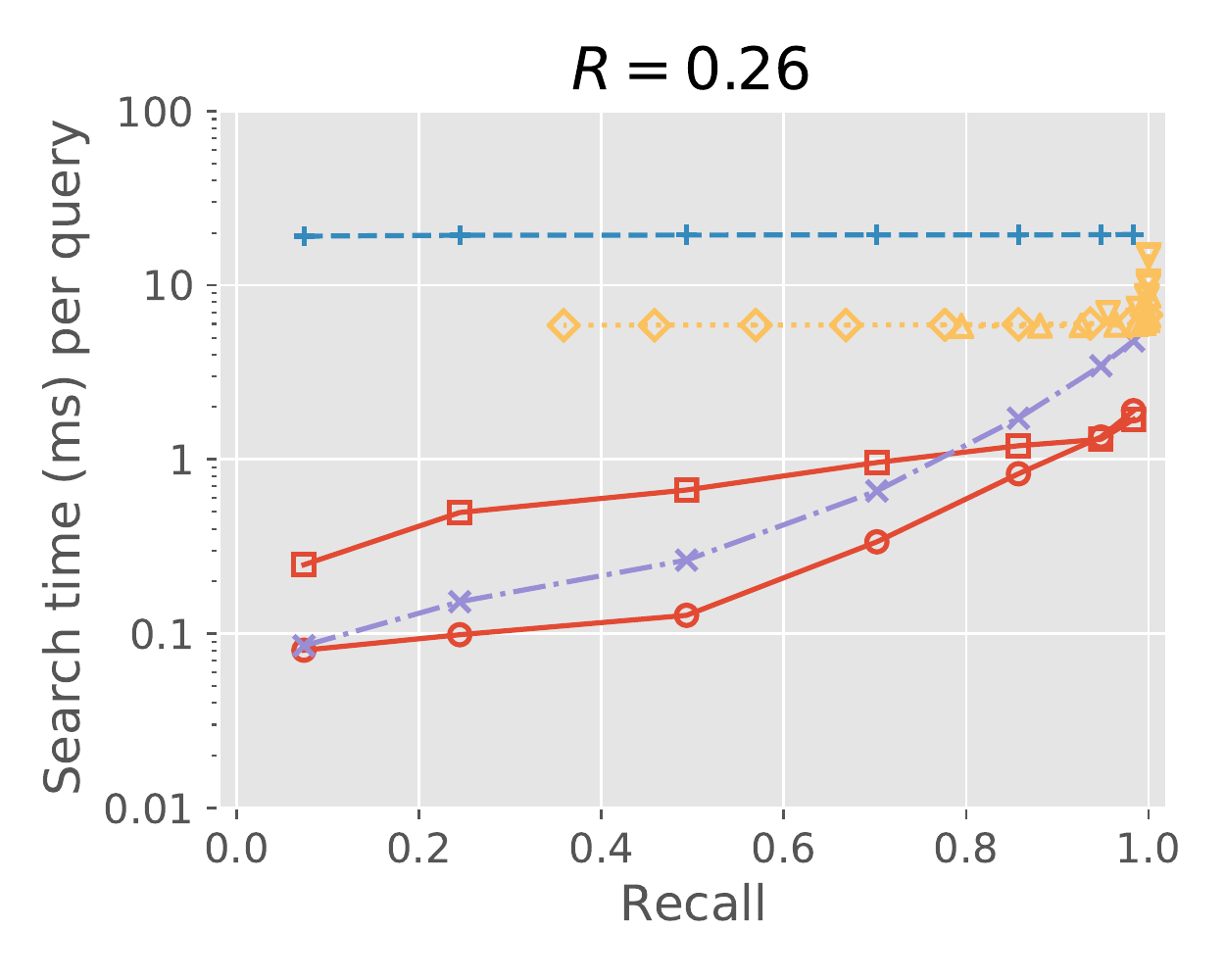}\\
        \includegraphics[width=\ChartWidth]{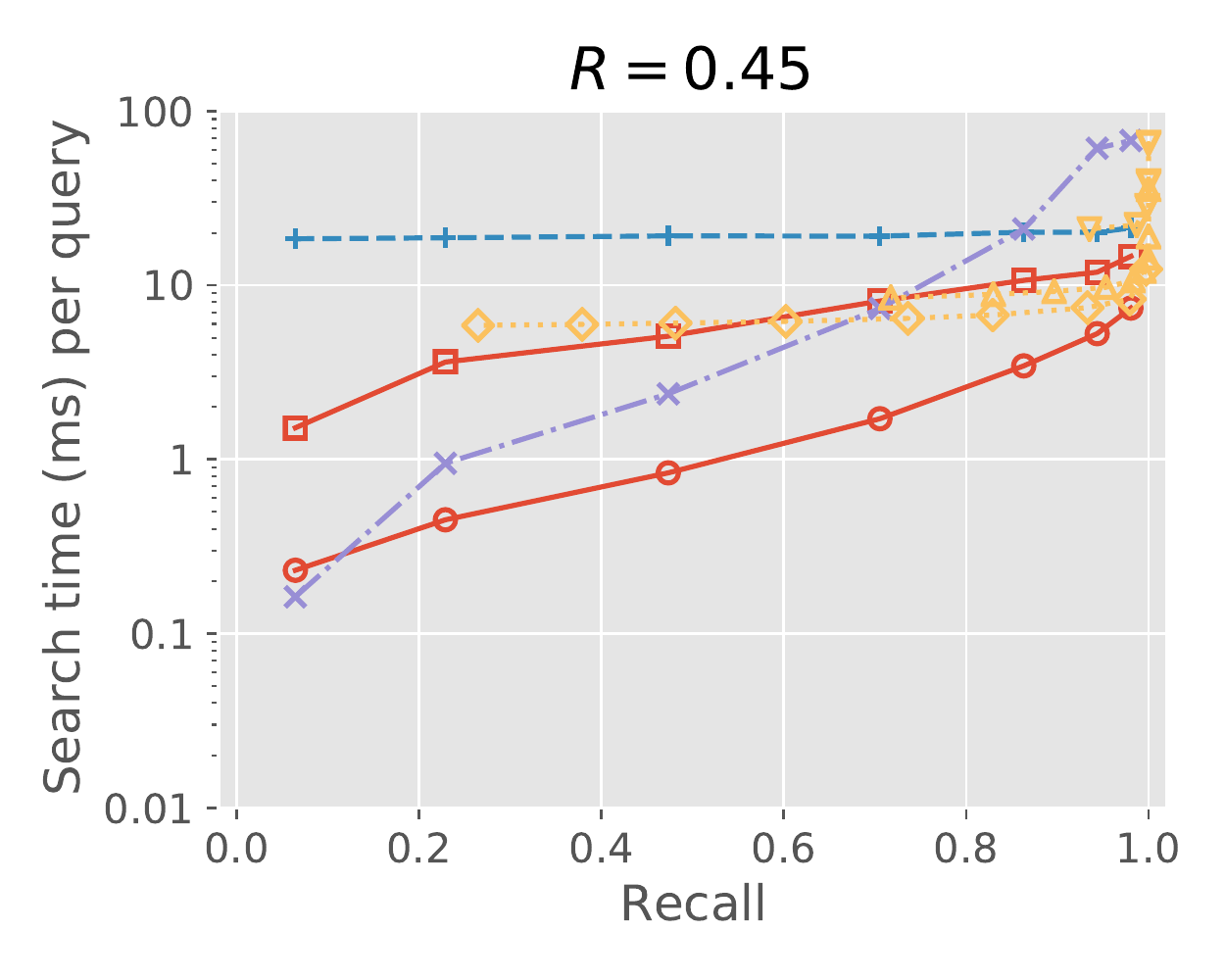}
        \end{tabular}
        \label{charts:search:NBA}
    }
    \subfloat[Search time on OSM]{
        \begin{tabular}{c}
        \includegraphics[width=\ChartWidth]{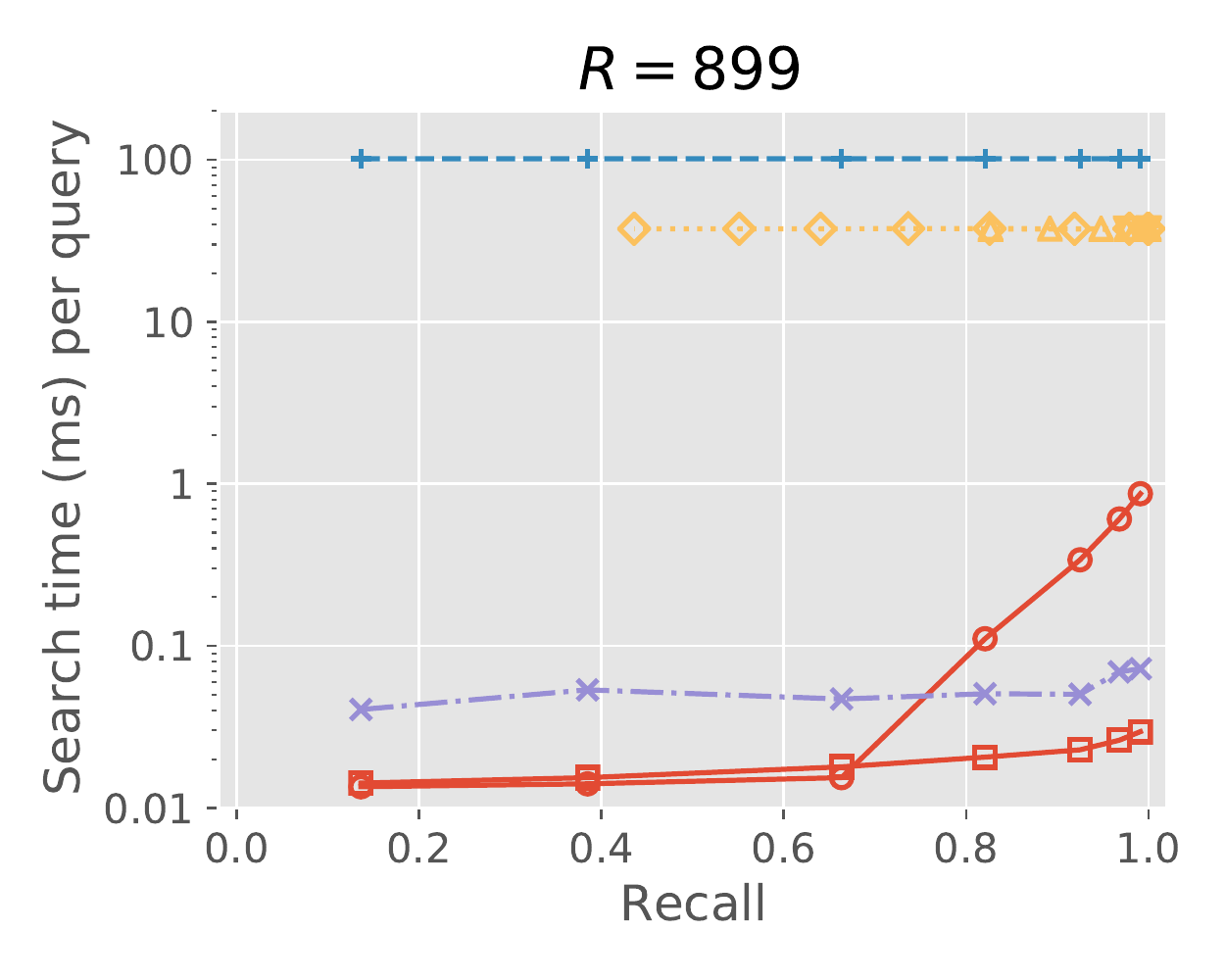}\\
        \includegraphics[width=\ChartWidth]{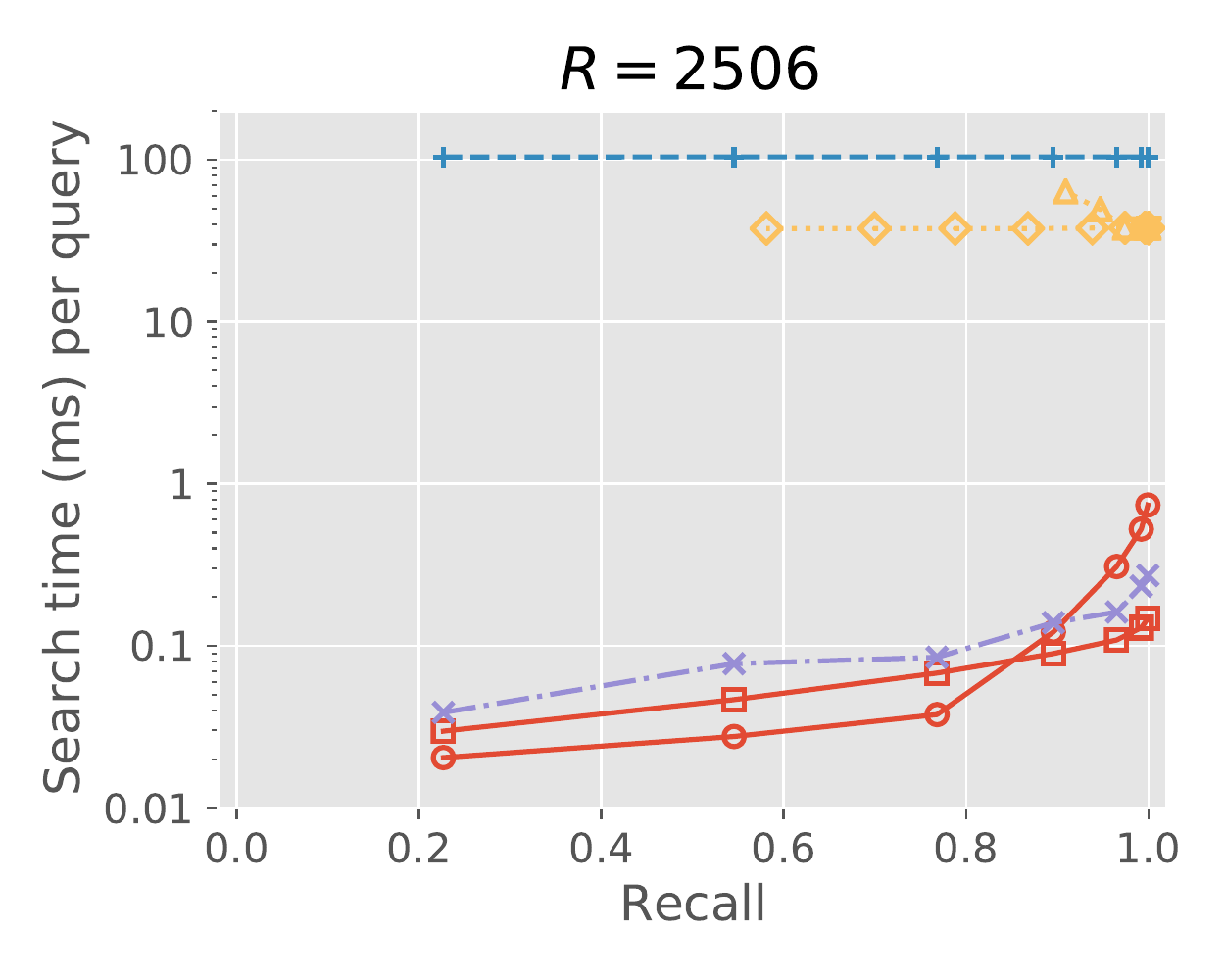}\\
        \includegraphics[width=\ChartWidth]{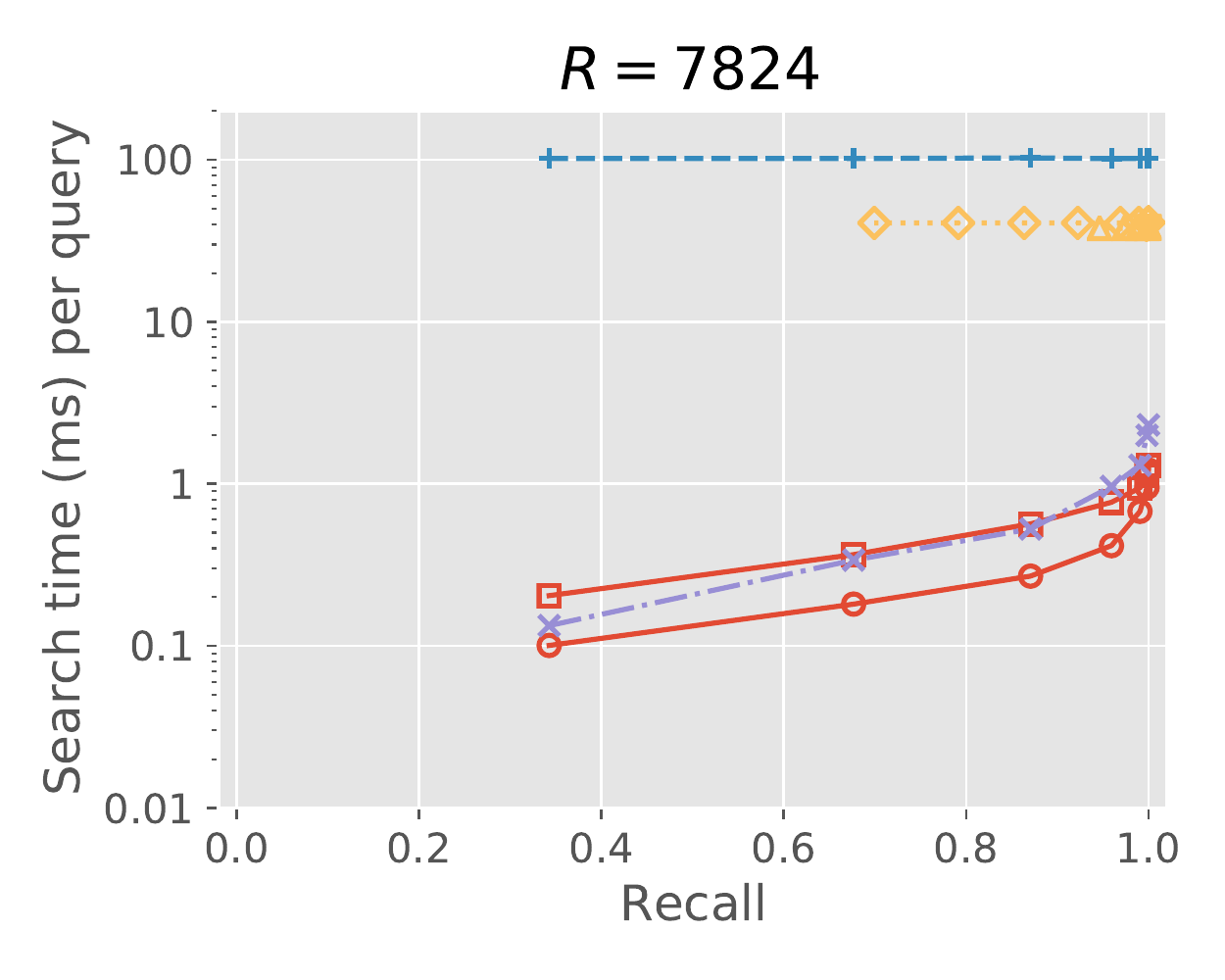}
        \end{tabular}
        \label{charts:search:OSM}
    }
    \subfloat[Search time on NBA (scalability)]{
        \begin{tabular}{c}
        \includegraphics[width=\ChartWidth]{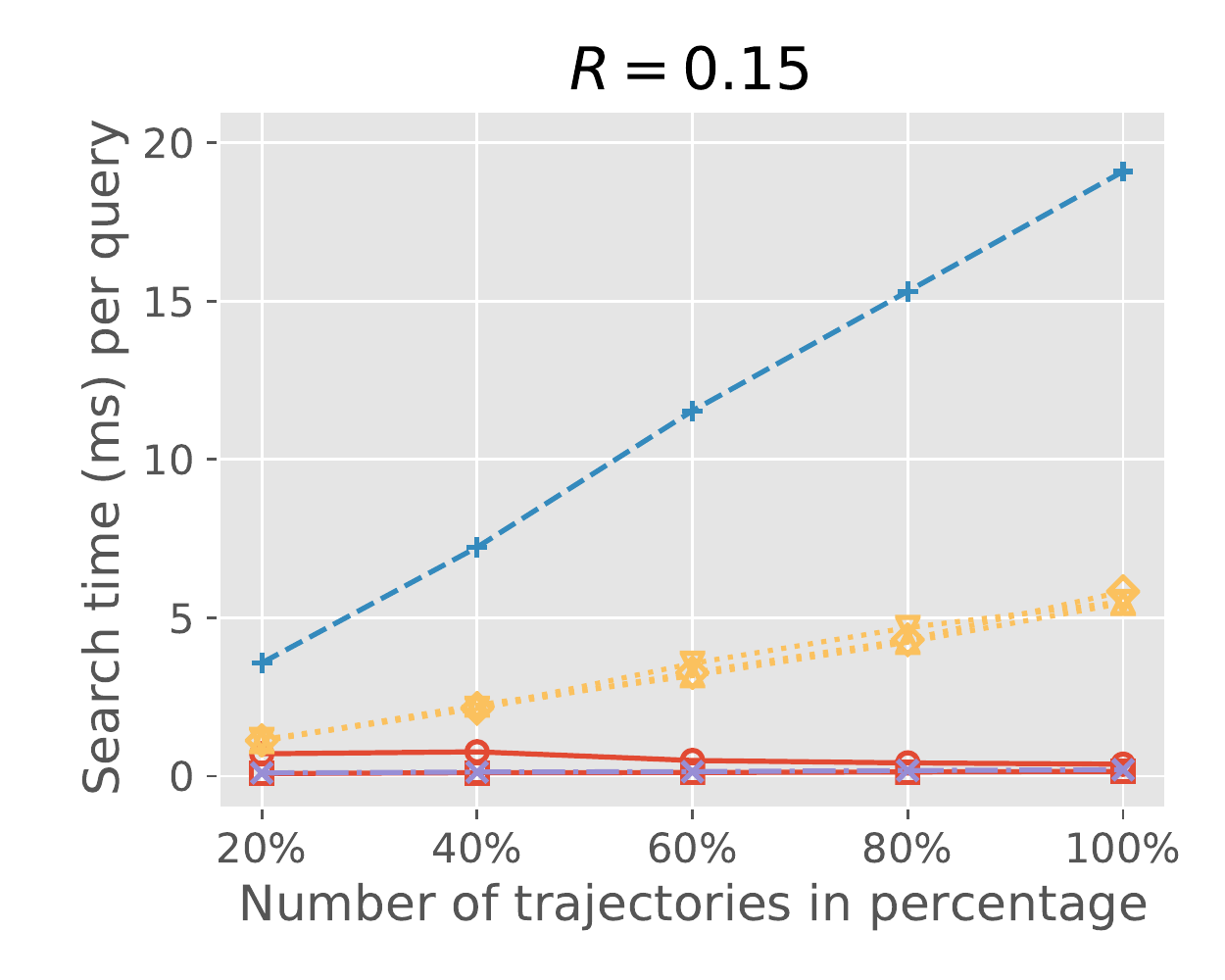}\\
        \includegraphics[width=\ChartWidth]{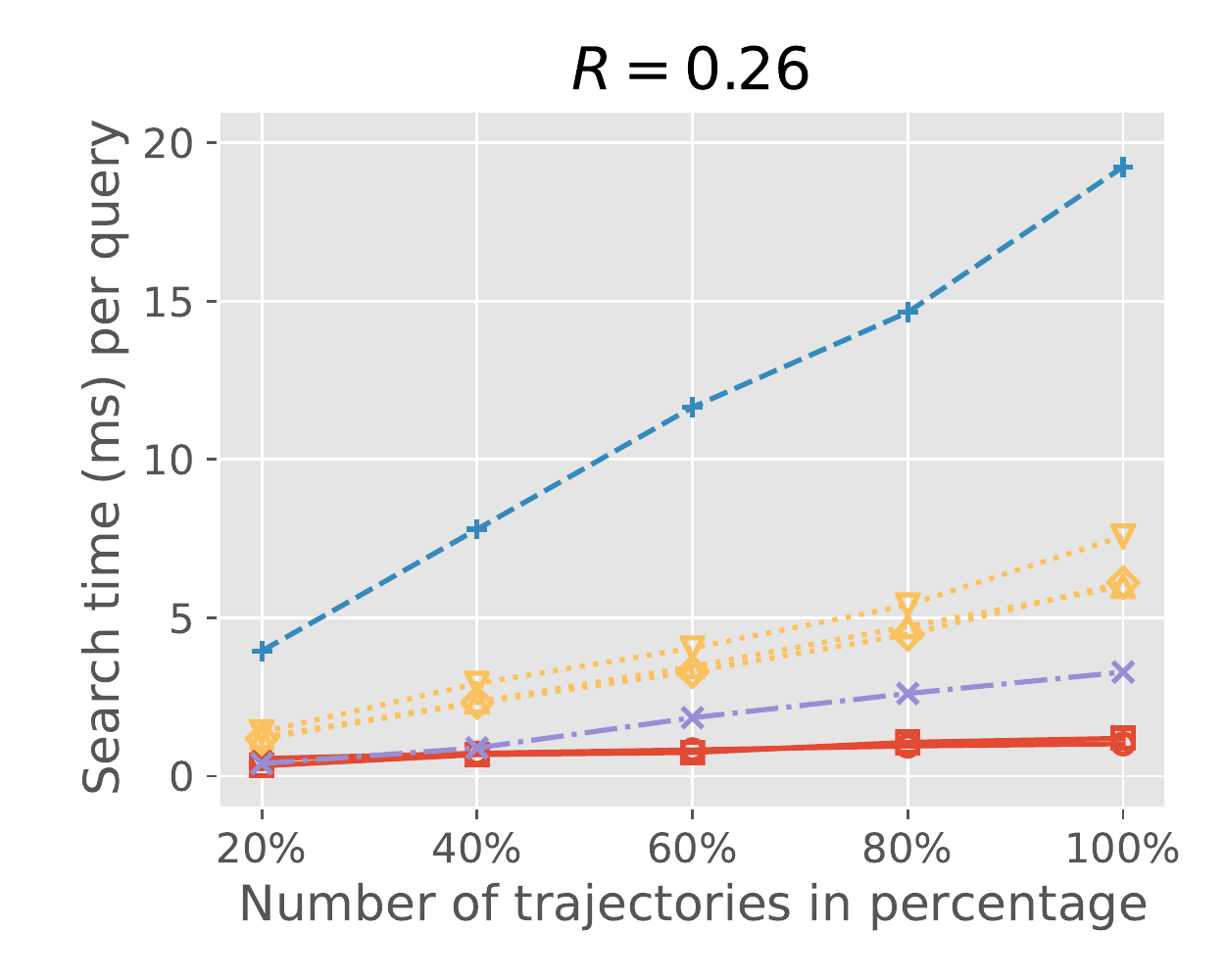}\\
        \includegraphics[width=\ChartWidth]{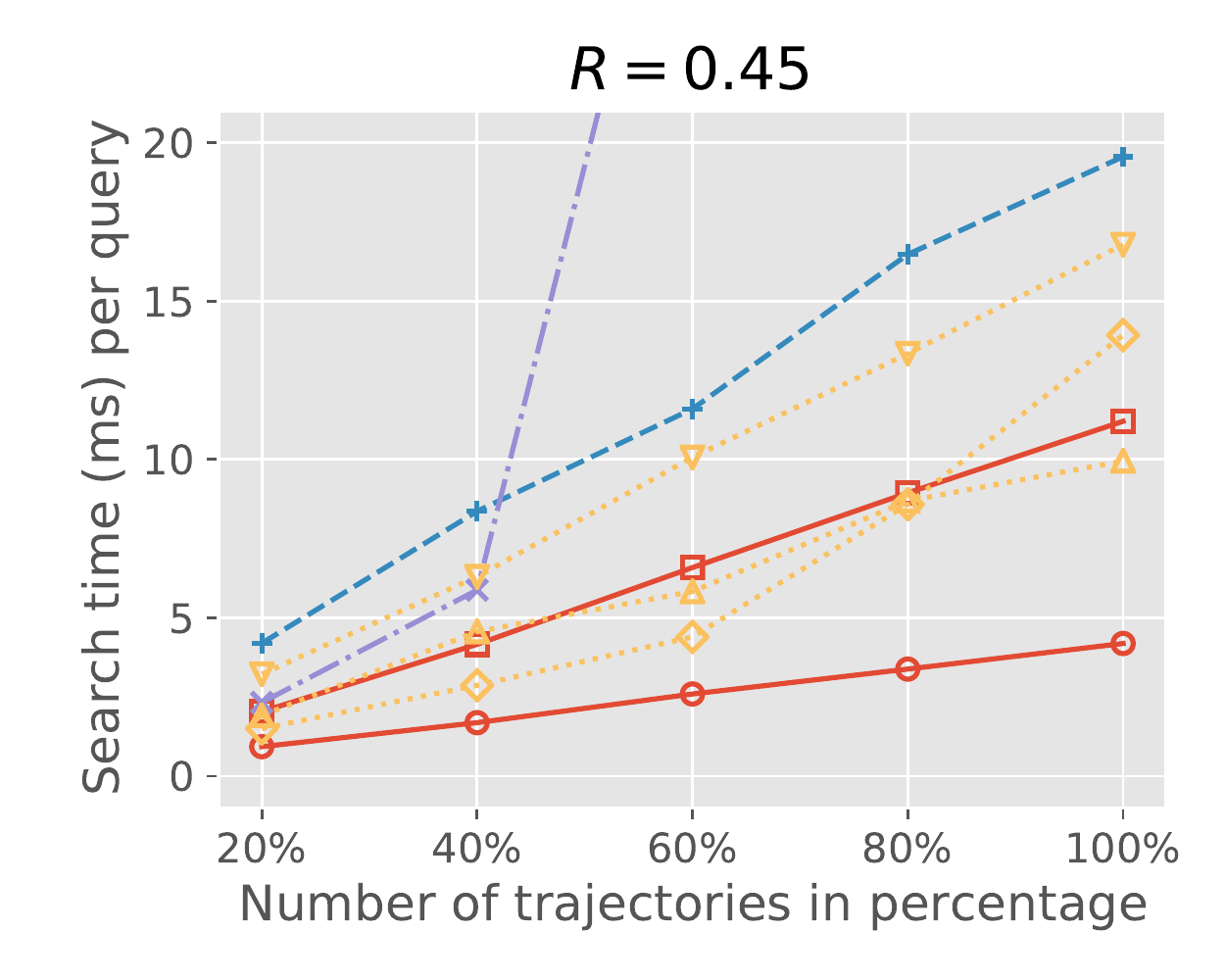}
        \end{tabular}
        \label{charts:search_scale:NBA}
    }\\
    \subfloat[Memory usage on Taxi]{
        \begin{tabular}{c}
        \includegraphics[width=\ChartWidth]{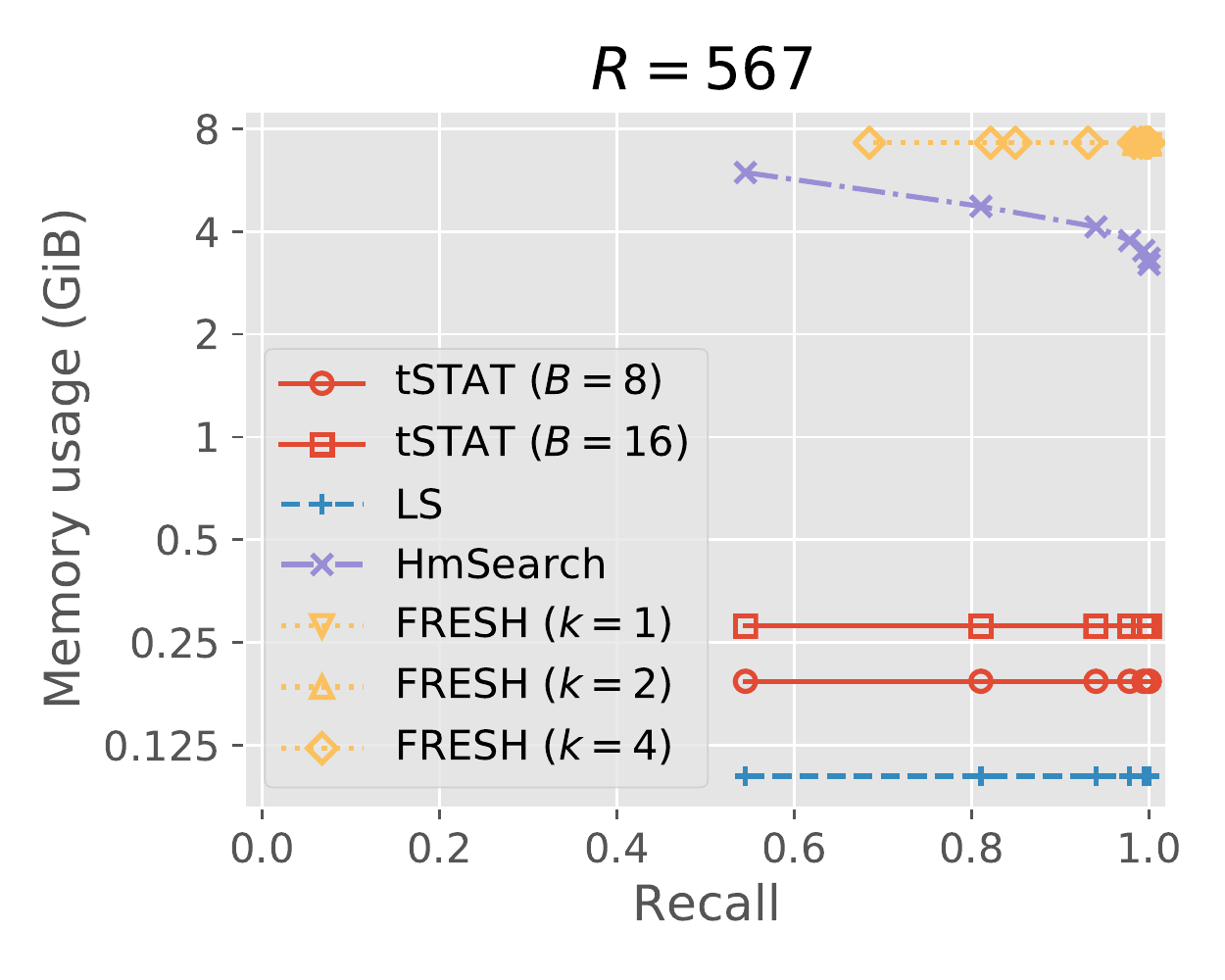}\\
        \end{tabular}
        \label{charts:memory:Taxi}
    }
    \subfloat[Memory usage on NBA]{
        \begin{tabular}{c}
        \includegraphics[width=\ChartWidth]{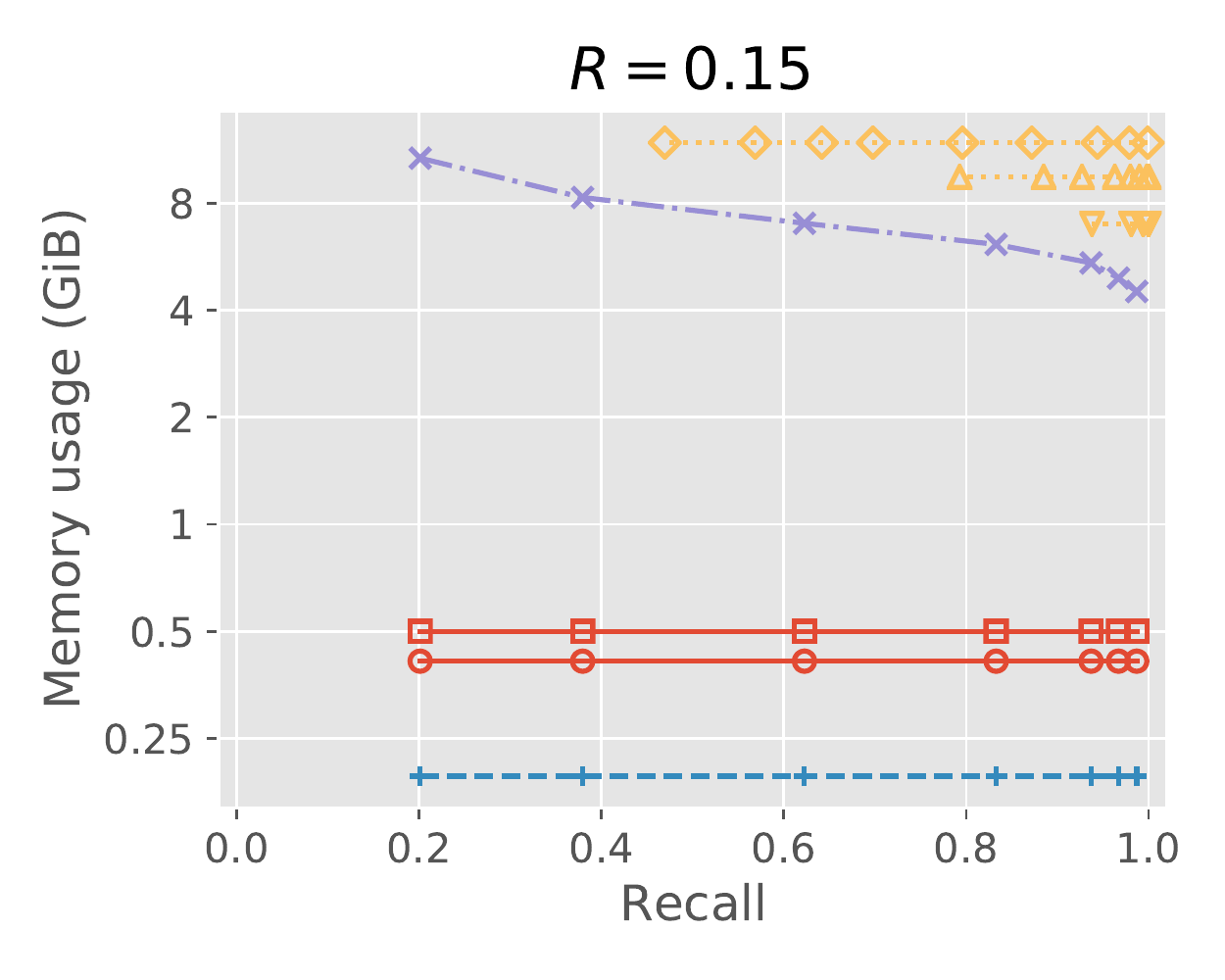}\\
        \end{tabular}
        \label{charts:memory:NBA}
    }
    \subfloat[Memory usage on OSM]{
        \begin{tabular}{c}
        \includegraphics[width=\ChartWidth]{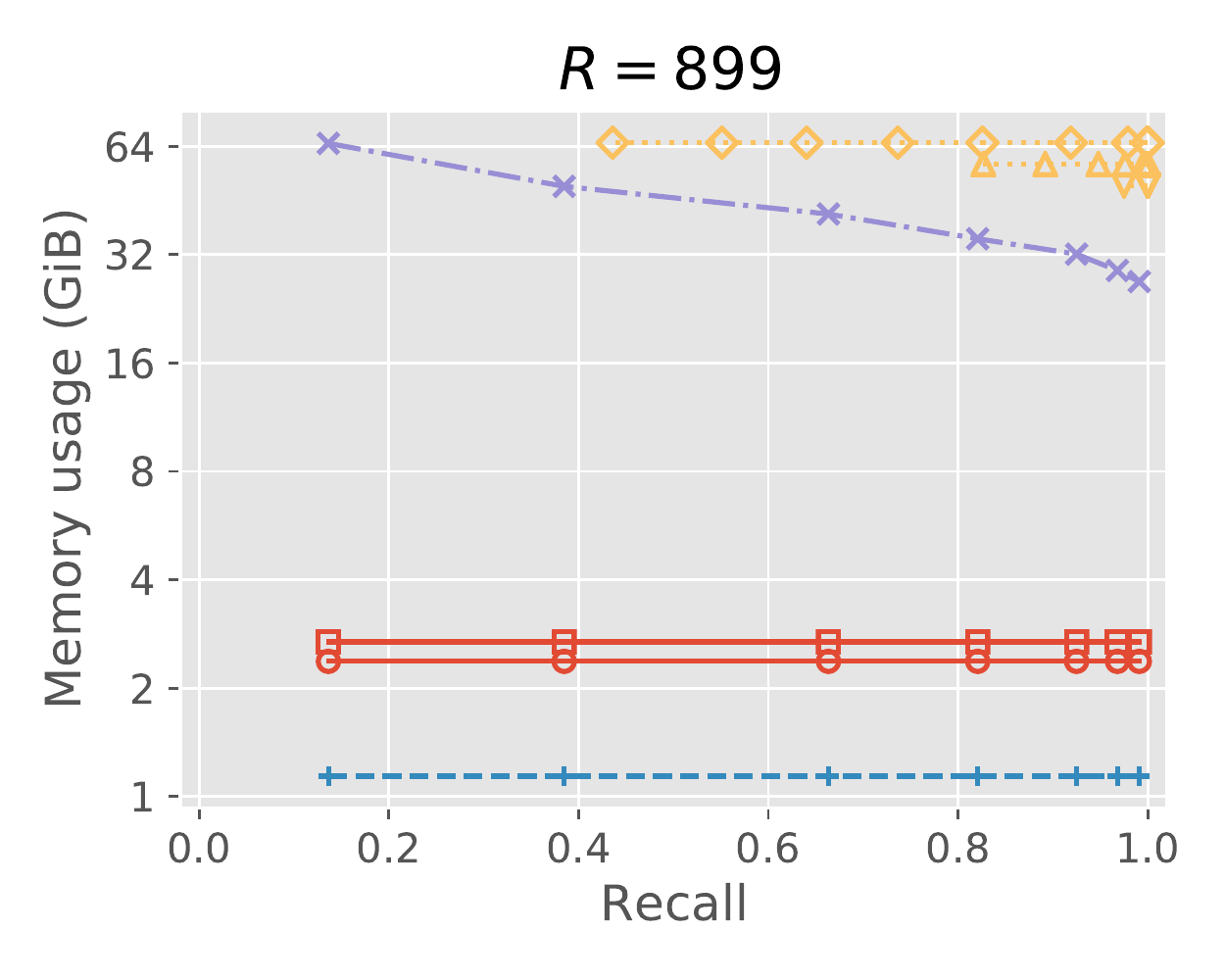}\\
        \end{tabular}
        \label{charts:memory:OSM}
    }
    \subfloat[Construction time on OSM]{
        \begin{tabular}{c}
        \includegraphics[width=\ChartWidth]{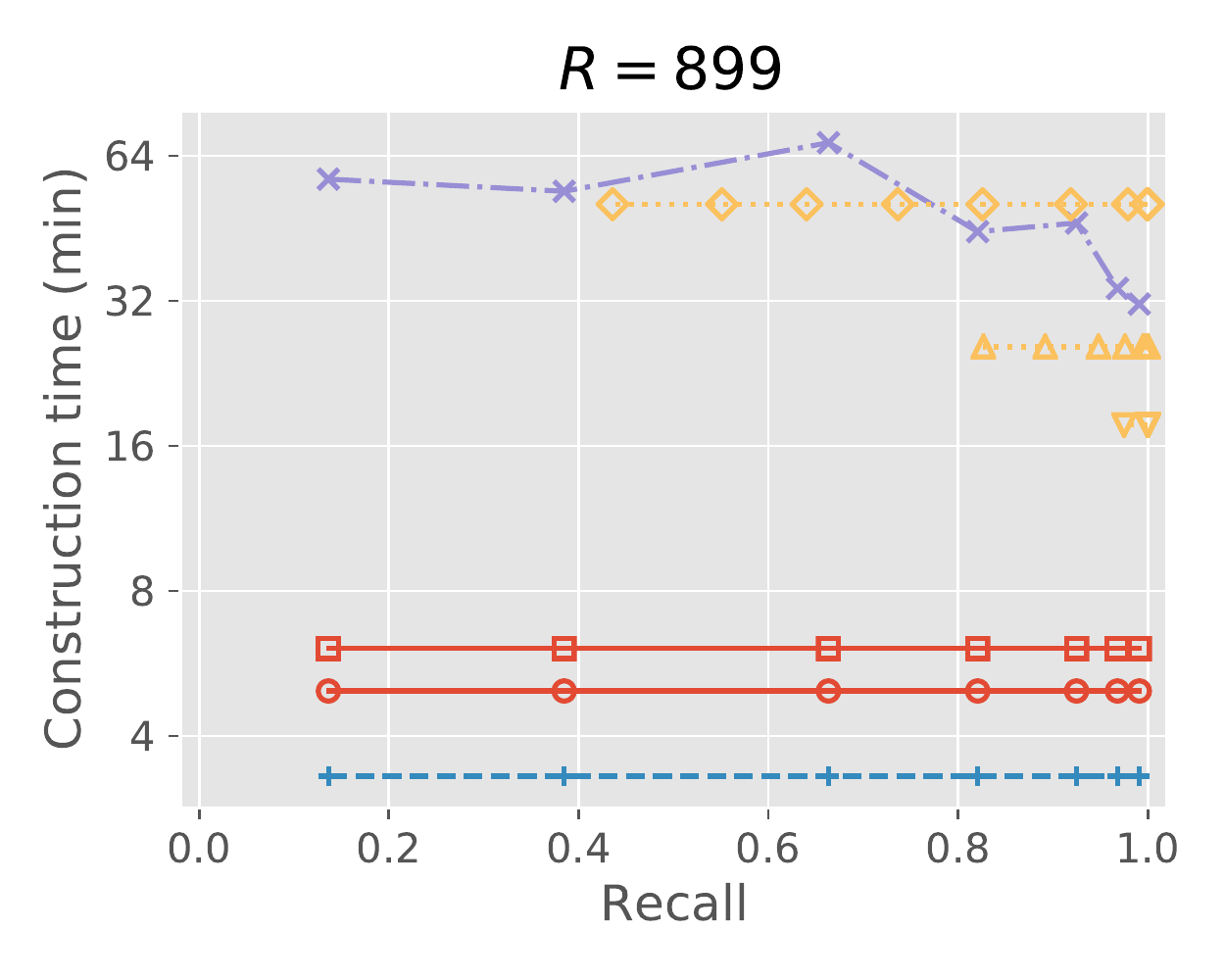}\\
        \end{tabular}
        \label{charts:constr:OSM}
    }
    \caption{Performance evaluation of tSTAT, LS, HmSearch, and FRESH. Each of charts (a)--(c) shows average search times per query in milliseconds (ms) for varying recalls. Chart (d) shows average search times per query in ms for varying the number of trajectories, and it demonstrates scalabilities of similarity search methods. Each of charts (e)--(g) shows memory usages in GiB for varying recalls. Chart (h) shows construction times in minutes for varying recalls. The charts except (d) are plotted in the logarithmic scale.}
    \label{charts:results}
\end{figure*}

We compared tSTAT with HmSearch \cite{zhang2013hmsearch}, FRESH \cite{ceccarello2019fresh}, and linear search (LS) in the computation of $\IdSet^\prime$.
LS is a strawman baseline that computes the Hamming distance between a sketch in $\Dict$ and a query one-by-one in $\BigO{Ln}$ time and $\BigO{Ln \log \sigma}$ bits of space.
Regarding exact solutions (without LSH), one possible baseline would be the winner of ACM SIGSPATIAL Cup 2017 \cite{baldus2017fast,bringmann2019walking}.
We did not include the winner in our competitors, since the experimental comparison between the winner and FRESH has been done in \cite{ceccarello2019fresh} and FRESH often outperformed the winner.

We fixed $L=64$.
The parameters of LSH for \Frechet{} distance were set to $\sigma = 2^8$, $\delta = 8dR$, $k = 1$, and $K = 2,4,6,\dots,12,14$, which were used in tSTAT, LS, and HmSearch.
Following the original paper \cite{ceccarello2019fresh}, FRESH was tested with the following parameter settings: $\sigma = 2^{32}$, $\delta = 4dR$, $k = 1,2,4$, and $K = 31, 35, 39, \dots, 59, 63$. 
For a fair comparison between different methods, 
we measured the time performance and the memory usage for varying recall for solution set $\IdSet^\prime$. 
The implementations of HmSearch and FRESH used in these experiments are downlodable from \url{https://github.com/kampersanda/hmsearch} and \url{https://github.com/Cecca/FRESH}, respectively. 
\fref{charts:results} shows experimental results.
A part of the experimental results are presented in \aref{appx:ex}.

\ftfref{charts:search:Taxi}{charts:search:OSM} show the search time for varying the recall up to $100\%$ for each method.
The result showed tSTAT was the fastest for most results and demonstrated the efficiency of our search algorithm on tries with the multi-index approach. 
FRESH was slow because the implementation takes $\BigO{n}$ time for $C_\mathrm{fresh}$.
For over 90\% recalls, tSTAT was at most \Times{60}, \Times{34}, and \Times{1600} faster than FRESH on Taxi, NBA, and OSM, respectively.
For over 90\% recalls, tSTAT was at most \Times{3.4}, \Times{12}, and \Times{2.3} faster than HmSearch on Taxi, NBA, and OSM, respectively.

We also evaluated the scalability of the methods by varying the collection size $n$.
We randomly sampled 20\% to 100\% of the trajectories from each dataset.
\fref{charts:search_scale:NBA} shows the fastest search times with more than 90\% recall for each method on NBA.
tSTAT was much faster than LS and FRESH for a wide range of collection sizes $n$ and \Frechet{} distance thresholds $R$.
While HmSearch was also fast for small $R$, the time efficiency was decreased for large $R$.
This is because, as described in \sref{sect:related:dm}, HmSearch stores and retrieves many sketches additionally generated, resulting in amplifying the number of candidate sketches that need to be verified (i.e., the cost $C_\mathrm{hms}$ in \tref{tab:complexity}) for large $R$.

\ftfref{charts:memory:Taxi}{charts:memory:OSM} show the memory usage for varying the recall up to $100\%$ for each method.
Various memories consumed by HmSearch were observed for different recalls because 
different data structures in HmSearch are built according to Hamming distance threshold $K$. 
Those results demonstrated a high memory efficiency of tSTAT. 
In fact, while tSTAT consumed only 2.4--2.7 GiB of memory on OSM, HmSearch and FRESH consumed 27--65 GiB and 50--65 GiB of memories, respectively.

\fref{charts:constr:OSM} shows the construction time for varying the recall up to $100\%$ for each method on OSM.
We measured the execution time for producing sketches from input trajectories and building the index from the sketches.
tSTAT took 5.0--6.1 minutes in total while LS took 3.3 minutes.
On the other hand, the time needed for only building STATs from sketches was only 1.7--2.8 minutes, which demonstrated fast algorithms for constructing STATs.
HmSearch and FRESH took 32--57 minutes and 18--51 minutes, respectively, which were much larger than the construction times of tSTAT. 

Considering the time and memory efficiency of tSTAT, the result demonstrated the feasibility of finding similar trajectories from large collections for a query, which would be beneficial to analyze trajectories that are similar to a query of interest.

\begin{figure*}[tb]
\centering
\includegraphics[scale=0.36]{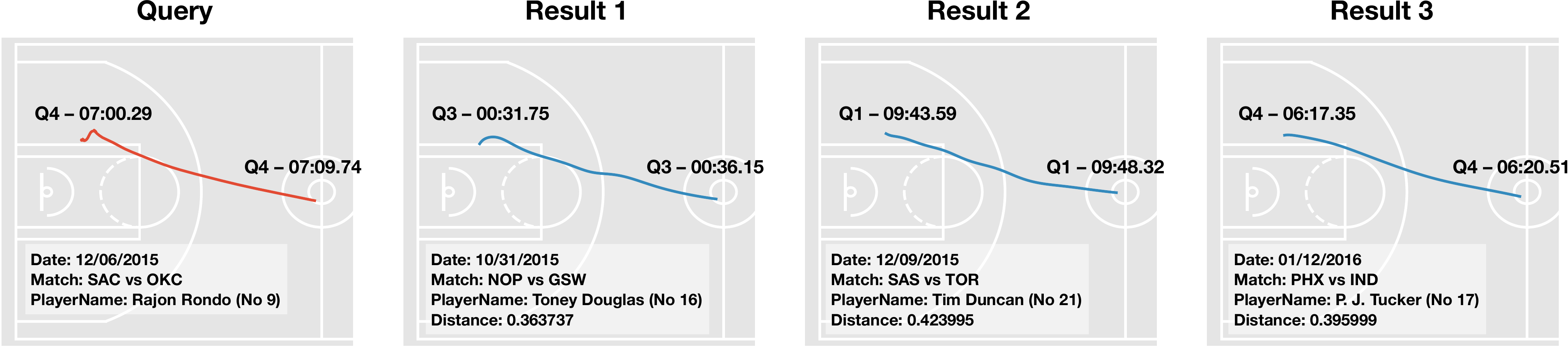}
\caption{Example of querying NBA trajectories using tSTAT with $R=0.45$.}
\label{fig:NBA_example}
\end{figure*}

\fref{fig:NBA_example} shows an example of querying NBA trajectories using tSTAT with $R=0.45$. 
Similar movements of NBA players for a short movement of Rajon Rondo in the match between Sacramento Kings and Oklahoma City Thunder on December 6, 2015 were retrieved from a large collection of 3.3 million trajectories in the NBA dataset. 
tSTAT successfully found similar movements of NBA players such as a movement of Tim Duncan in the match between San Antonio Spurs and Toronto Raptors on December 9, 2015. 
The result demonstrated an effectiveness of tSTAT for discovering similar movements of NBA payers, which 
would be beneficial to analyze various movements in sports data analyses \cite{10.1145/3054132,sha2016chalkboarding,DBLP:journals/corr/abs-1710-02255}. 

\section{Conclusion}
\label{sect:conc}

We presented tSTAT, a novel method for fast and memory-efficient trajectory similarity search under \Frechet{} distance.
Experimental results on real-world large datasets demonstrated that tSTAT was faster and more memory-efficient than state-of-the-art similarity search methods.

Our method is applicable to other distance measures such as continuous \Frechet{} distance and dynamic time warping distance on trajectories, and 
several methods was presented for those distance measures~\cite{driemel2017locality,ceccarello2019fresh}.
Thus, One important future work is to develop similarity searches for other distance measures by extending the idea behind tSTAT. 
This would be beneficial for users analyzing massive trajectories in research and industry.

\begin{acks}
KF's work was supported by JSPS KAKENHI (Grant Numbers 19H04941 and 20H04075).
We would like to thank Yoichi Sasaki for introducing succinct data structures on trits to us.
We also would like to thank Giulio Ermanno Pibiri for useful comments.
\end{acks}

\bibliographystyle{plain}
\bibliography{bibfiles/library_Lv3,bibfiles/datasets}

\begin{thebibliography}{10}

\bibitem{datasets:NBA}
{NBA Advanced Stats}.
\newblock \url{https://stats.nba.com}.

\bibitem{datasets:OSM}
{OpenStreetMap Data Extracts}.
\newblock \url{http://download.geofabrik.de/index.html}.

\bibitem{datasets:PortoTaxi}
{Taxi Service Trajectory (TST) Prediction Challenge 2015}.
\newblock \url{http://www.geolink.pt/ecmlpkdd2015-challenge/index.html}.

\bibitem{alt1995computing}
Helmut Alt and Michael Godau.
\newblock {Computing the Fr{\'{e}}chet distance between two polygonal curves}.
\newblock {\em Int. J. Comput. Geom. Appl.}, 5(01n02):75--91, 1995.

\bibitem{arslan2002dictionary}
Abdullah~N Arslan and {\"{O}}mer Eğecioğlu.
\newblock {Dictionary look-up within small edit distance}.
\newblock In {\em COCOON}, pages 127--136, 2002.

\bibitem{astefanoaei2018multi}
Maria Astefanoaei, Paul Cesaretti, Panagiota Katsikouli, Mayank Goswami, and
  Rik Sarkar.
\newblock {Multi-resolution sketches and locality sensitive hashing for fast
  trajectory processing}.
\newblock In {\em SIGSPATIAL}, pages 279--288, 2018.

\bibitem{baldus2017fast}
Julian Baldus and Karl Bringmann.
\newblock {A fast implementation of near neighbors queries for Fr{\'{e}}chet
  distance (GIS Cup)}.
\newblock In {\em SIGSPATIAL}, page~99, 2017.

\bibitem{benoit2005representing}
David Benoit, Erik~D. Demaine, J.~Ian Munro, Rajeev Raman, Venkatesh Raman, and
  S.~Srinivasa Rao.
\newblock {Representing trees of higher degree}.
\newblock {\em Algorithmica}, 43(4):275--292, 2005.

\bibitem{bringmann2014walking}
Karl Bringmann.
\newblock {Why walking the dog takes time: Fr{\'{e}}chet distance has no
  strongly subquadratic algorithms unless SETH fails}.
\newblock In {\em FOCS}, pages 661--670, 2014.

\bibitem{bringmann2019walking}
Karl Bringmann, Marvin K{\"{u}}nnemann, and Andr{\'{e}} Nusser.
\newblock {Walking the dog fast in practice: Algorithm engineering of the
  Fr{\'{e}}chet distance}.
\newblock In {\em SoCG}, volume 129, pages 17:1----17:21, 2019.

\bibitem{buchin2011detecting}
Kevin Buchin, Maike Buchin, Joachim Gudmundsson, Maarten L{\"{o}}ffler, and Jun
  Luo.
\newblock {Detecting commuting patterns by clustering subtrajectories}.
\newblock {\em Int. J. Comput. Geom. Appl.}, 21(03):253--282, 2011.

\bibitem{buchin2017efficient}
Kevin Buchin, Yago Diez, Tom van Diggelen, and Wouter Meulemans.
\newblock {Efficient trajectory queries under the Fr{\'{e}}chet distance (GIS
  Cup)}.
\newblock In {\em SIGSPATIAL}, page 101, 2017.

\bibitem{campbell2015clustering}
Jonathan~C Campbell, Jonathan Tremblay, and Clark Verbrugge.
\newblock {Clustering player paths}.
\newblock In {\em FDG}, 2015.

\bibitem{ceccarello2019fresh}
Matteo Ceccarello, Anne Driemel, and Francesco Silvestri.
\newblock {FRESH: Fr{\'{e}}chet similarity with hashing}.
\newblock In {\em WADS}, pages 254--268, 2019.

\bibitem{driemel2017locality}
Anne Driemel and Francesco Silvestri.
\newblock {Locality-sensitive hashing of curves}.
\newblock In {\em SoCG}, 2017.

\bibitem{dutsch2017filter}
Fabian D{\"{u}}tsch and Jan Vahrenhold.
\newblock {A filter-and-refinement-algorithm for range queries based on the
  Fr{\'{e}}chet distance (GIS Cup)}.
\newblock In {\em SIGSPATIAL}, page 100, 2017.

\bibitem{eiter1994computing}
Thomas Eiter and Heikki Mannila.
\newblock {Computing discrete Fr{\'{e}}chet distance}.
\newblock Technical report, TU Vienna, 1994.

\bibitem{fischer2016glouds}
Johannes Fischer and Daniel Peters.
\newblock {GLOUDS: Representing tree-like graphs}.
\newblock {\em J. Discrete Algorithm.}, 36:39--49, 2016.

\bibitem{fredkin1960trie}
Edward Fredkin.
\newblock {Trie memory}.
\newblock {\em Commun. ACM}, 3(9):490--499, 1960.

\bibitem{gog2014theory}
Simon Gog, Timo Beller, Alistair Moffat, and Matthias Petri.
\newblock {From theory to practice: Plug and play with succinct data
  structures}.
\newblock In {\em SEA}, pages 326--337, 2014.

\bibitem{gog2016fast}
Simon Gog and Rossano Venturini.
\newblock {Fast and compact Hamming distance index}.
\newblock In {\em SIGIR}, pages 285--294, 2016.

\bibitem{gonzalez2005practical}
Rodrigo Gonz{\'{a}}lez, Szymon Grabowski, Veli M{\"{a}}kinen, and Gonzalo
  Navarro.
\newblock {Practical implementation of rank and select queries}.
\newblock In {\em WEA}, pages 27--38, 2005.

\bibitem{greene1994multi}
Dan Greene, Michal Parnas, and Frances Yao.
\newblock {Multi-index hashing for information retrieval}.
\newblock In {\em FOCS}, pages 722--731, 1994.

\bibitem{10.1145/3054132}
Joachim Gudmundsson and Michael Horton.
\newblock {Spatio-temporal analysis of team sports}.
\newblock {\em ACM Comput. Surv.}, 50(2), 2017.

\bibitem{indyk2002approximate}
Piotr Indyk.
\newblock {Approximate nearest neighbor algorithms for Fr{\'{e}}chet distance
  via product metrics}.
\newblock In {\em SCG}, pages 102--106, 2002.

\bibitem{indyk1998approximate}
Piotr Indyk and Rajeev Motwani.
\newblock {Approximate nearest neighbors: towards removing the curse of
  dimensionality}.
\newblock In {\em STOC}, pages 604--613, 1998.

\bibitem{jacobson1989space}
Guy Jacobson.
\newblock {Space-efficient static trees and graphs}.
\newblock In {\em FOCS}, pages 549--554, 1989.

\bibitem{kanda2017compressed}
Shunsuke Kanda, Kazuhiro Morita, and Masao Fuketa.
\newblock {Compressed double-array tries for string dictionaries supporting
  fast lookup}.
\newblock {\em Knowl. Inf. Syst.}, 51(3):1023--1042, 2017.

\bibitem{konzack2017visual}
Maximilian Konzack, Thomas McKetterick, Tim Ophelders, Maike Buchin, Luca
  Giuggioli, Jed Long, Trisalyn Nelson, Michel~A Westenberg, and Kevin Buchin.
\newblock {Visual analytics of delays and interaction in movement data}.
\newblock {\em Int. J. Geogr. Inf. Sci.}, 31(2):320--345, 2017.

\bibitem{krogh2016efficient}
Benjamin Krogh, Christian~S Jensen, and Kristian Torp.
\newblock {Efficient In-memory indexing of network-constrained trajectories}.
\newblock In {\em SIGSPATIAL}, 2016.

\bibitem{li2008efficient}
Chen Li, Jiaheng Lu, and Yiming Lu.
\newblock {Efficient merging and filtering algorithms for approximate string
  searches}.
\newblock In {\em ICDE}, pages 257--266, 2008.

\bibitem{liu2011large}
Alex~X Liu, Ke~Shen, and Eric Torng.
\newblock {Large scale hamming distance query processing}.
\newblock In {\em ICDE}, pages 553--564, 2011.

\bibitem{luo2013finding}
Wuman Luo, Haoyu Tan, Lei Chen, and Lionel~M Ni.
\newblock {Finding time period-based most frequent path in big trajectory
  data}.
\newblock In {\em SIGMOD}, pages 713--724, 2013.

\bibitem{manku2007detecting}
Gurmeet~Singh Manku, Arvind Jain, and Anish {Das Sarma}.
\newblock {Detecting near-duplicates for web crawling}.
\newblock In {\em WWW}, pages 141--150, 2007.

\bibitem{norouzi2014fast}
Mohammad Norouzi, Ali Punjani, and David~J Fleet.
\newblock {Fast exact search in Hamming space with multi-index hashing}.
\newblock {\em IEEE Trans. Pattern Anal. Mach. Intell.}, 36(6):1107--1119,
  2014.

\bibitem{patrascu2008succincter}
Mihai Patrascu.
\newblock {Succincter}.
\newblock In {\em FOCS}, pages 305--313, 2008.

\bibitem{pibiri2017efficient}
Giulio~Ermanno Pibiri and Rossano Venturini.
\newblock {Efficient data structures for massive n-gram datasets}.
\newblock In {\em SIGIR}, pages 615--624, 2017.

\bibitem{pibiri2019handling}
Giulio~Ermanno Pibiri and Rossano Venturini.
\newblock {Handling massive N-gram datasets efficiently}.
\newblock {\em ACM Trans. Inf. Syst.}, 37(2), 2019.

\bibitem{qin2019generalizing}
Jianbin Qin, Chuan Xiao, Yaoshu Wang, and Wei Wang.
\newblock {Generalizing the pigeonhole principle for similarity search in
  Hamming space}.
\newblock {\em IEEE Trans. Knowl. Data Eng.}, 2019.

\bibitem{rayatidamavandi2017comparison}
Maede Rayatidamavandi, Yu~Zhuang, and Mahshid Rahnamay-Naeini.
\newblock {A comparison of hash-based methods for trajectory clustering}.
\newblock In {\em DASC}, pages 107--112, 2017.

\bibitem{sanchez2016fast}
Ivan Sanchez, Zay Maung~Maung Aye, Benjamin I.~P. Rubinstein, and Kotagiri
  Ramamohanarao.
\newblock {Fast trajectory clustering using hashing methods}.
\newblock In {\em IJCNN}, pages 3689--3696, 2016.

\bibitem{sha2016chalkboarding}
Long Sha, Patrick Lucey, Yisong Yue, Peter Carr, Charlie Rohlf, and Iain
  Matthews.
\newblock {Chalkboarding: A new spatiotemporal query paradigm for sports play
  retrieval}.
\newblock In {\em IUI}, pages 336--347, 2016.

\bibitem{DBLP:journals/corr/abs-1710-02255}
Long Sha, Patrick Lucey, Stephan Zheng, Taehwan Kim, Yisong Yue, and Sridha
  Sridharan.
\newblock {Fine-grained retrieval of sports plays using tree-based alignment of
  trajectories}.
\newblock {\em CoRR}, abs/1710.0, 2017.

\bibitem{shang2014personalized}
Shuo Shang, Ruogu Ding, Kai Zheng, Christian~S Jensen, Panos Kalnis, and
  Xiaofang Zhou.
\newblock {Personalized trajectory matching in spatial networks}.
\newblock {\em The VLDB Journal}, 23(3):449--468, 2014.

\bibitem{shang2018dita}
Zeyuan Shang, Guoliang Li, and Zhifeng Bao.
\newblock {DITA: Distributed in-memory trajectory analytics}.
\newblock In {\em SIGMOD}, pages 725--740, 2018.

\bibitem{song2014press}
Renchu Song, Weiwei Sun, Baihua Zheng, and Yu~Zheng.
\newblock {PRESS: A novel framework of trajectory compression in road
  networks}.
\newblock {\em PVLDB}, 7(9):661--672, 2014.

\bibitem{sriraghavendra2007frechet}
E~Sriraghavendra, K~Karthik, and Chiranjib Bhattacharyya.
\newblock {Fr{\'{e}}chet distance based approach for searching online
  handwritten documents}.
\newblock In {\em ICDAR}, volume~1, pages 461--465, 2007.

\bibitem{toohey2015trajectory}
Kevin Toohey and Matt Duckham.
\newblock {Trajectory similarity measures}.
\newblock {\em SIGSPATIAL Special}, 7(1):43--50, 2015.

\bibitem{vigna2008broadword}
Sebastiano Vigna.
\newblock {Broadword implementation of rank/select queries}.
\newblock In {\em WEA}, pages 154--168, 2008.

\bibitem{wang2018torch}
Sheng Wang, Zhifeng Bao, J~Shane Culpepper, Zizhe Xie, Qizhi Liu, and Xiaolin
  Qin.
\newblock {Torch: A search engine for trajectory data}.
\newblock In {\em SIGIR}, pages 535--544, 2018.

\bibitem{werner2018acm}
Martin Werner and Dev Oliver.
\newblock {ACM SIGSPATIAL GIS Cup 2017: Range queries under Fr{\'{e}}chet
  distance}.
\newblock {\em SIGSPATIAL Special}, 10(1):24--27, 2018.

\bibitem{wylie2013protein}
Tim Wylie and Binhai Zhu.
\newblock {Protein chain pair simplification under the discrete Fr{\'{e}}chet
  distance}.
\newblock {\em IEEE/ACM Trans. Comput. Biol. Bioinf.}, 10(6):1372--1383, 2013.

\bibitem{xie2017distributed}
Dong Xie, Feifei Li, and Jeff~M Phillips.
\newblock {Distributed trajectory similarity search}.
\newblock {\em PVLDB}, 10(11):1478--1489, 2017.

\bibitem{yuan2019distributed}
H~Yuan and G~Li.
\newblock {Distributed In-memory trajectory similarity search and join on road
  network}.
\newblock In {\em ICDE}, pages 1262--1273, 2019.

\bibitem{zhang2018surf}
Huanchen Zhang, Hyeontaek Lim, Viktor Leis, David~G Andersen, Michael Kaminsky,
  Kimberly Keeton, and Andrew Pavlo.
\newblock {SuRF: Practical range query filtering with fast succinct tries}.
\newblock In {\em SIGMOD}, pages 323--336, 2018.

\bibitem{zhang2013hmsearch}
Xiaoyang Zhang, Jianbin Qin, Wei Wang, Yifang Sun, and Jiaheng Lu.
\newblock {HmSearch: An efficient Hamming distance query processing algorithm}.
\newblock In {\em SSDBM}, page~19, 2013.

\bibitem{zheng2015trajectory}
Yu~Zheng.
\newblock {Trajectory data mining: an overview}.
\newblock {\em ACM Trans. Intell. Syst. Technol.}, 6(3):29, 2015.

\bibitem{zhu2010mining}
Haohan Zhu, Jun Luo, Hang Yin, Xiaotao Zhou, Joshua~Zhexue Huang, and
  F~Benjamin Zhan.
\newblock {Mining trajectory corridors using Fr{\'{e}}chet distance and meshing
  grids}.
\newblock In {\em SIGKDD}, pages 228--237, 2010.

\end{thebibliography}

\clearpage

\appendix
\newcommand{\ChartWidthApp}{41mm}

\section{Implementation Details}
\label{appx:impl}

\subsection{Succinct Data Structures on Trits}

Although many practical implementations of succinct data structures on bits have been developed and are available online (e.g., \cite{vigna2008broadword,gog2014theory}), there is no any available implementation on trits as far as we know.
Thus, we design a practical implementation of the data structure on trits as follows.

\paragraph{Compact Implementation of Trit Array}

Given a trit array $A$ of length $M$, a straightforward compact implementation represents each trit using two bits, which consumes $2M$ bits of memory.
However, this memory usage is not close to the theoretically-optimal one of $\Ceil{M\log_2 3} = \Ceil{1.58M}$ bits \cite{patrascu2008succincter}.

To more compactly handle trits on byte-aligned memory, we pack five trits $c_1,c_2,\dots,c_5$ into one chunk $\sum_{i=1}^{5}{c_i 3^{i-1}}$ (called \emph{tryte}) and store the tryte using one byte, as in \cite{fischer2016glouds}.
For example, five trits $1$, $2$, $2$, $0$ and $1$ are packed into tryte $1 \cdot 3^0 + 2 \cdot 3^1 + 2 \cdot 3^2 + 0 \cdot 3^3 + 1 \cdot 3^4 = 106$, which can be stored within one byte.
For a trit array $A$ of length $M$, we pack five trits each and produce a tryte array $A'$ of length $\Ceil{M/5}$ in which each element is represented as a byte.
Then, $A[i]$ can be extracted by getting the corresponding tryte $t \gets A'[\Floor{i/5}]$ and computing $\Floor{t/3^{i \bmod{} 5}} \bmod 3$.
The memory usage of $A'$ is $8 \Ceil{M/5} = \Ceil{1.6M}$ bits and is close to the optimal $\Ceil{1.58M}$ bits.

\paragraph{Rank Data Structure}

We implement $\Rank$ data structures on $A$ using a two-layer structure, in a similar manner for bit arrays \cite{gonzalez2005practical,vigna2008broadword}.
The two-layer structure partitions $A$ into blocks of a fixed-length and stores precomputed $\Rank$ results for the blocks.
The query is solved by obtaining the $\Rank$ value for the target block using the structure and computing the $\Rank$ value for the target position within the block.

\begin{figure}[tb]
\centering
\includegraphics[scale=0.26]{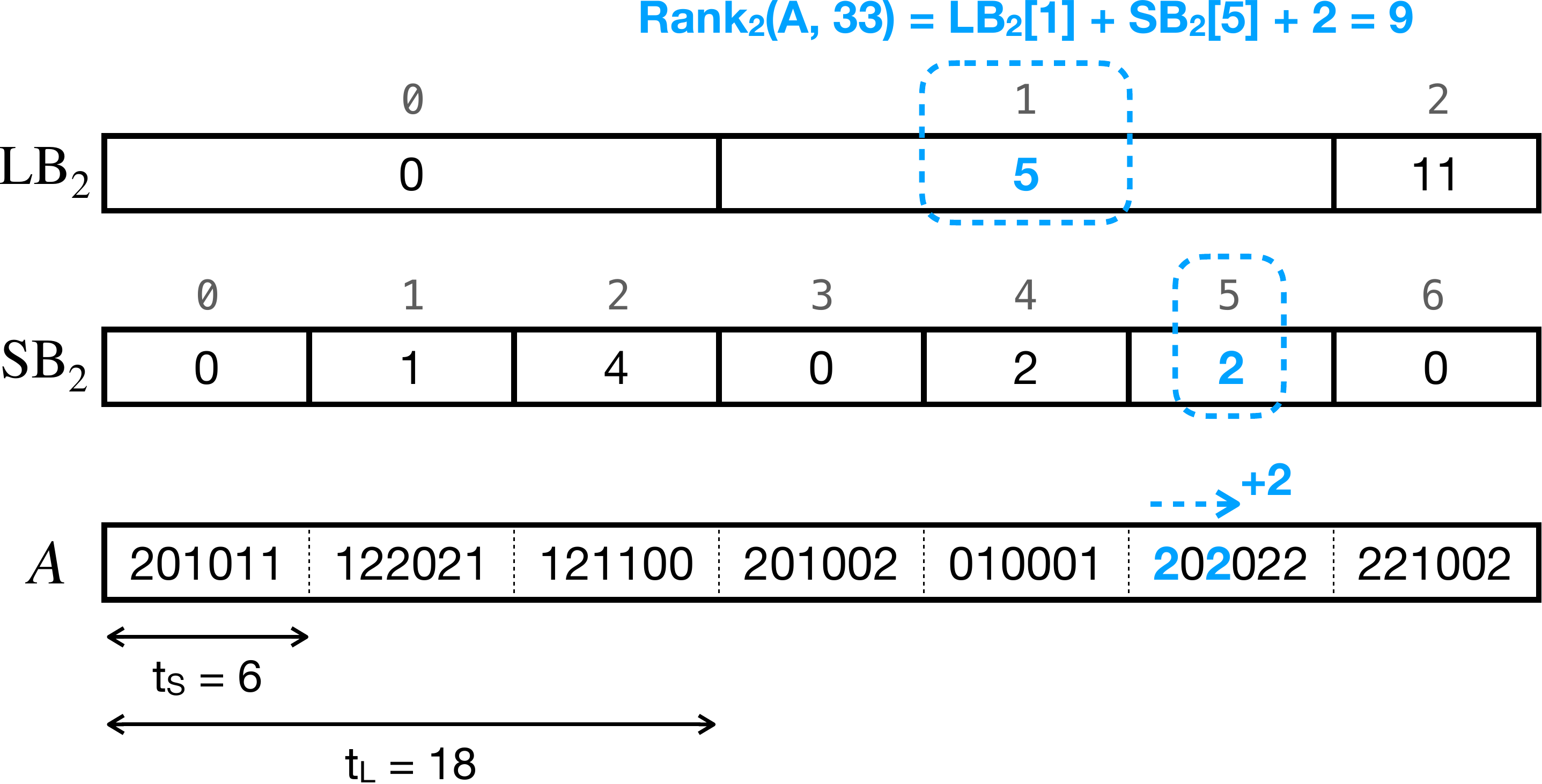}
\caption{
Example of a $\Rank_2$ data structure when $t_L = 18$ and $t_S = 6$.
$\Rank_2(A, 33)$ is computed as follows.
We first compute the Rank value at the beginning position of the corresponding small block by $\LB_2[\Floor{33/18}] + \SB_2[\Floor{33/6}] = \LB_2[1] + \SB_2[5] = 5 + 2 = 7$.
Then, we scan the remaining $33 \bmod 6 = 3$ trits in the fifth small block and find two trits of value $2$.
Thus, the result is $7 + 2 = 9$.
}
\label{fig:rank}
\end{figure}

The two-layer structure consists of two arrays $\LB_c$ and $\SB_c$ for the $\Rank_c$ operation on $A$, which are built as follows.
We first partition $A$ into subarrays consisting of $t_{L}$ trits, calling these subarrays \emph{large blocks}.
$\LB_c[j]$ stores the $\Rank_c$ value at the beginning position of the $j$-th large block, i.e., $\LB_c[j] = \Rank_c(A, t_{L} \cdot j)$.
Then, we partition $A$ into subarrays consisting of $t_{S}$ trits, calling these subarrays \emph{small blocks}, such that the $k$-th small block belongs to the $\Floor{t_{S} \cdot k / t_{L}}$-th large block.
$\SB_c[k]$ stores the $\Rank_c$ value at the beginning position of the $k$-th small block relative to the corresponding large block, i.e., $\SB_c[k] = \Rank_c(A, t_{S} \cdot k) - \LB_c[\Floor{t_{S} \cdot k / t_{L}}]$.
\fref{fig:rank} shows an example of a $\Rank_2$ structure when $t_L = 18$ and $t_S = 6$.

By using the arrays $\LB_c$ and $\SB_c$, $\Rank_c(A,i)$ is computed as follows.
We first compute the Rank value at the beginning position of the corresponding small block by $\LB_c[\Floor{i/t_L}] + \SB_c[\Floor{i/t_S}]$.
Then, we scan the remaining $(i \bmod t_S)$ trits in the small block and counts the number of $c$s.
\fref{fig:rank} shows an example of the computation of $\Rank_2(A, 33)$.
When we set $t_L = 65550$ and $t_S = 50$, $\LB_c$ and $\SB_c$ can be implemented as 64-bit and 16-bit arrays, respectively.
Then, $\LB_c$ and $\SB_c$ uses $64 \Ceil{M/65550} + 16\Ceil{M/50} \approx 0.32M$ bits of space.

In the same manner as bit arrays, $\Select_c(A,i)$ can also be implemented by binary search on $\LB_c$ and $\SB_c$.
Since tSTAT does not use the $\Select$ operation on trits, we do not mention it further.
Our implementation of succinct data structures on trits is available on \url{https://github.com/kampersanda/succinctrits}.

\subsection{Fast Hamming Distance Computation}
\label{appx:impl:ham}

We consider to compute the Hamming distance between integer sketches $S$ and $T$ of length $L$.
The computation time with a na\"{i}ve comparison approach is $\BigO{L}$, assuming that two integers can be compared in $\BigO{1}$ time.

Zhang et al. \cite{zhang2013hmsearch} proposed a fast approach by exploiting a vertical format and bit-parallelism offered by CPUs.
This approach encodes $S$ into $\hat{S}$ in a \emph{vertical} format, i.e., the $i$-th significant $L$ bits of each character of $S$ are stored to $\hat{S}[i]$ of consecutive $L$ bits.
Given sketches $\hat{S}$ and $\hat{T}$ in the vertical format, we can compute $\Ham{S,T}$ as follows.
Initially, we prepare a bitmap $bits$ of $L$ bits in which all the bits are set to zero.
For each $i = 0,1,\ldots,\Ceil{\log_2 \sigma}-1$, we iteratively perform $bits \gets bits \vee (\hat{S}[i] \oplus \hat{T}[i])$, where $\vee$ and $\oplus$ denote bitwise-OR and -XOR operations, respectively.
For the resulting $bits$, $\Popcnt{bits}$ corresponds to $\Ham{S,T}$, where $\Popcnt{\cdot}$ counts the number of 1s and belongs to the instruction sets of modern CPUs.
The operations $\vee$, $\oplus$ and $\Popcnt{\cdot}$ can be performed in $\BigO{1}$ time per machine word.
Let $w$ be the machine word size in bits. 
We can compute $\Ham{S,T}$ in $\BigO{\Ceil{\log \sigma} \cdot \Ceil{L/w}}$ time.
In practice, setting $L$ to $w$ is efficient because $\hat{S}[i]$ of $L$ bits can be represented within one machine word, i.e., $L = 64$ in our experimental environment.

\section{Experimental Results}
\label{appx:ex}

This section shows the other experimental results not presented in \sref{sect:ex}.

\begin{figure*}[p]
\centering
\setlength{\tabcolsep}{2mm}
\subfloat[Number of internal nodes]{
\includegraphics[width=\ChartWidthApp]{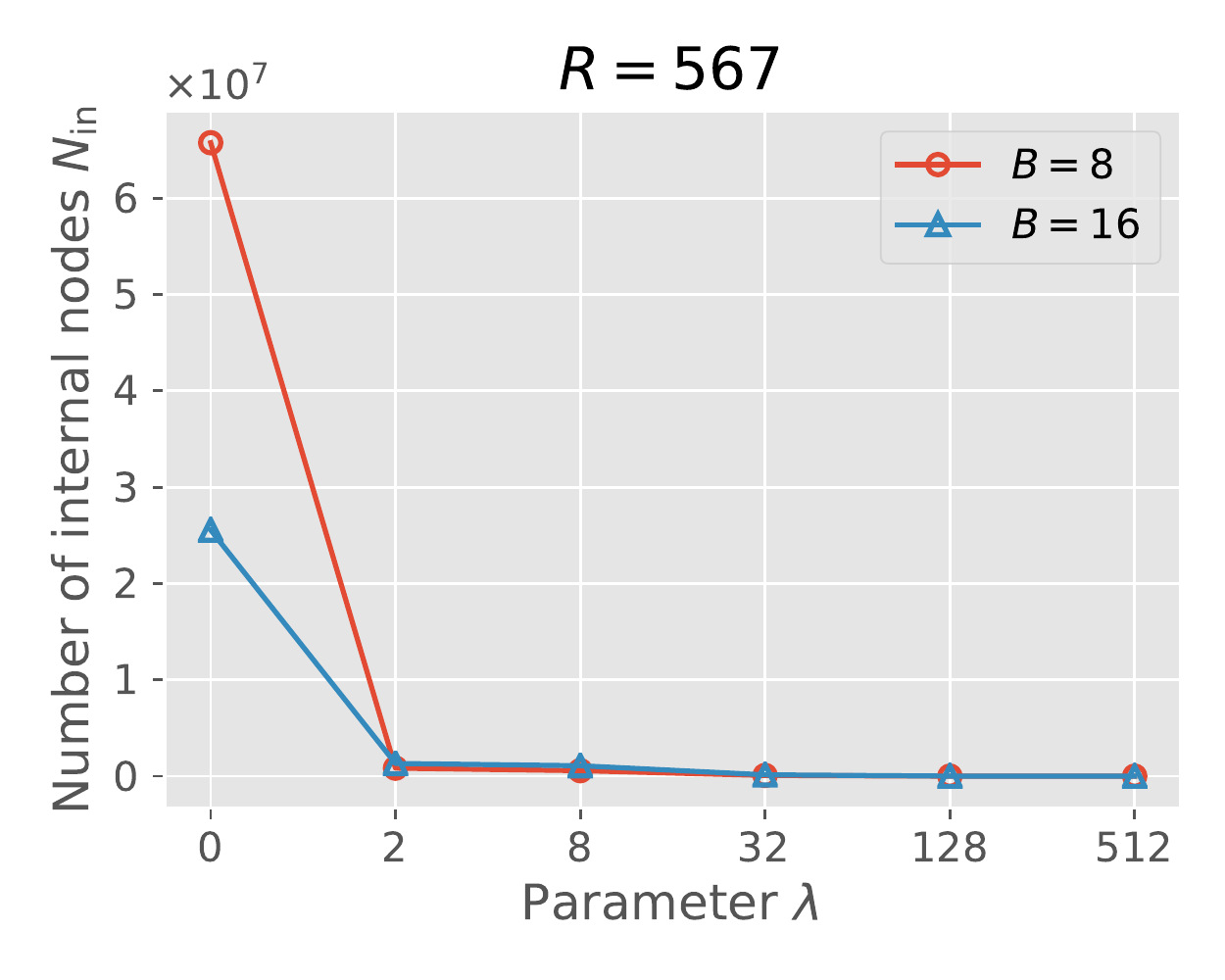}
\includegraphics[width=\ChartWidthApp]{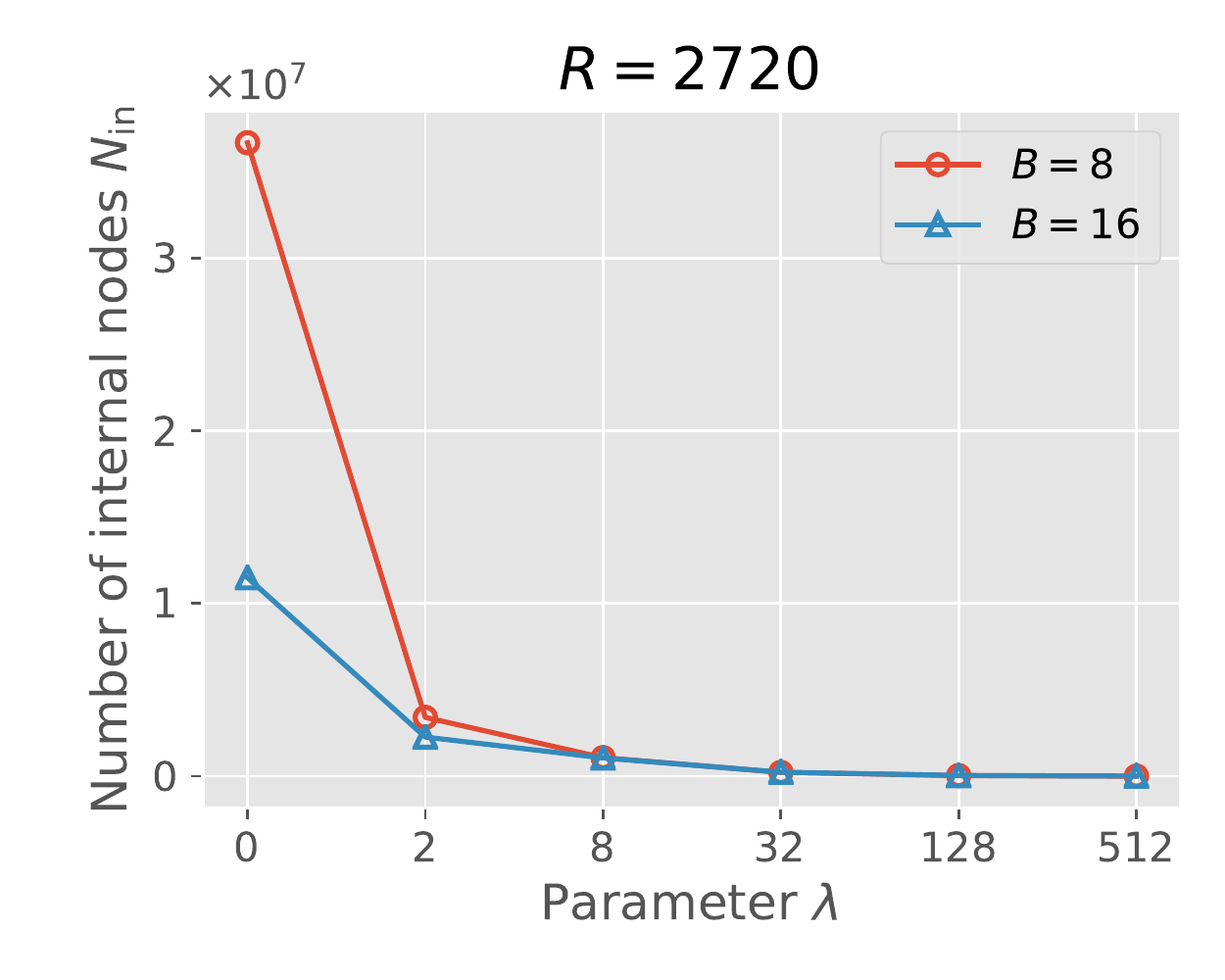}
\includegraphics[width=\ChartWidthApp]{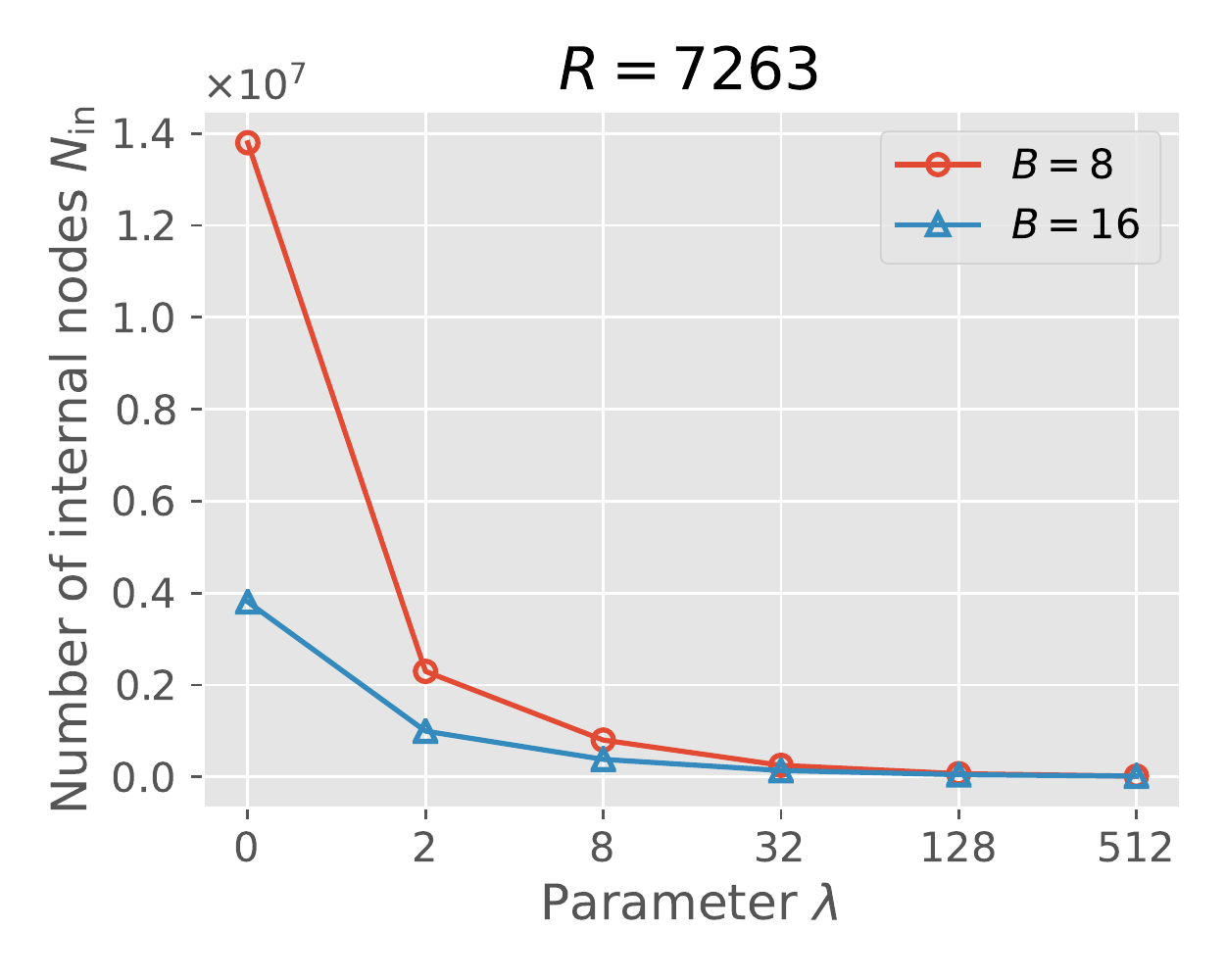}
}\\
\subfloat[Memory usage in GiB]{
\includegraphics[width=\ChartWidthApp]{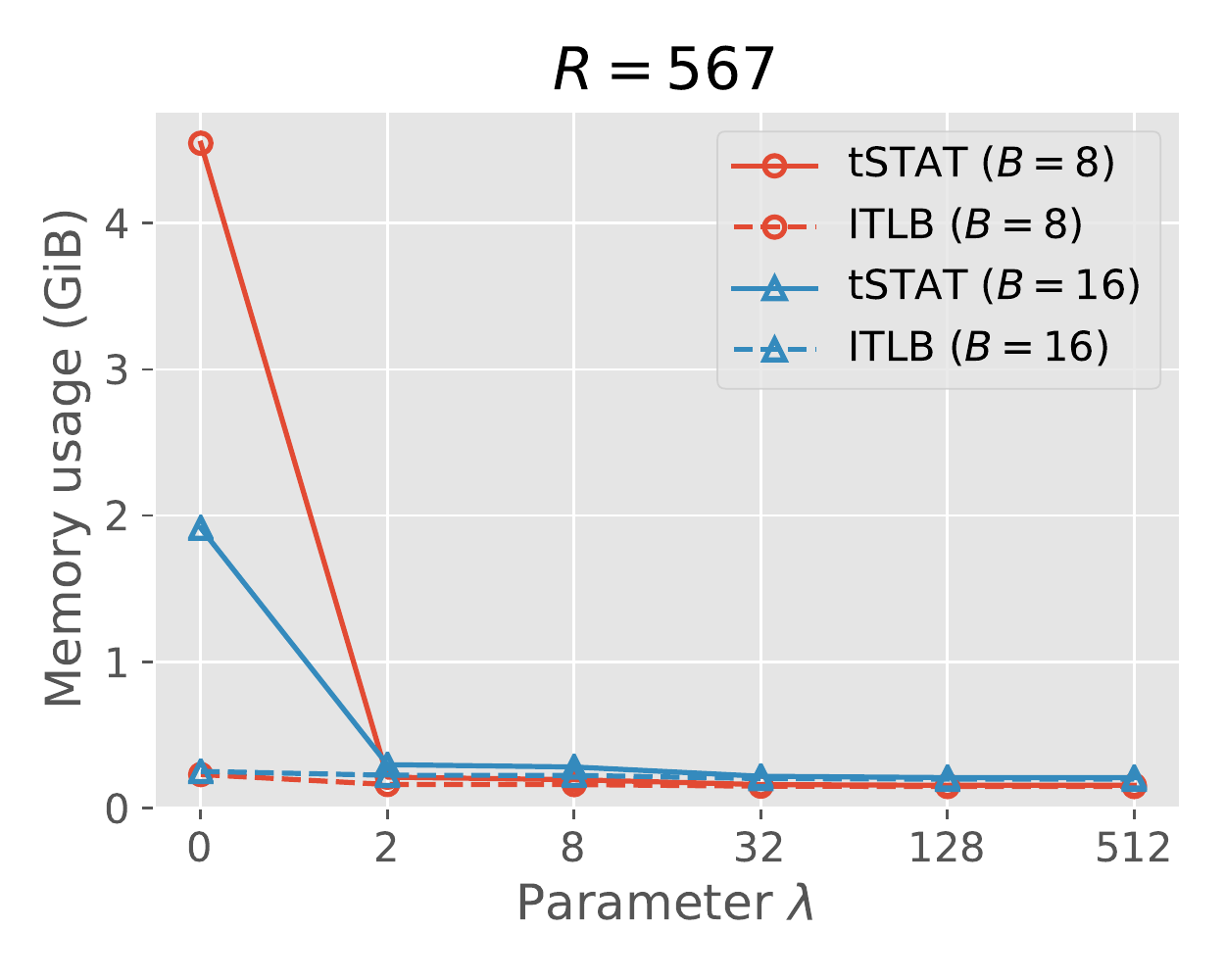}
\includegraphics[width=\ChartWidthApp]{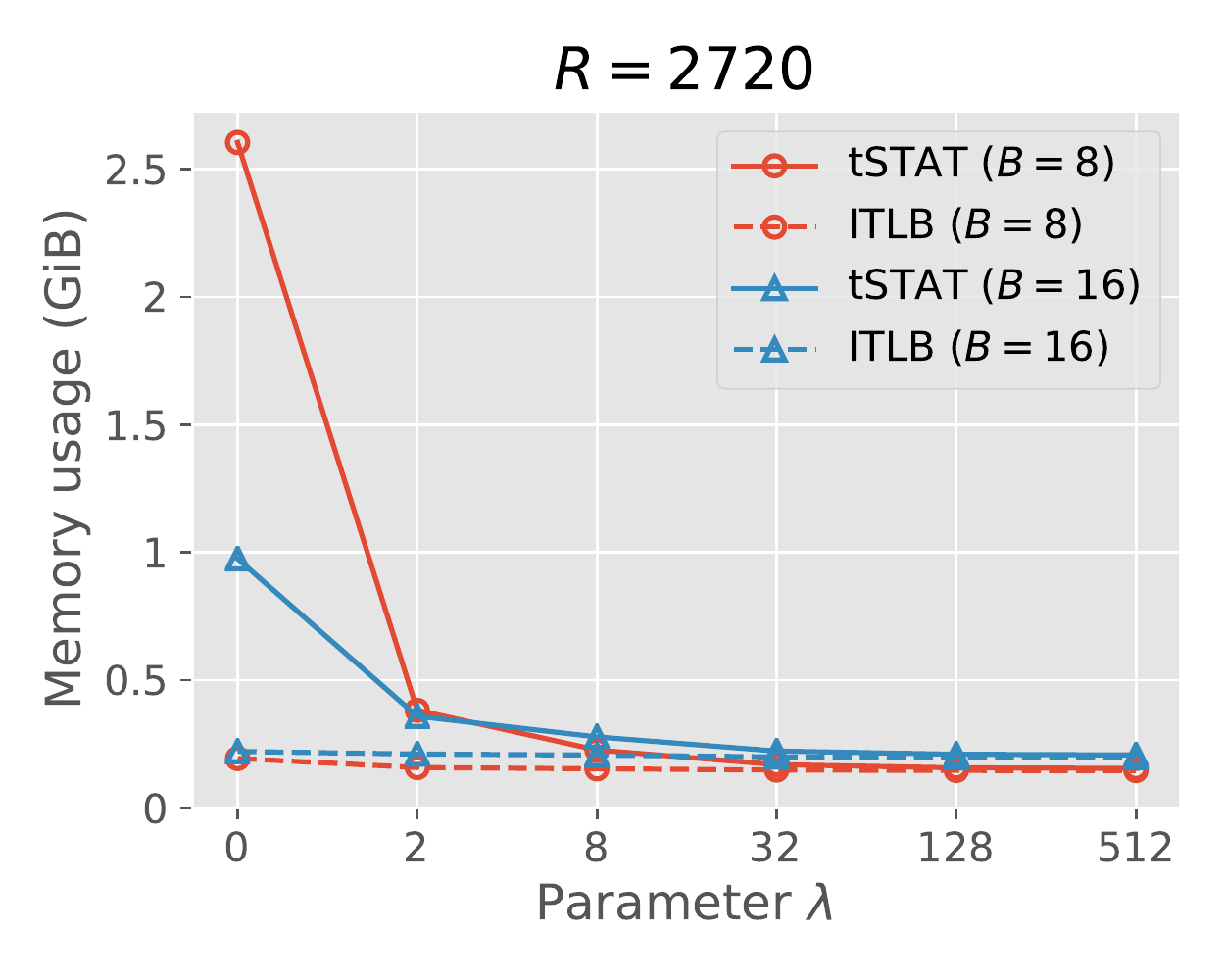}
\includegraphics[width=\ChartWidthApp]{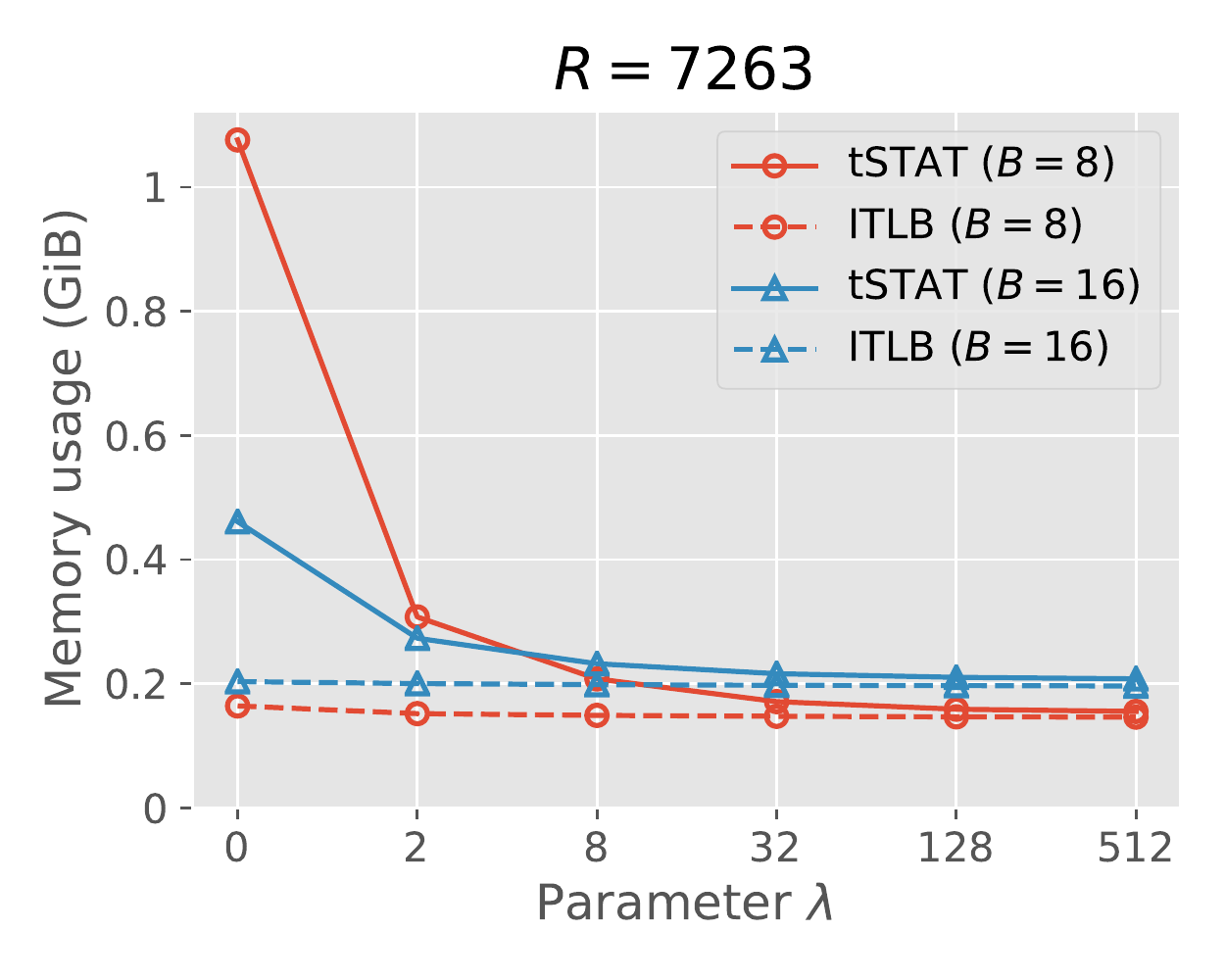}
}\\
\subfloat[Number of candidates $|\Cand|$]{
\begin{tabular}{c}
\includegraphics[width=\ChartWidthApp]{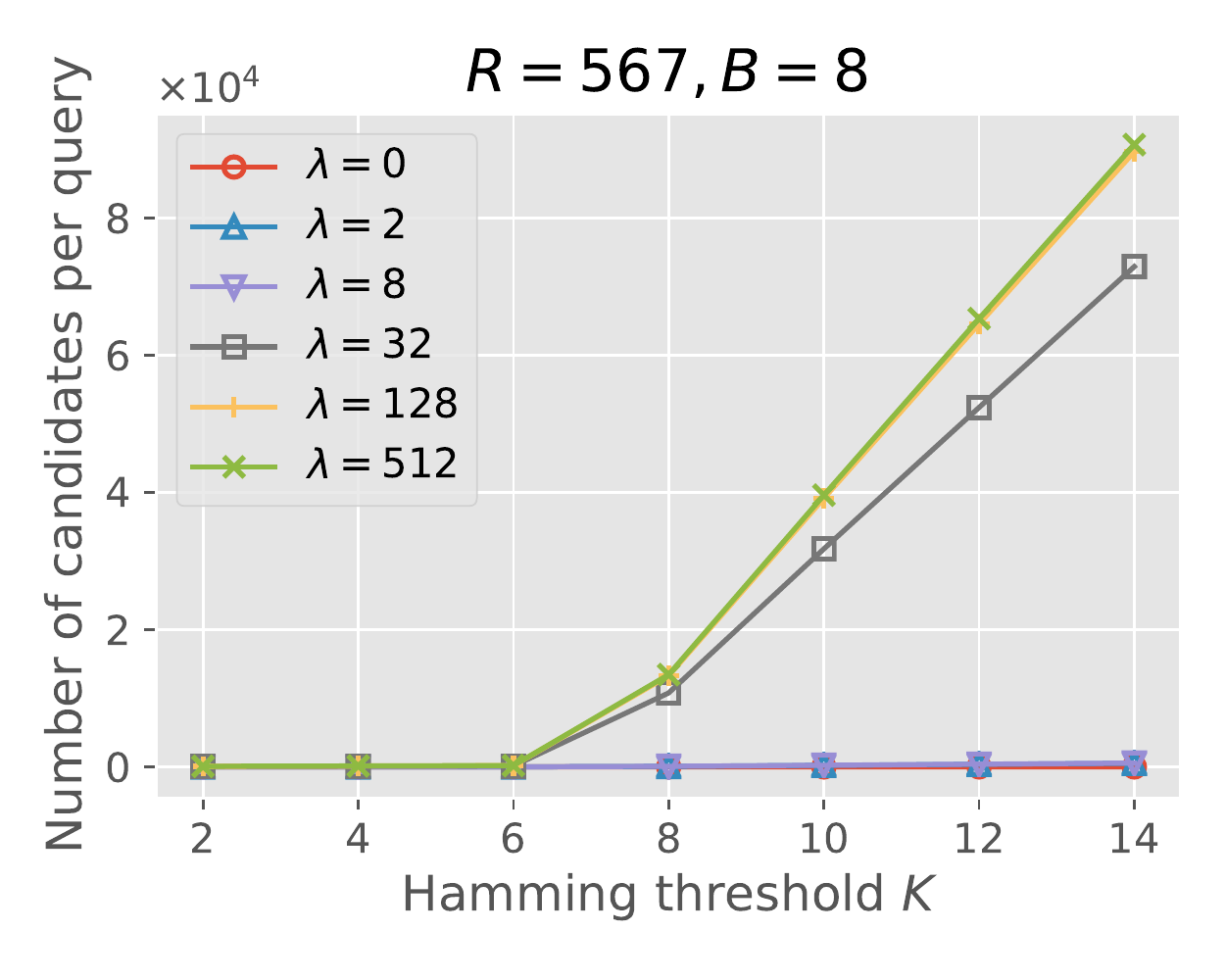}
\includegraphics[width=\ChartWidthApp]{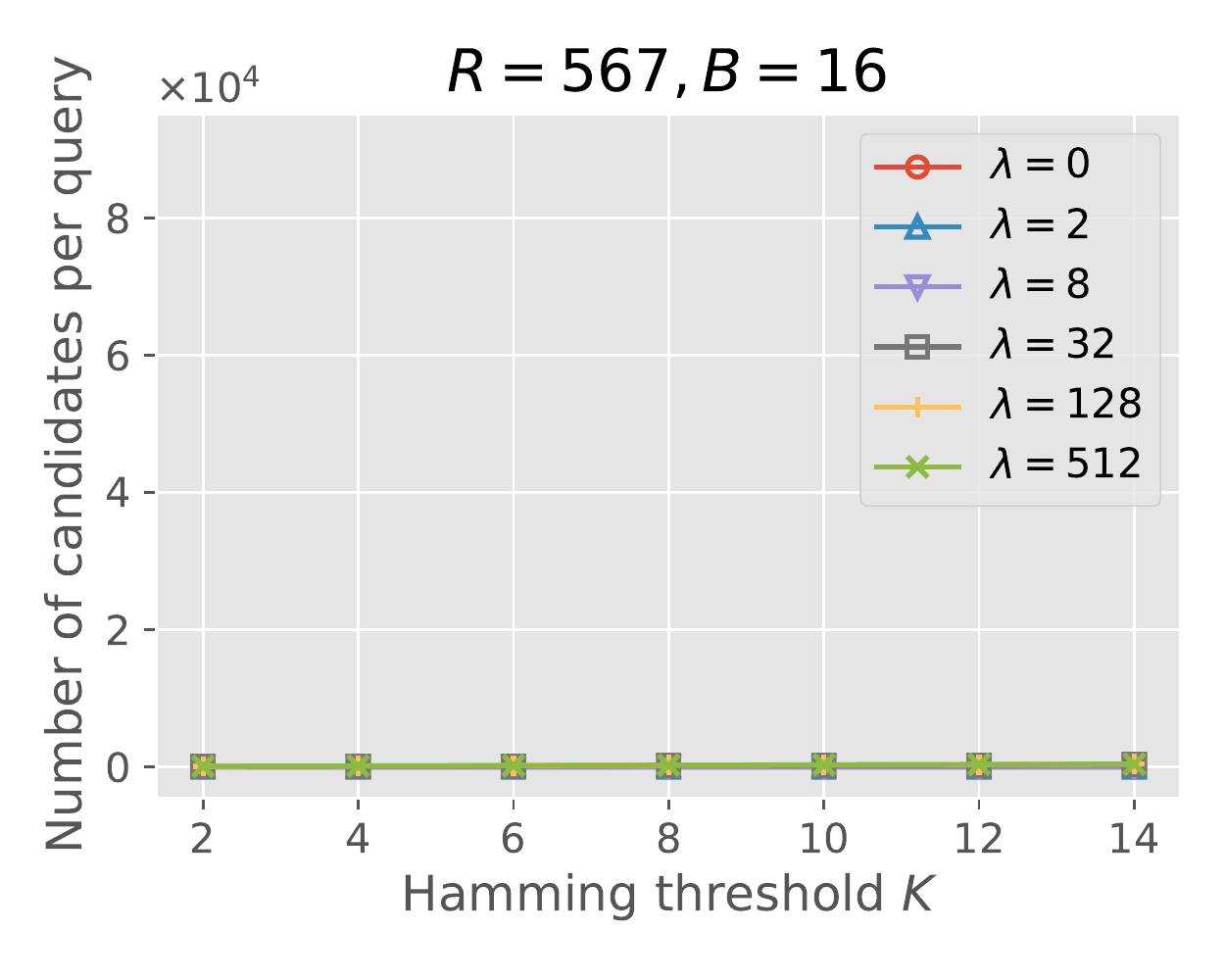}\\
\includegraphics[width=\ChartWidthApp]{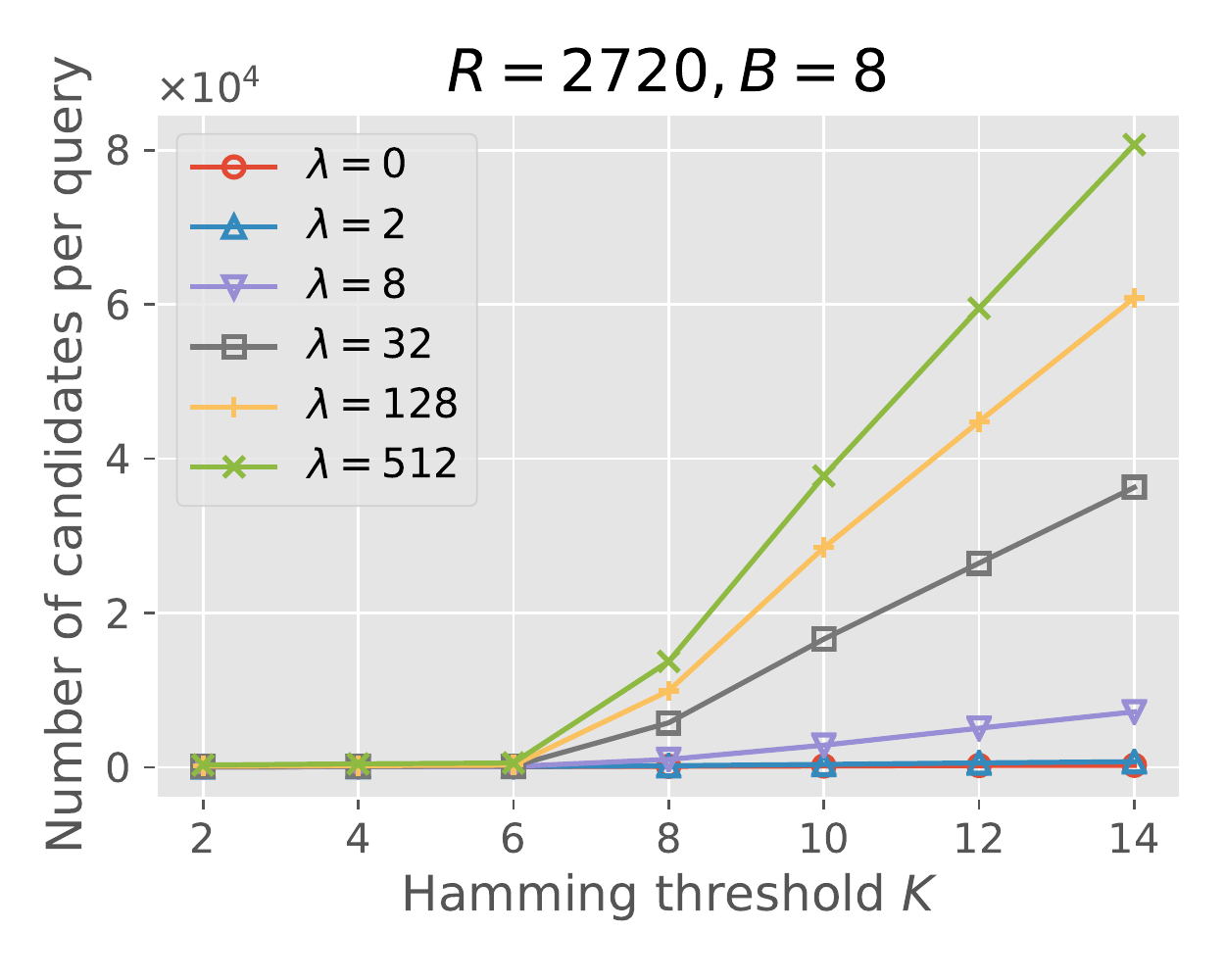}
\includegraphics[width=\ChartWidthApp]{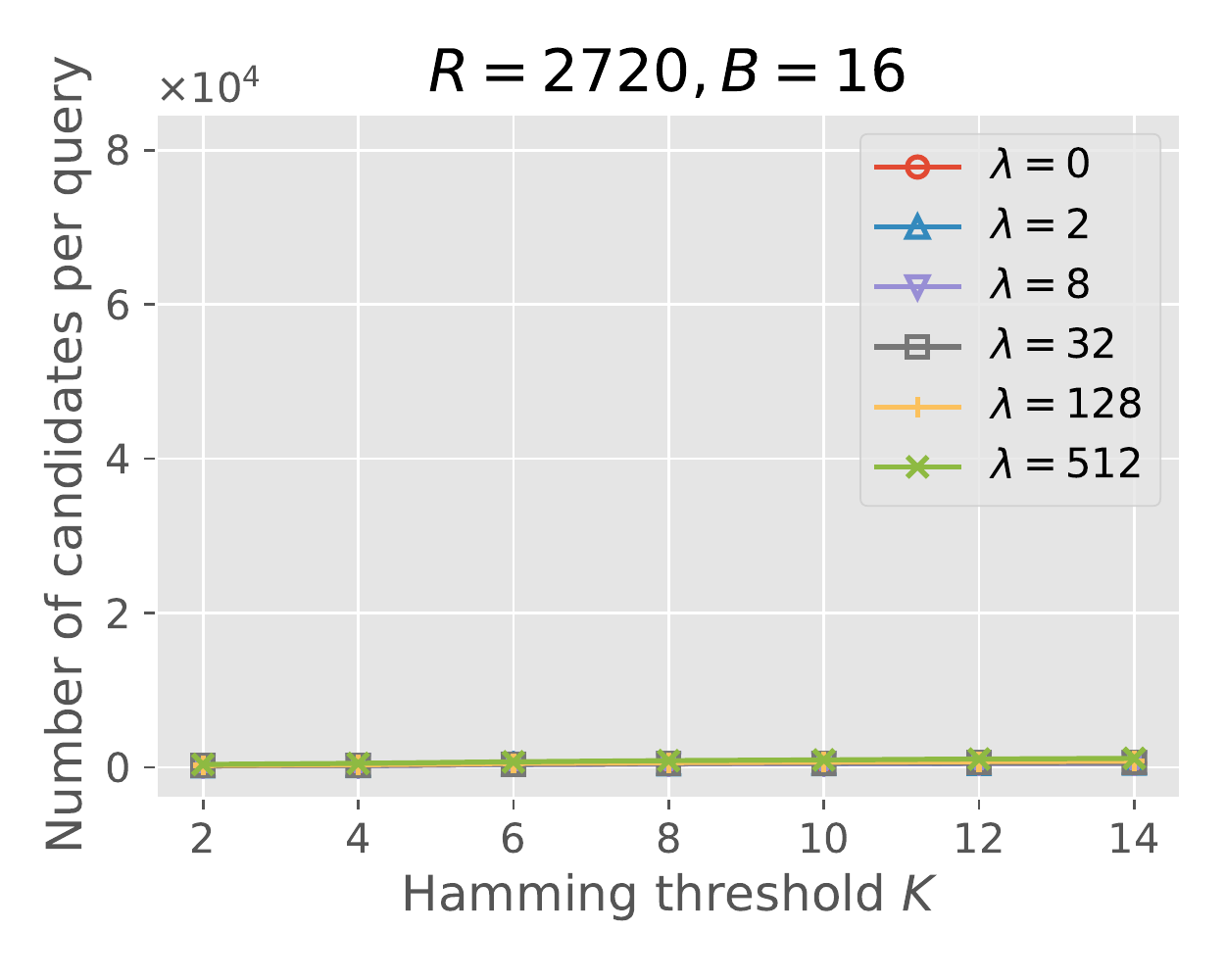}\\
\includegraphics[width=\ChartWidthApp]{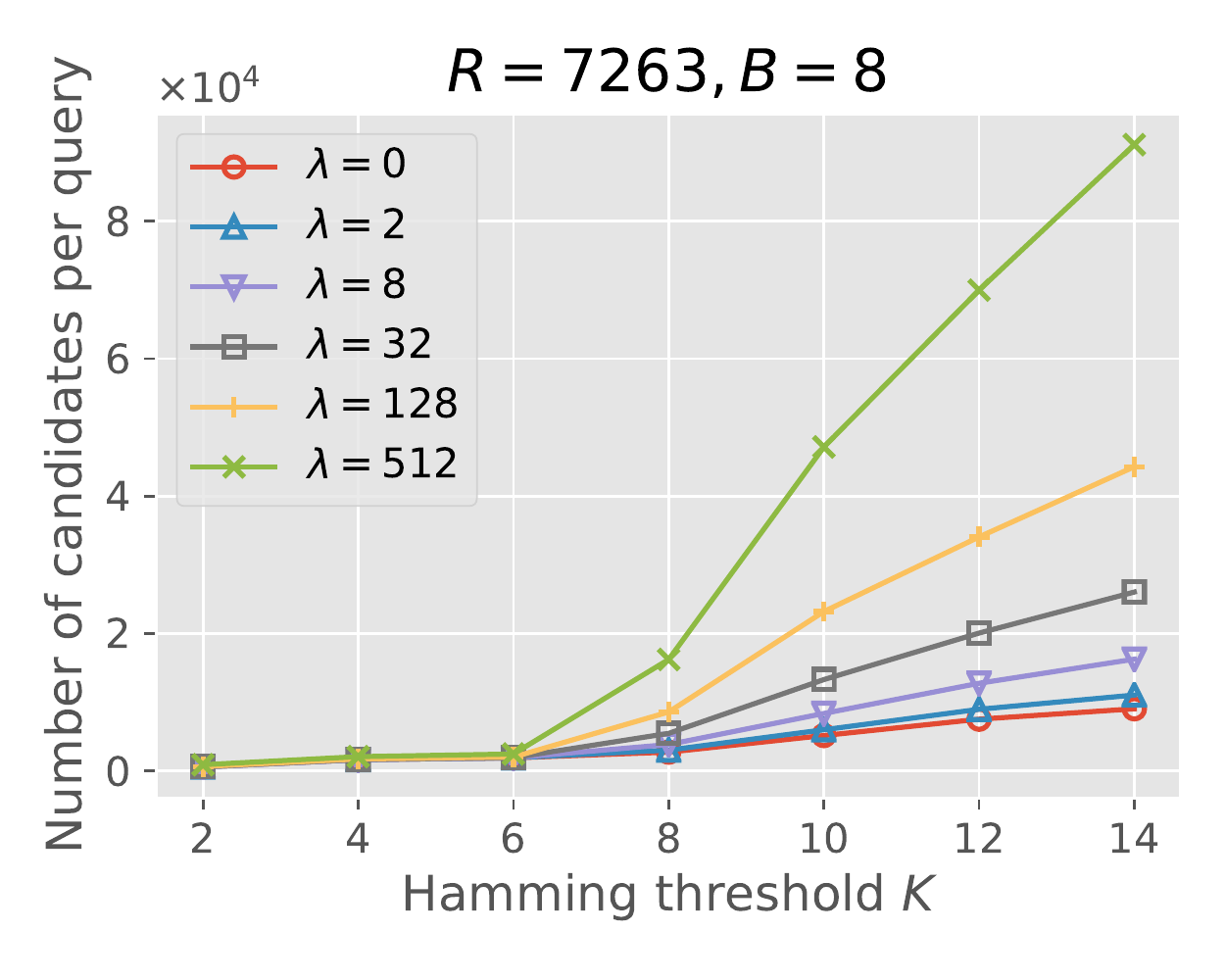}
\includegraphics[width=\ChartWidthApp]{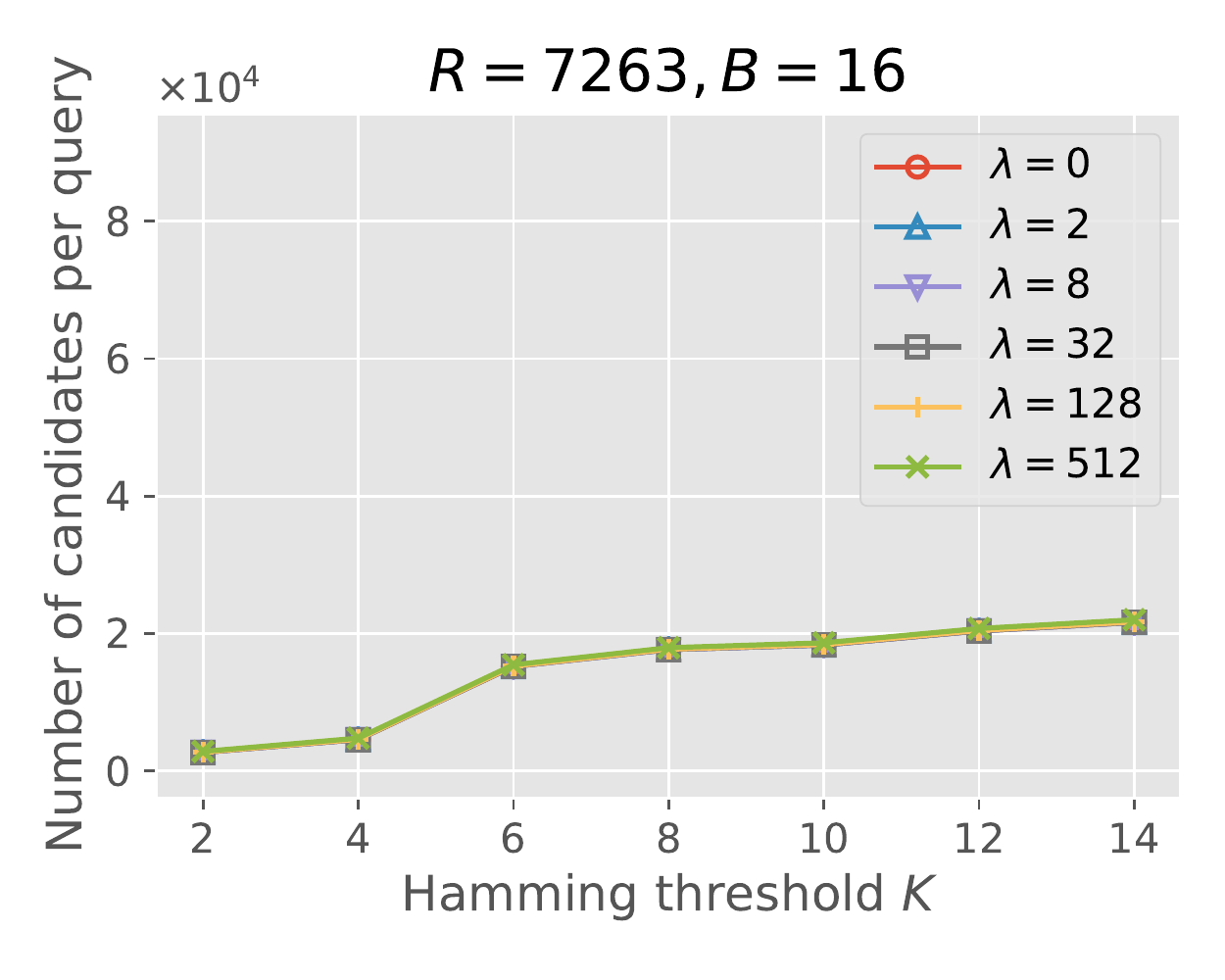}
\end{tabular}
}
\subfloat[Search time in milliseconds (ms) per query]{
\begin{tabular}{c}
\includegraphics[width=\ChartWidthApp]{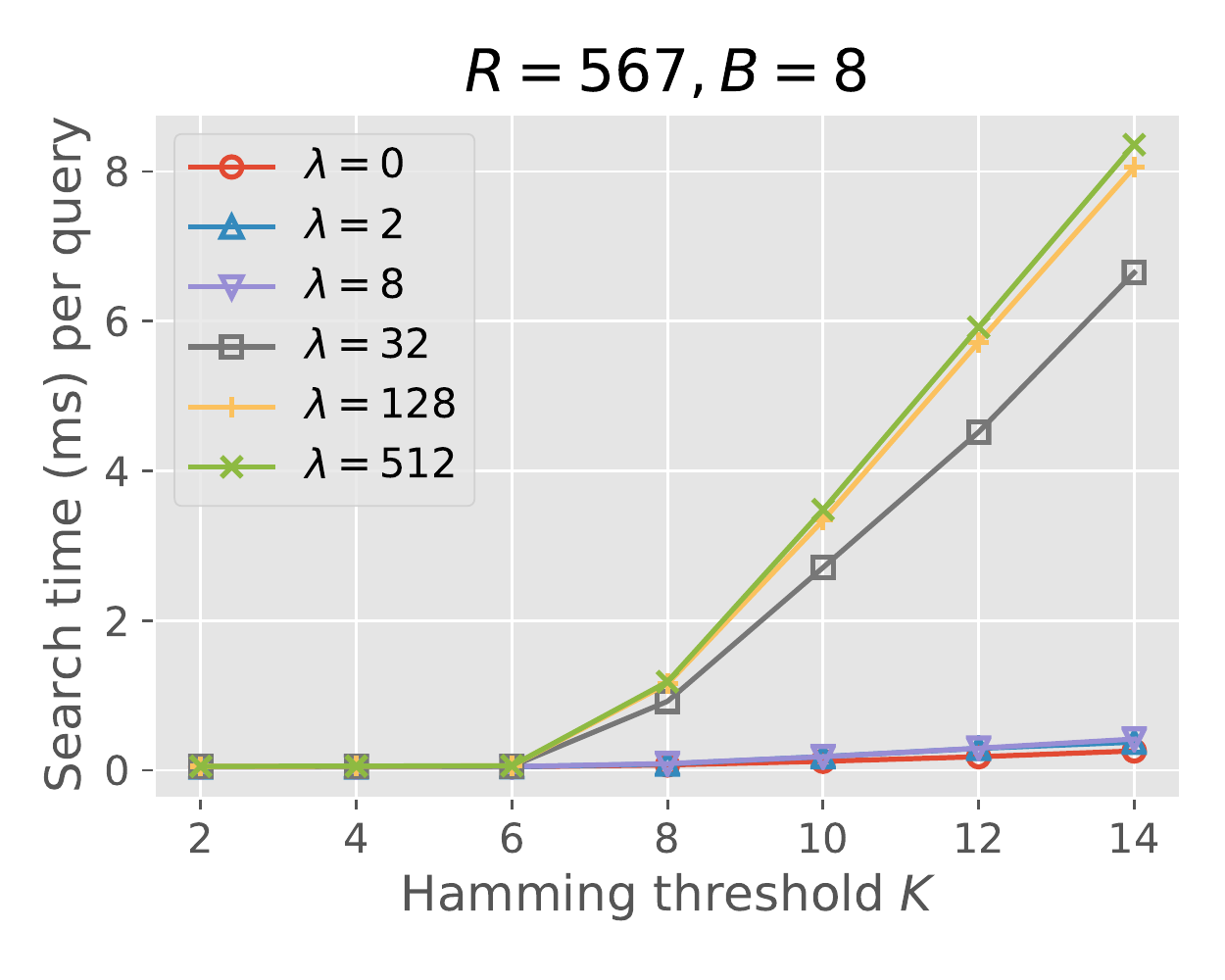}
\includegraphics[width=\ChartWidthApp]{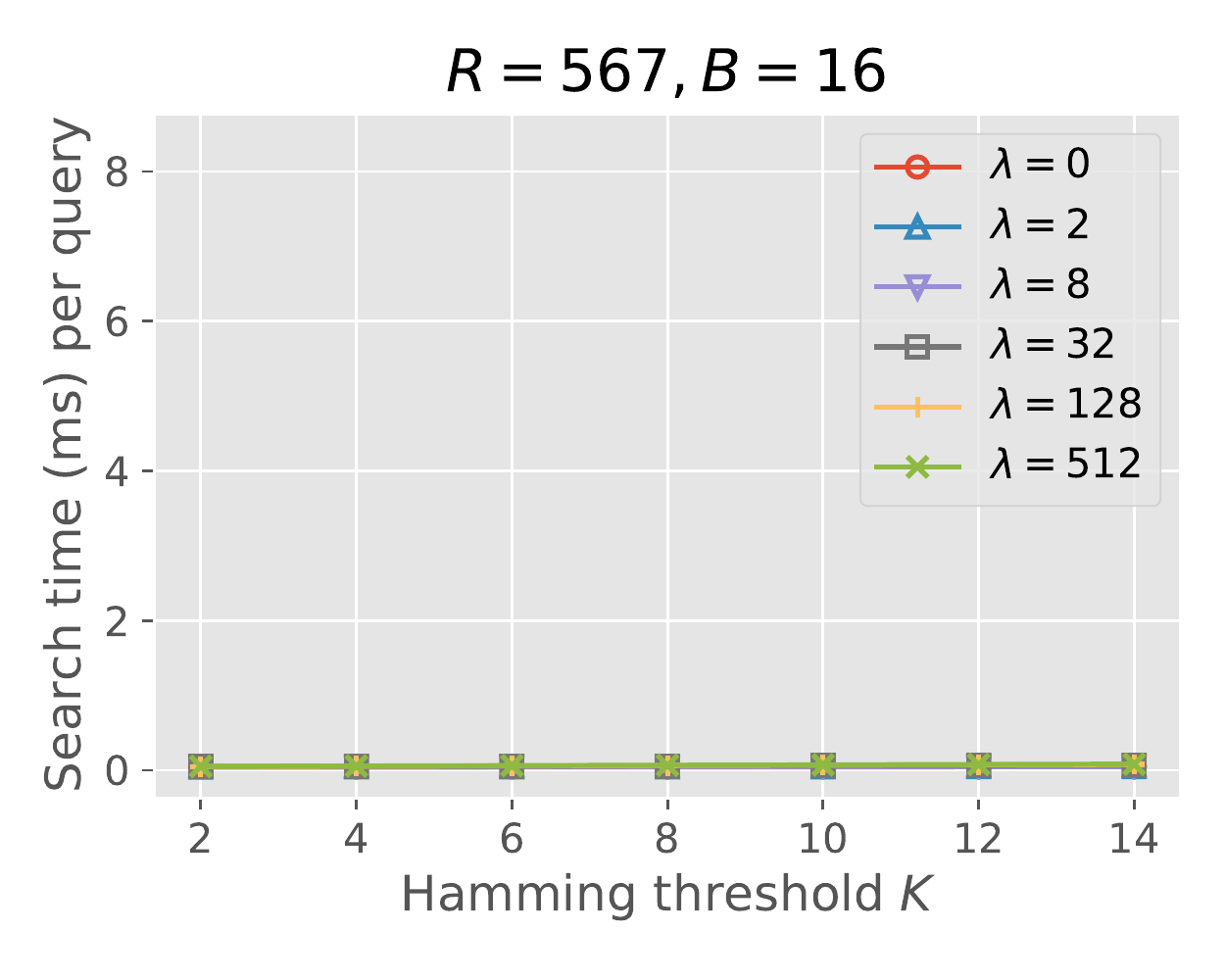}\\
\includegraphics[width=\ChartWidthApp]{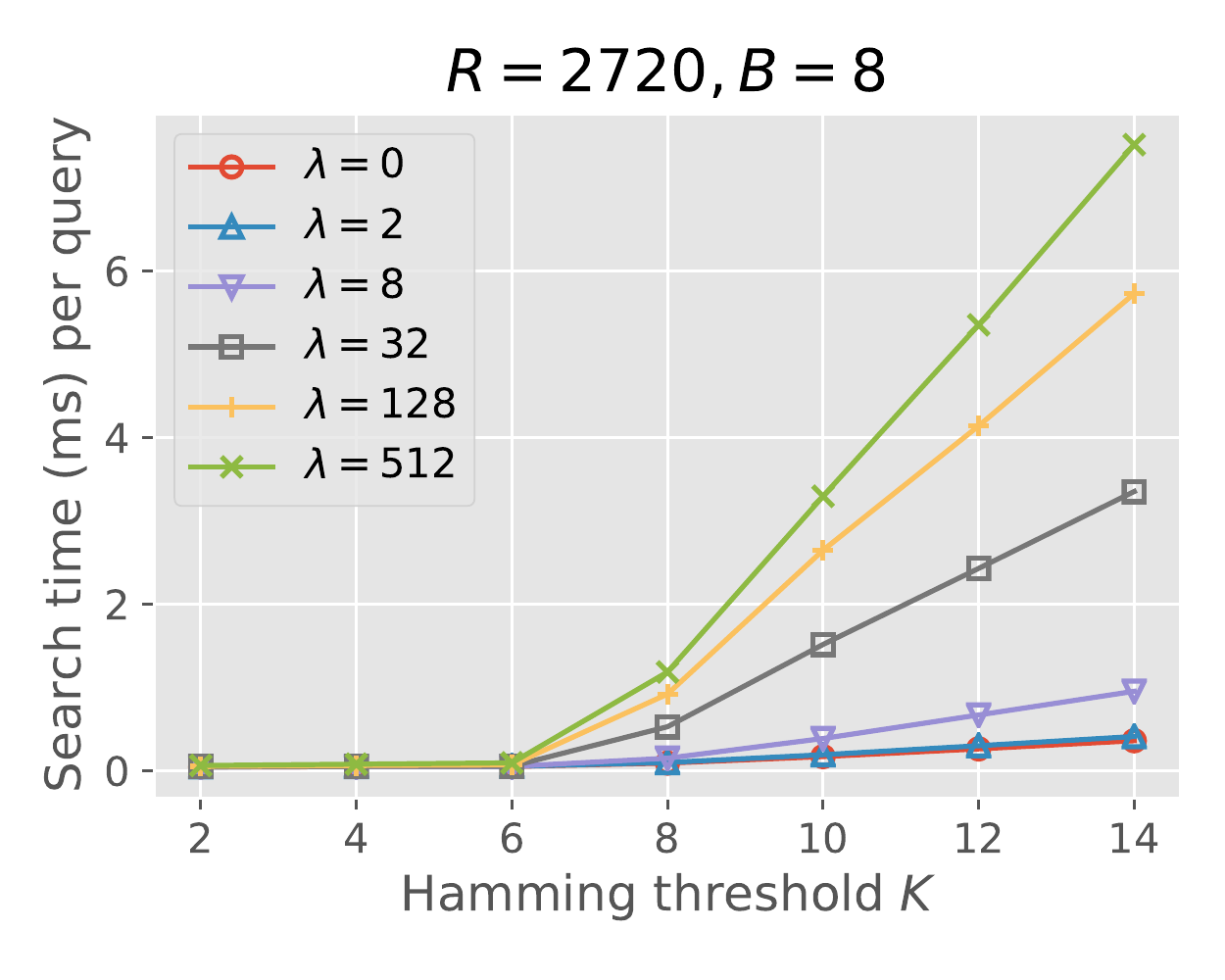}
\includegraphics[width=\ChartWidthApp]{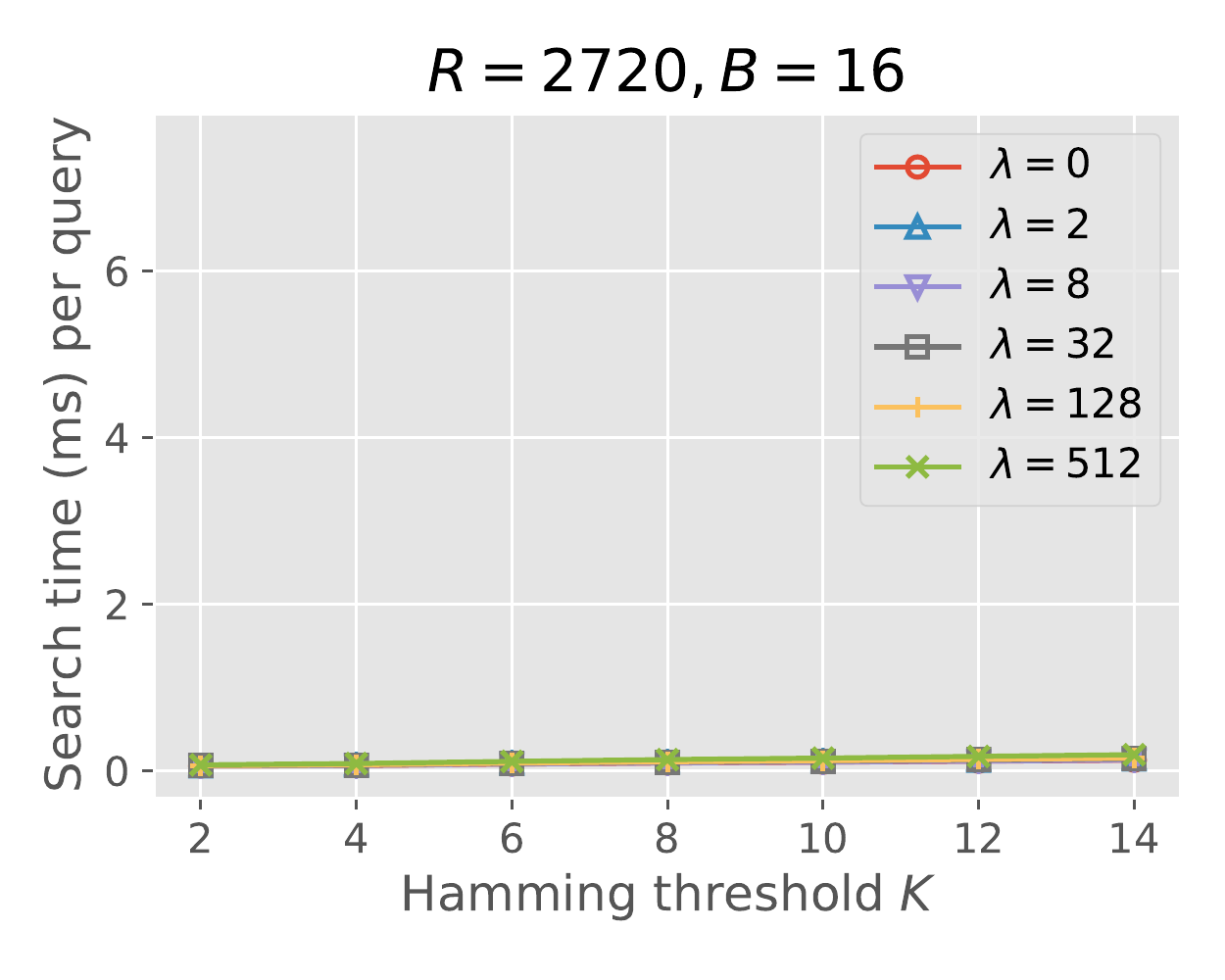}\\
\includegraphics[width=\ChartWidthApp]{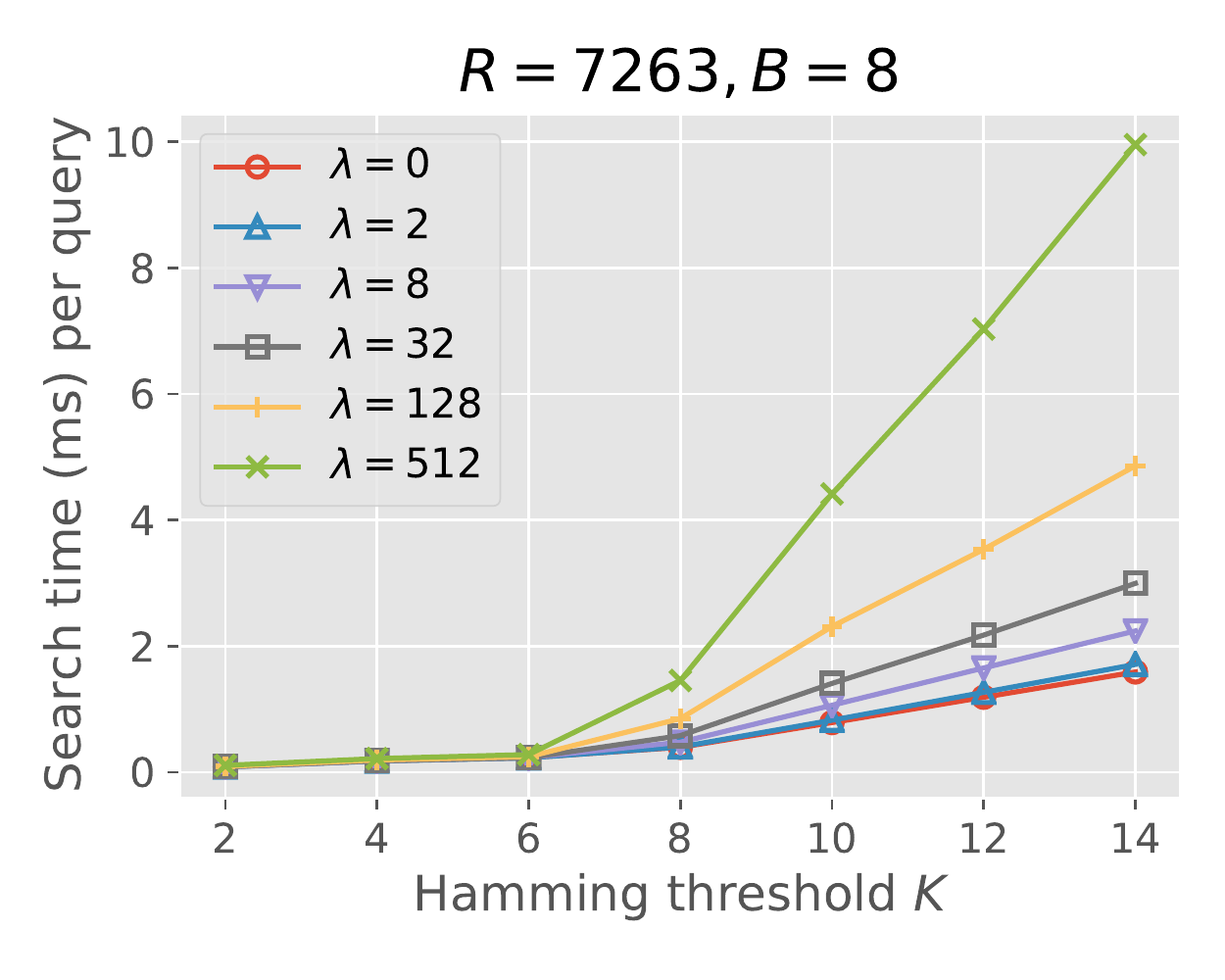}
\includegraphics[width=\ChartWidthApp]{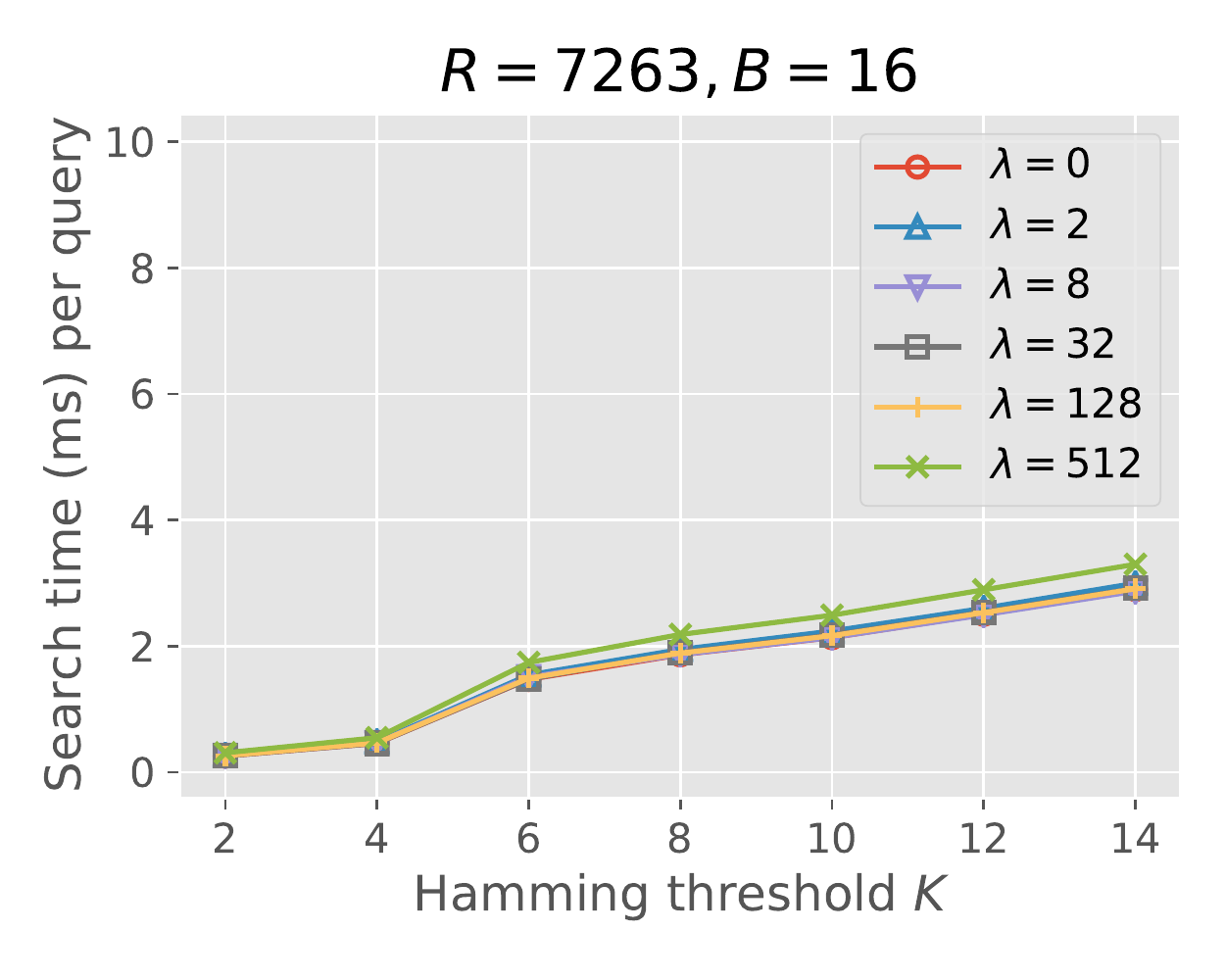}
\end{tabular}
}
\caption{Results of node reduction on Taxi.}
\label{charts:app:reduce:Taxi}
\end{figure*}

\ftfref{charts:app:reduce:Taxi}{charts:app:reduce:OSM} show the full experimental results not presented in \sref{sect:ex:node} for showing an efficiency of node reductions.
As observed in \sref{sect:ex:node}, setting $\lambda = 8,32$ was beneficial for achieving high search and memory efficiencies of tSTAT.

\fref{charts:app:results:search_scale} shows the search time for showing the scalability.
For all the datasets, the search times of tSTAT were always fast.
\fref{charts:app:results:constr} shows the construction time.
Also as observed in \sref{sect:ex:comp}, the time needed for only building STATs from sketches was not significant because the construction times of tSTAT was approximate to those of LS.

\clearpage

\begin{figure*}[p]
\centering
\setlength{\tabcolsep}{2mm}
\subfloat[Number of internal nodes]{
\includegraphics[width=\ChartWidthApp]{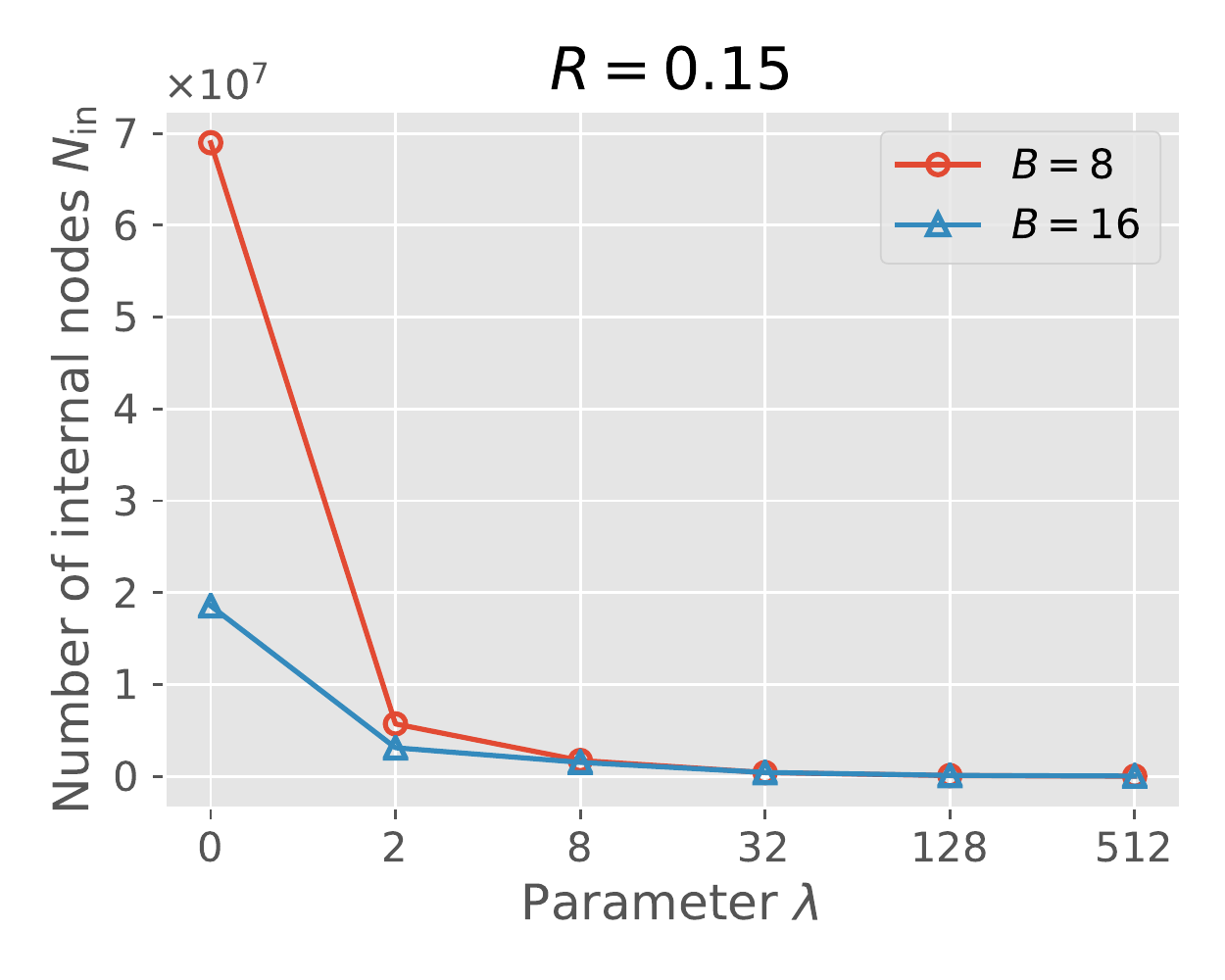}
\includegraphics[width=\ChartWidthApp]{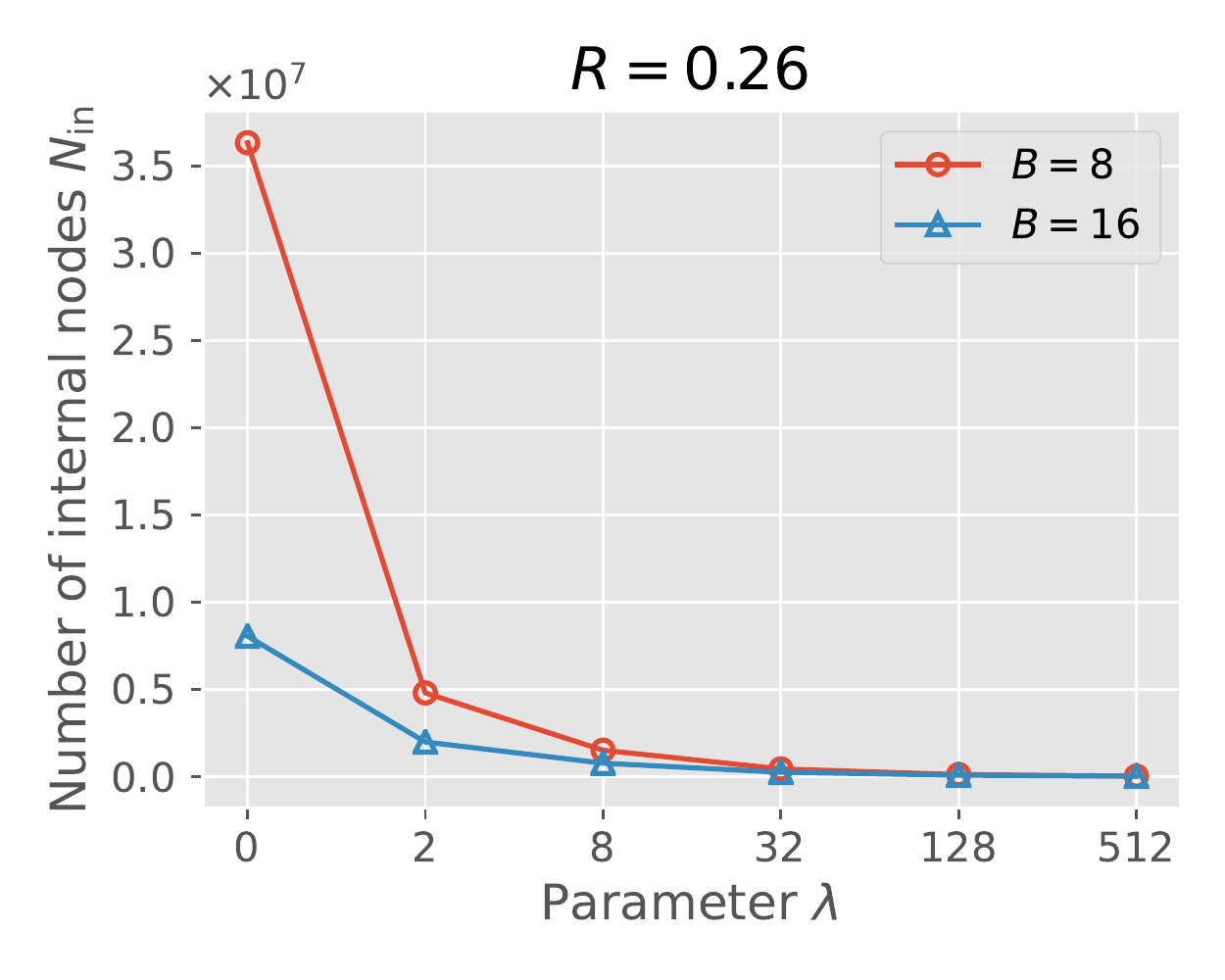}
\includegraphics[width=\ChartWidthApp]{charts/mstat_inner_nodes_perfs_T-NBA_base-0_4519r-8_16B-64L-8_0F.pdf}
}\\
\subfloat[Memory usage in GiB]{
\includegraphics[width=\ChartWidthApp]{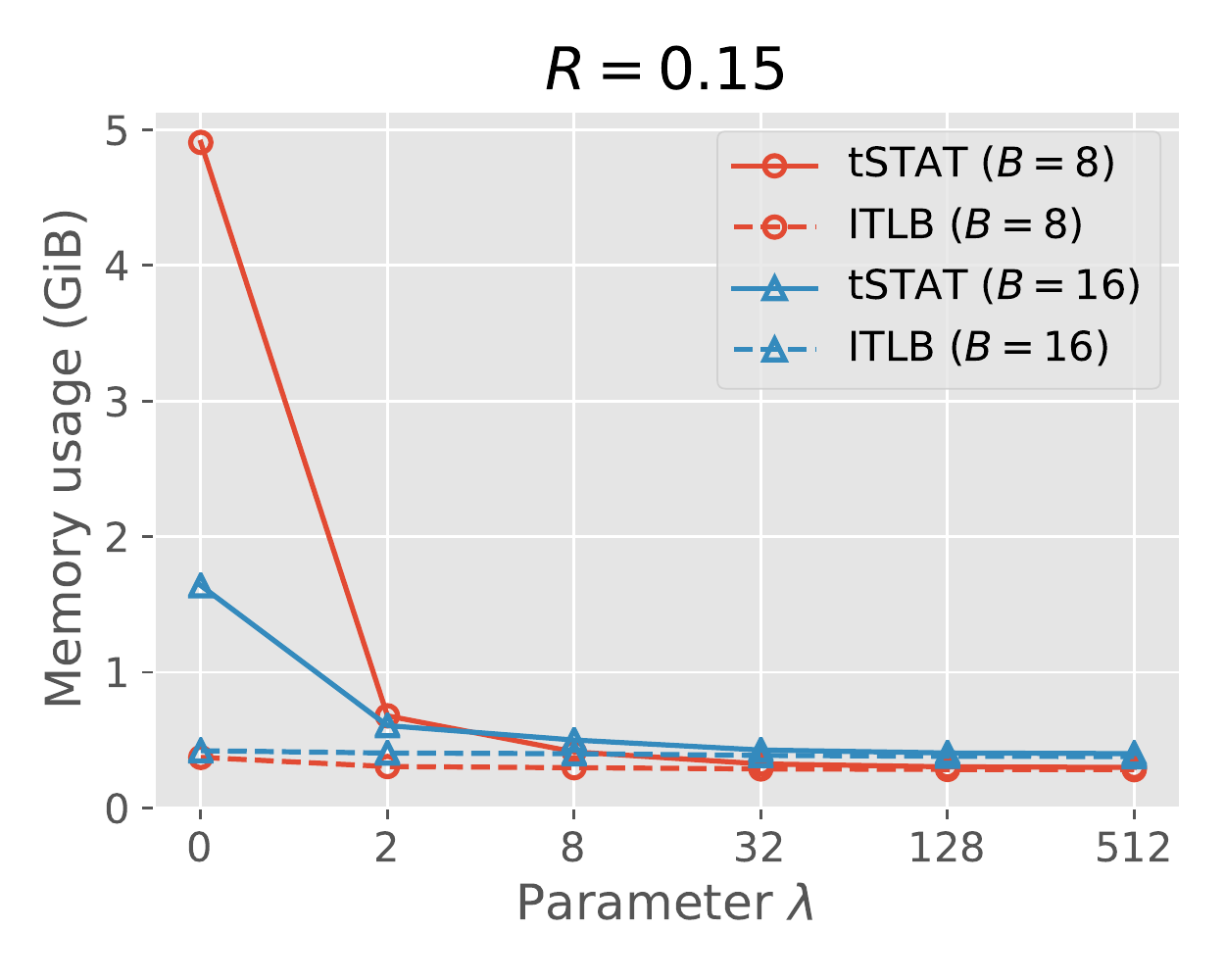}
\includegraphics[width=\ChartWidthApp]{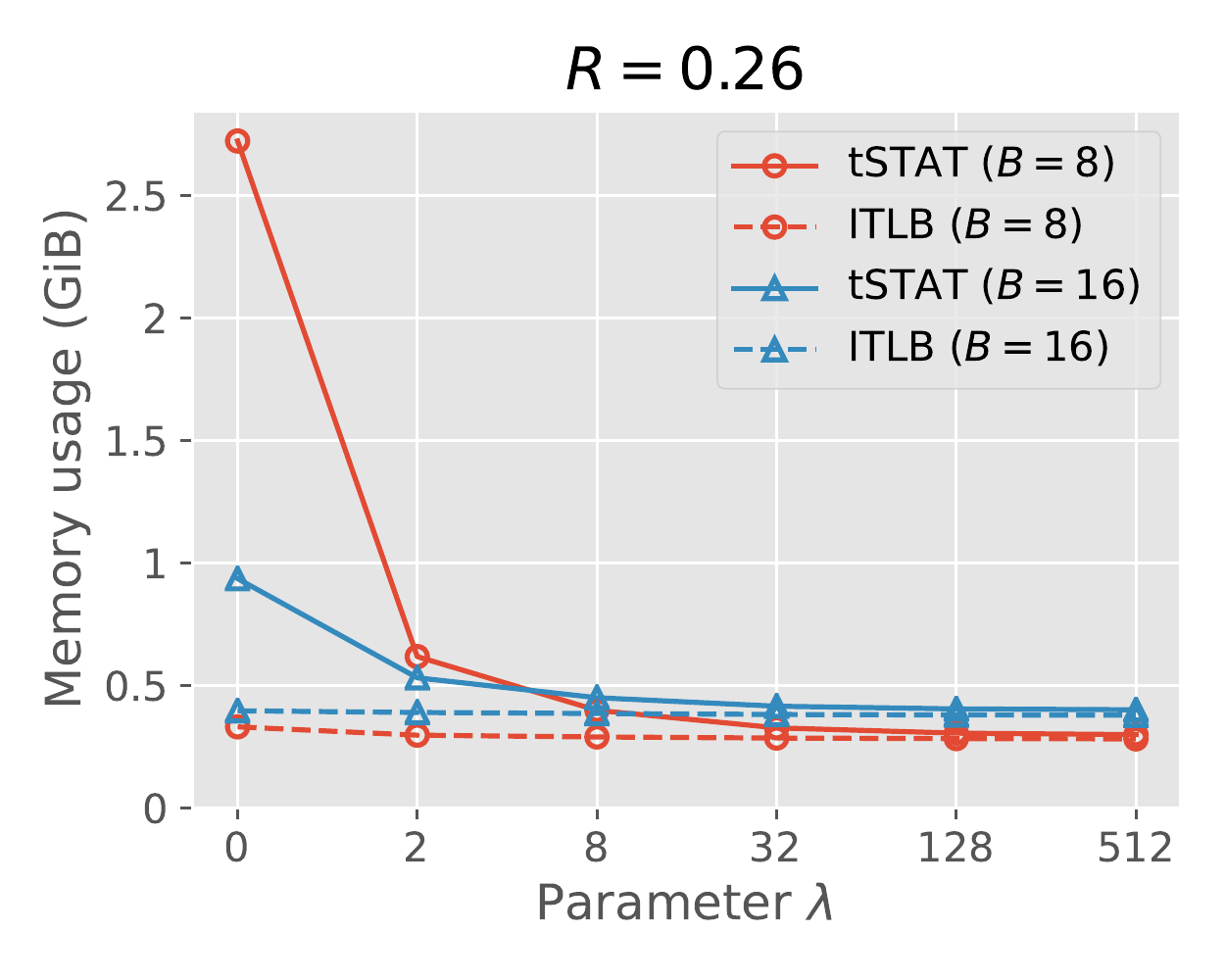}
\includegraphics[width=\ChartWidthApp]{charts/mstat_memory_perfs_T-NBA_base-0_4519r-8_16B-64L-8_0F.pdf}
}\\
\subfloat[Number of candidates $|\Cand|$]{
\begin{tabular}{c}
\includegraphics[width=\ChartWidthApp]{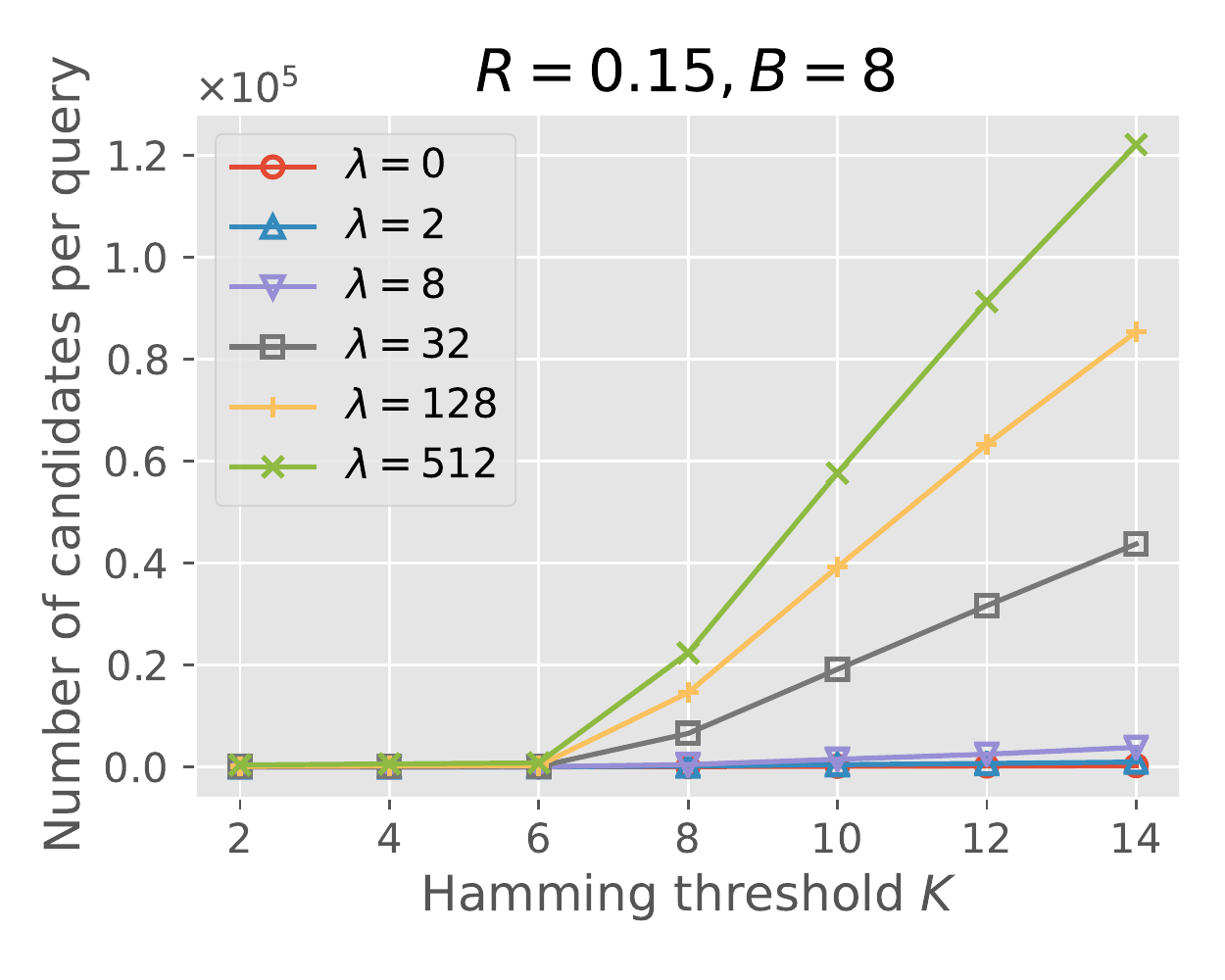}
\includegraphics[width=\ChartWidthApp]{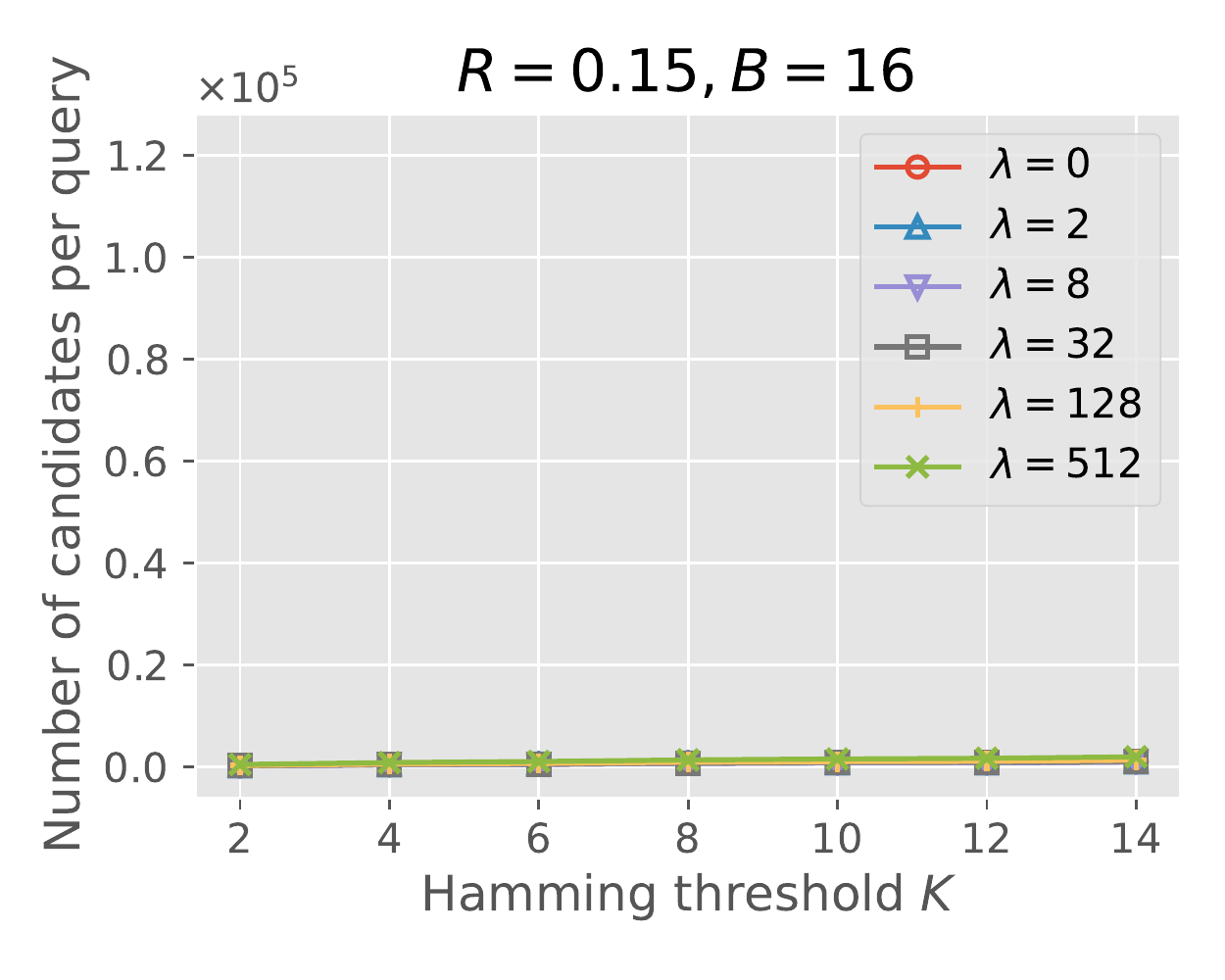}\\
\includegraphics[width=\ChartWidthApp]{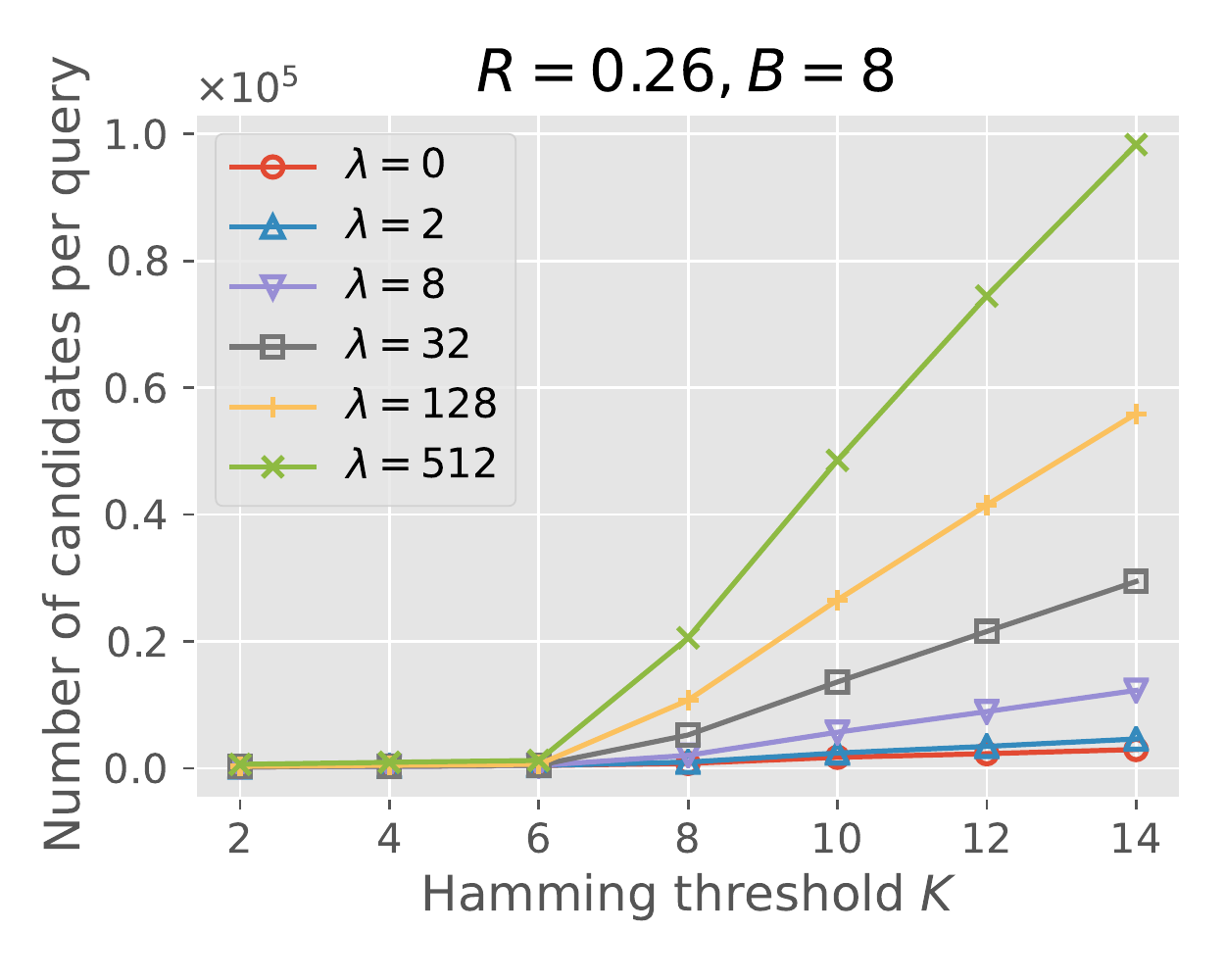}
\includegraphics[width=\ChartWidthApp]{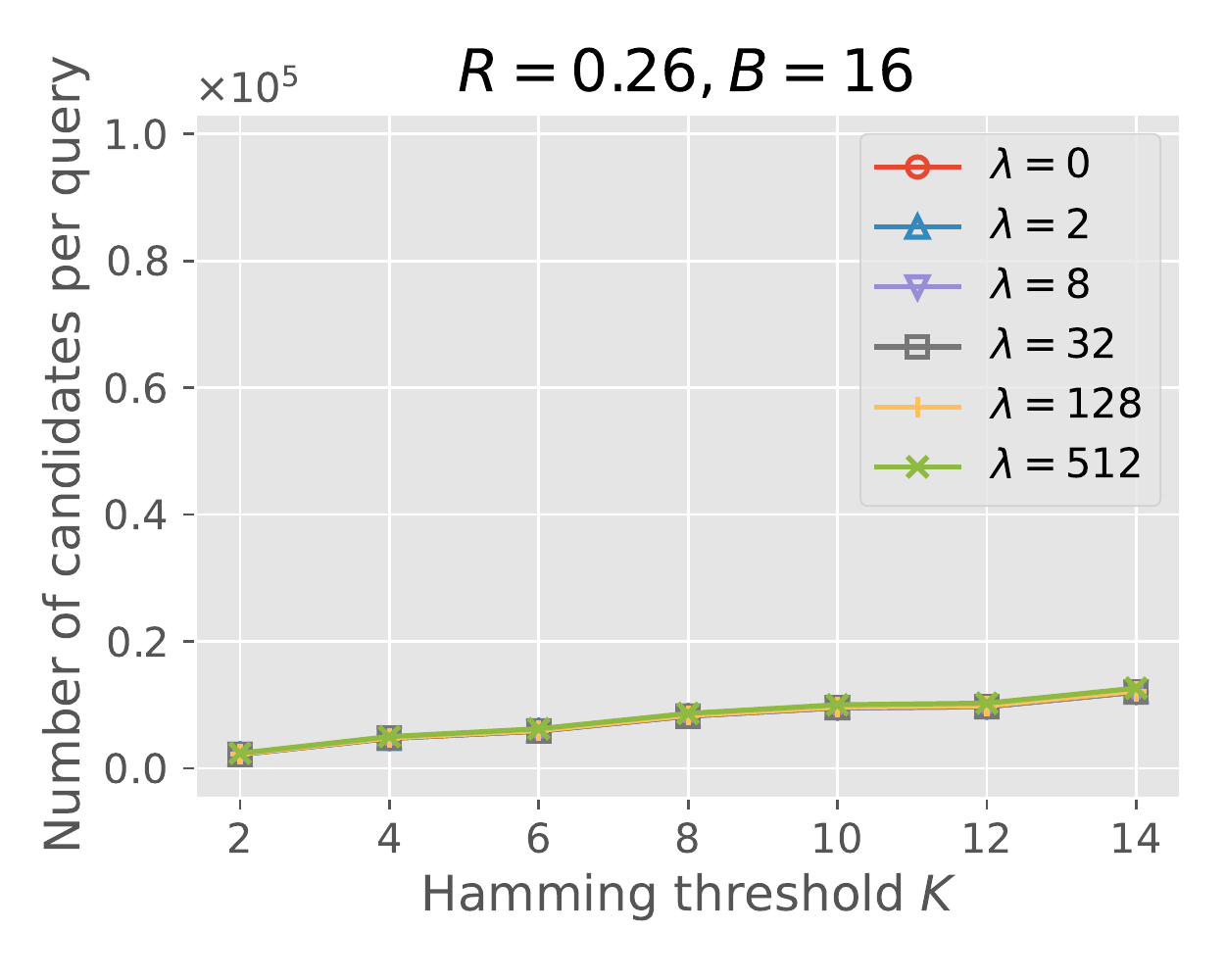}\\
\includegraphics[width=\ChartWidthApp]{charts/mstat_cand_perfs_T-NBA_base-NBA_query-0_4519r-8B-64L-8_0F.pdf}
\includegraphics[width=\ChartWidthApp]{charts/mstat_cand_perfs_T-NBA_base-NBA_query-0_4519r-16B-64L-8_0F.pdf}
\end{tabular}
}
\subfloat[Search time in milliseconds (ms) per query]{
\begin{tabular}{c}
\includegraphics[width=\ChartWidthApp]{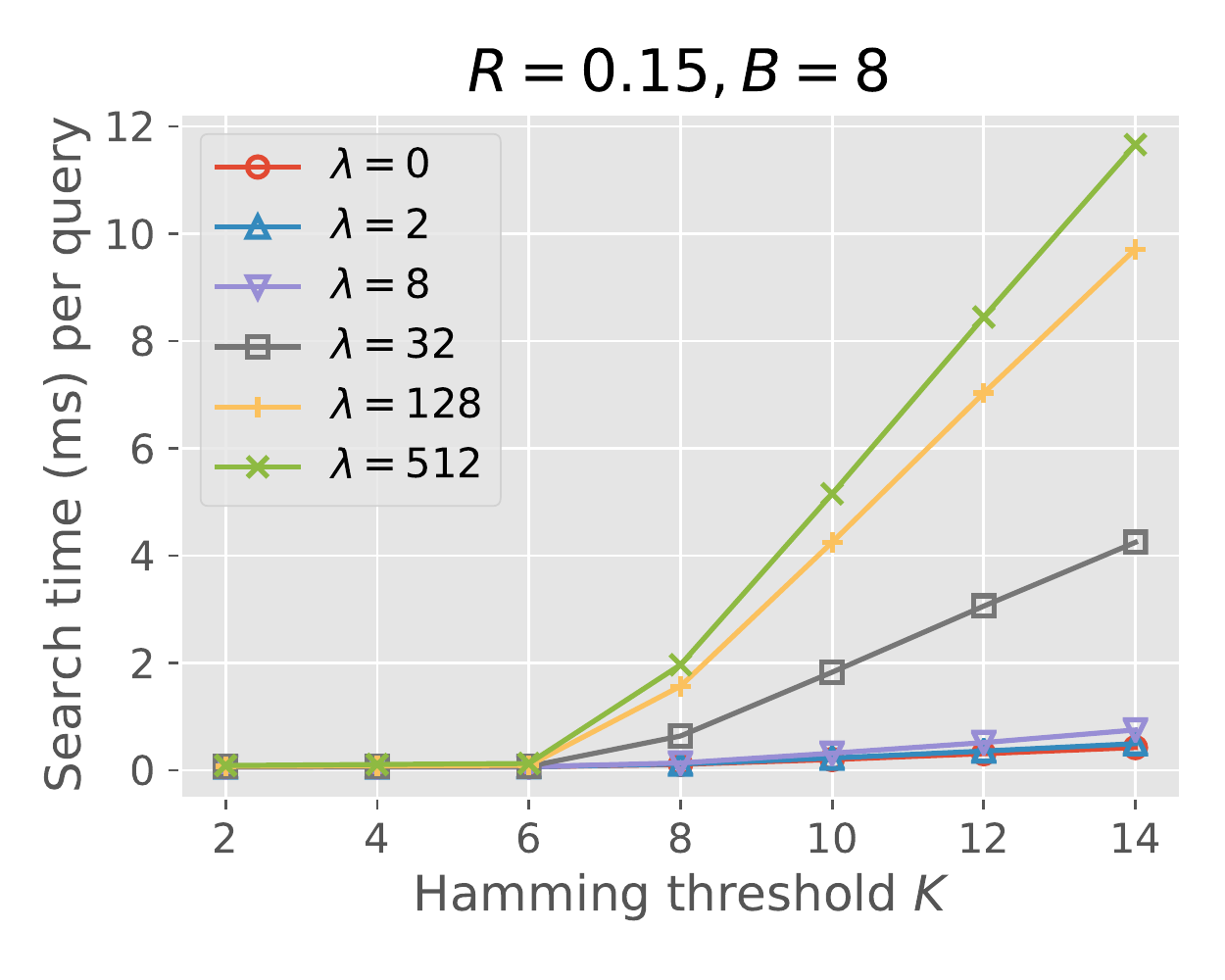}
\includegraphics[width=\ChartWidthApp]{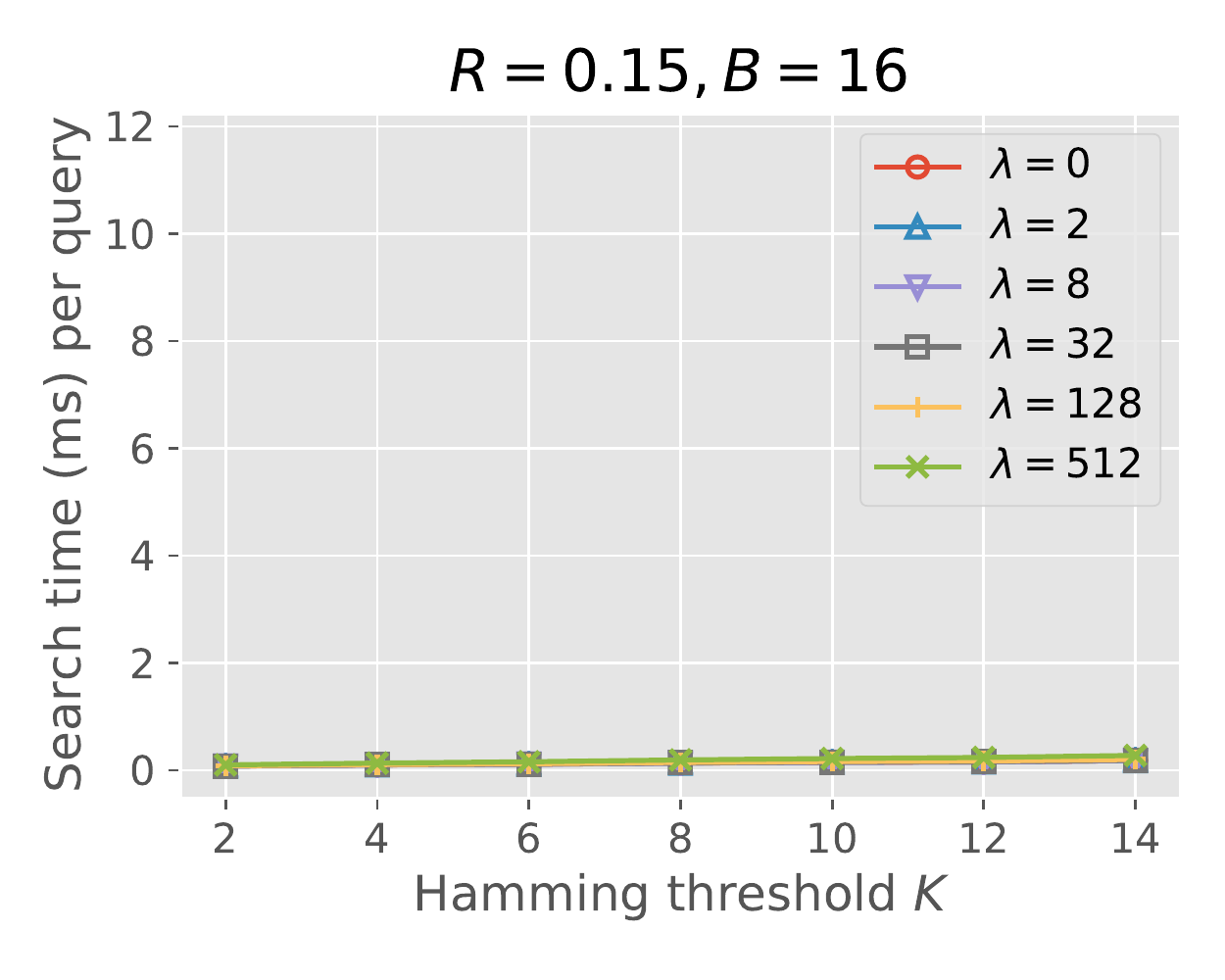}\\
\includegraphics[width=\ChartWidthApp]{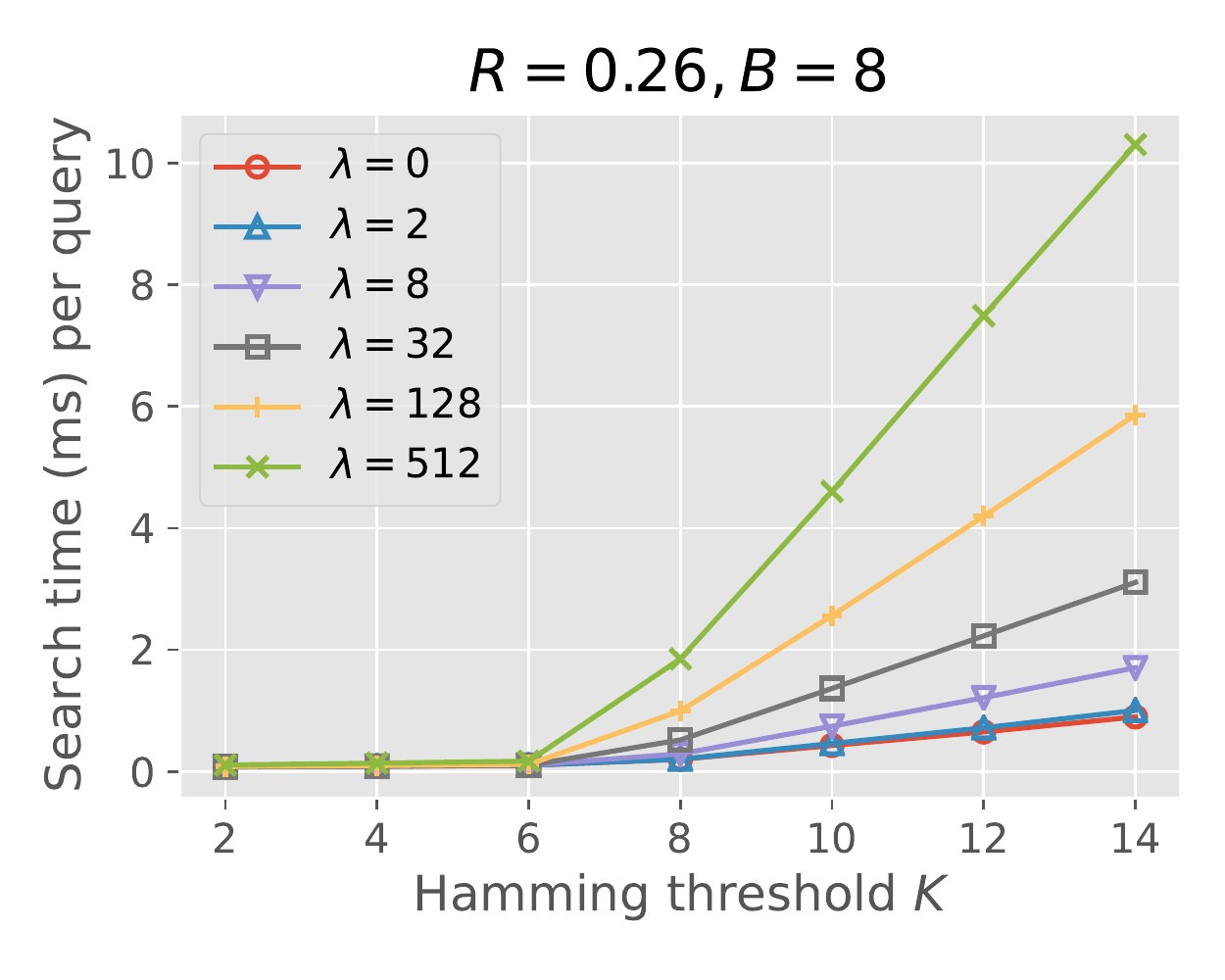}
\includegraphics[width=\ChartWidthApp]{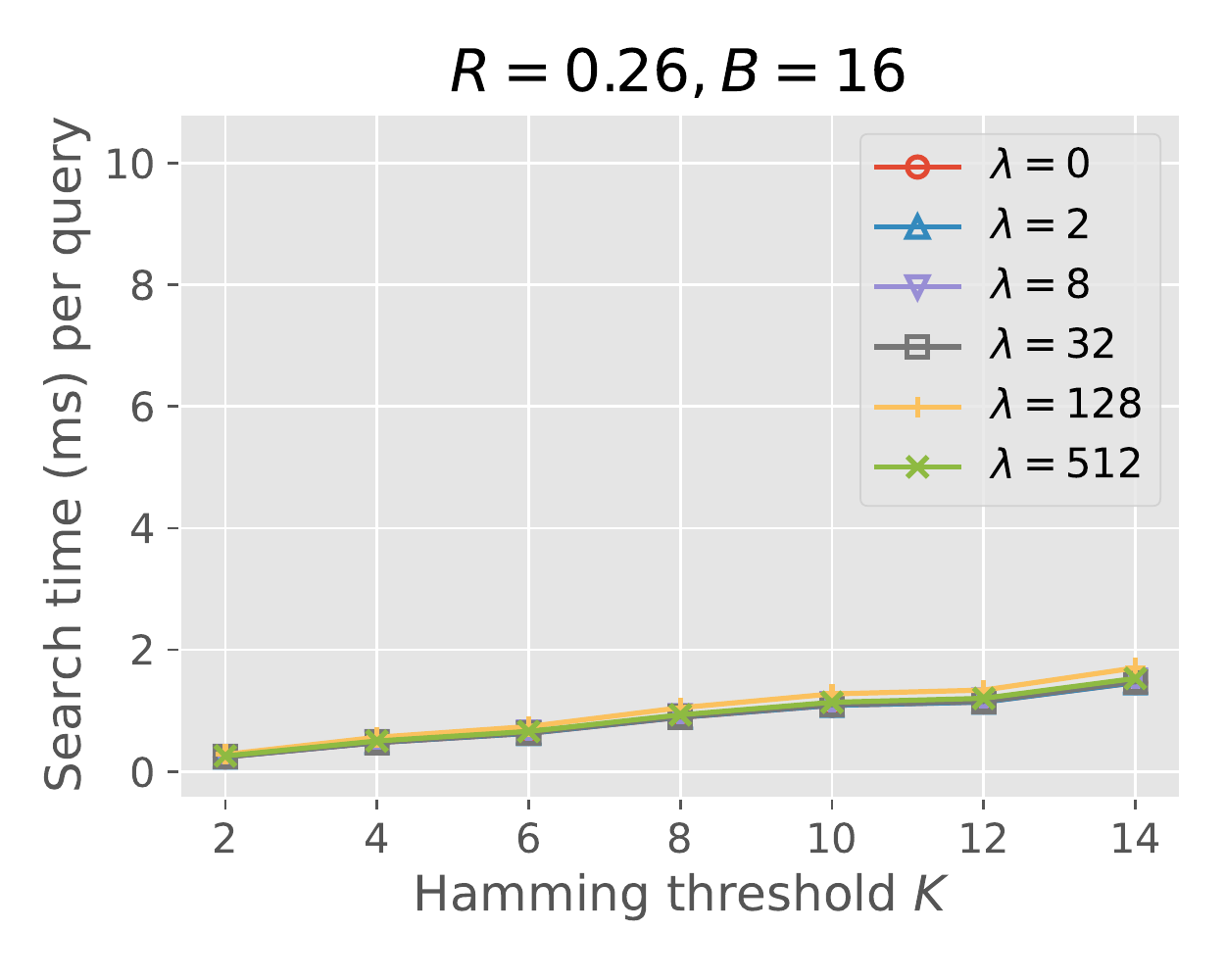}\\
\includegraphics[width=\ChartWidthApp]{charts/mstat_time_perfs_T-NBA_base-NBA_query-0_4519r-8B-64L-8_0F.pdf}
\includegraphics[width=\ChartWidthApp]{charts/mstat_time_perfs_T-NBA_base-NBA_query-0_4519r-16B-64L-8_0F.pdf}
\end{tabular}
}
\caption{Results of node reduction on NBA.}
\label{charts:app:reduce:NBA}
\end{figure*}

\begin{figure*}[p]
\centering
\setlength{\tabcolsep}{2mm}
\subfloat[Number of internal nodes]{
\includegraphics[width=\ChartWidthApp]{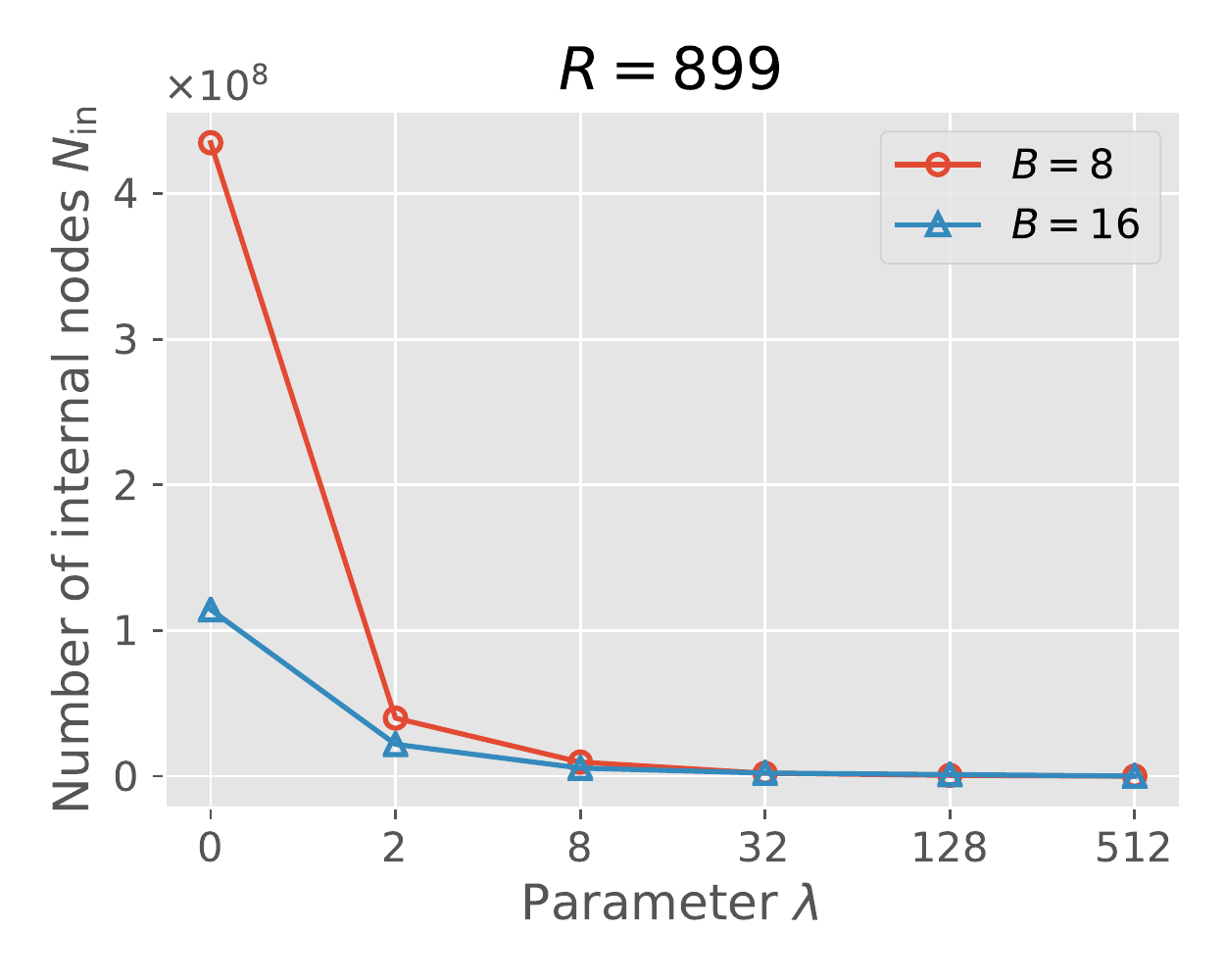}
\includegraphics[width=\ChartWidthApp]{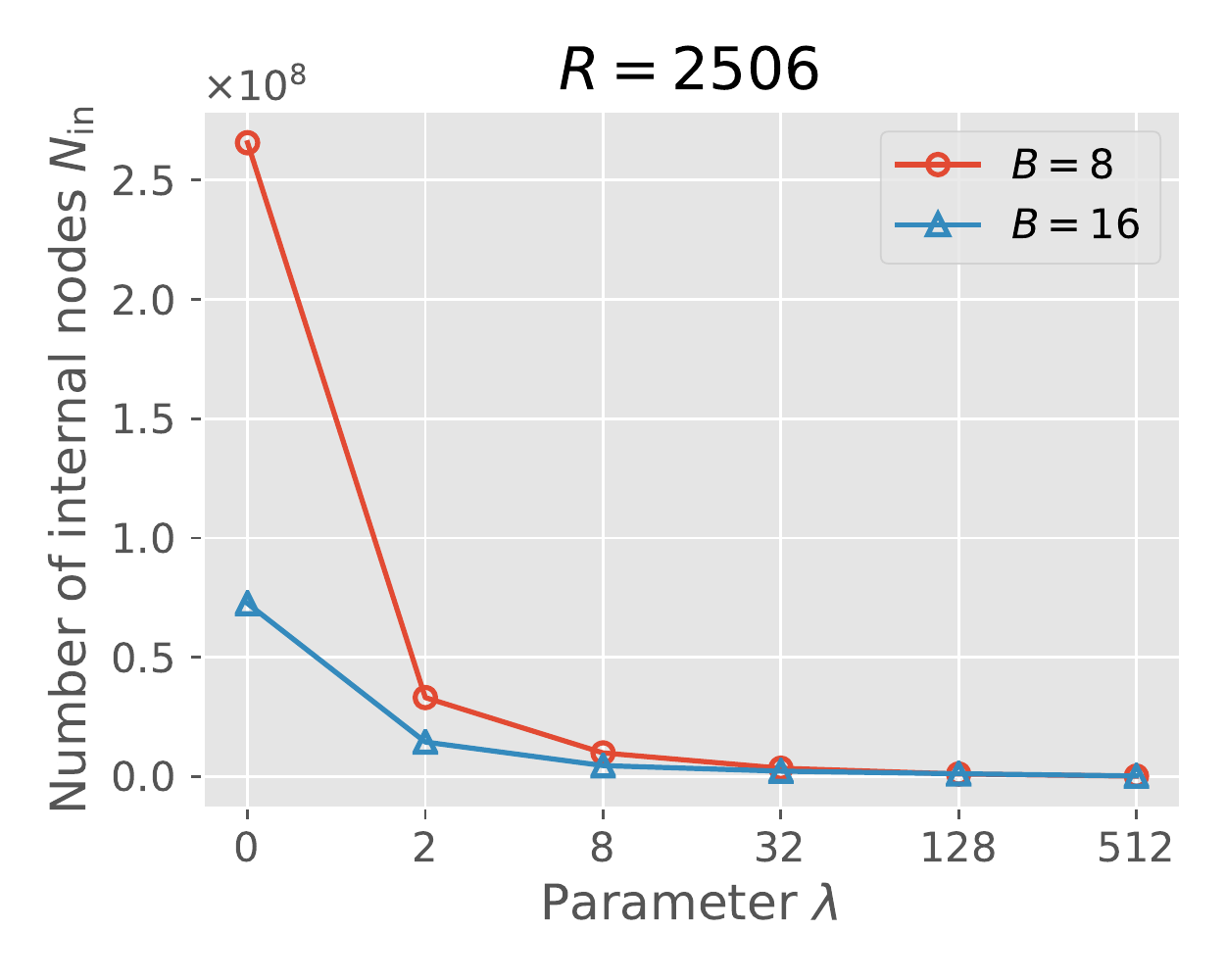}
\includegraphics[width=\ChartWidthApp]{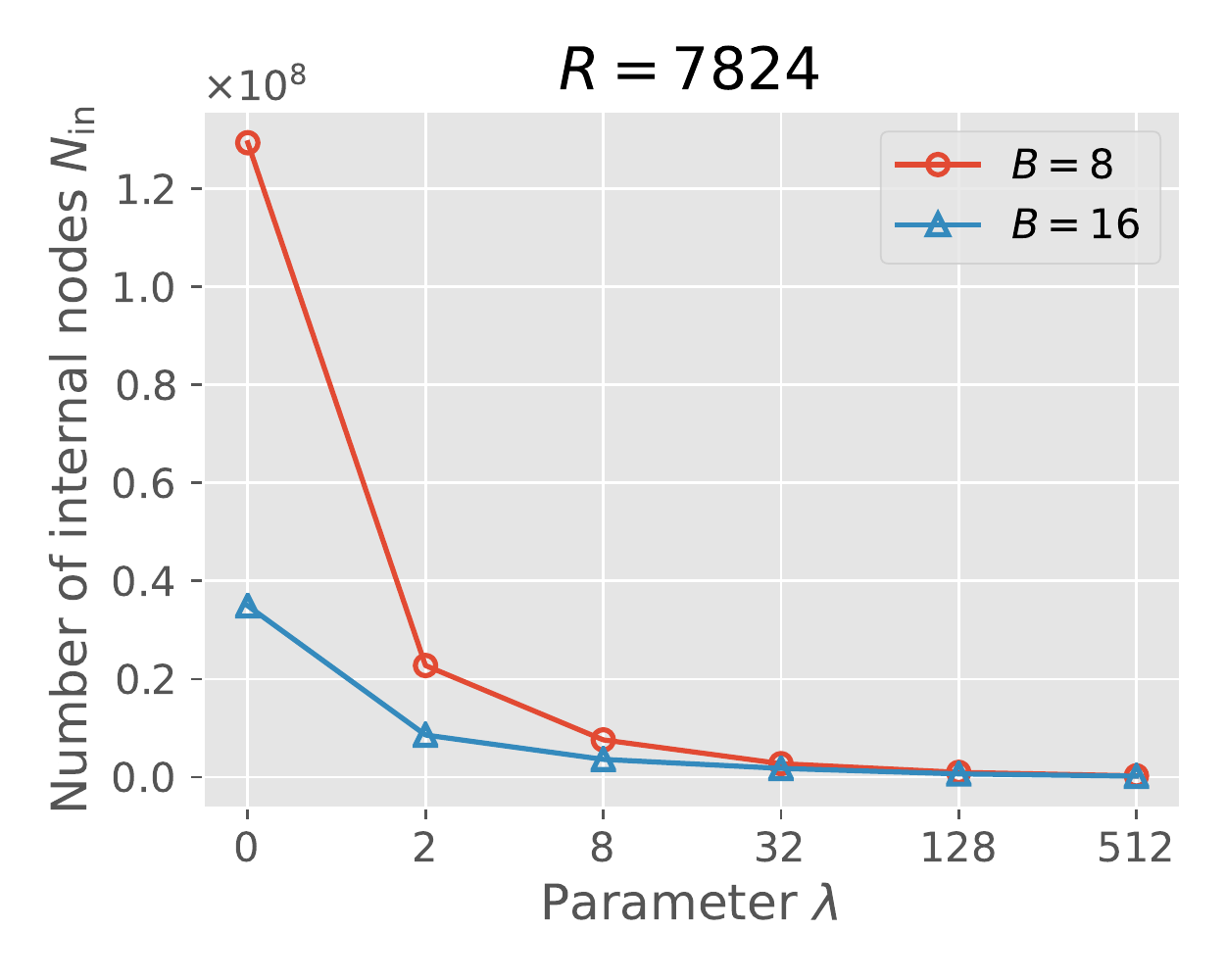}
}\\
\subfloat[Memory usage in GiB]{
\includegraphics[width=\ChartWidthApp]{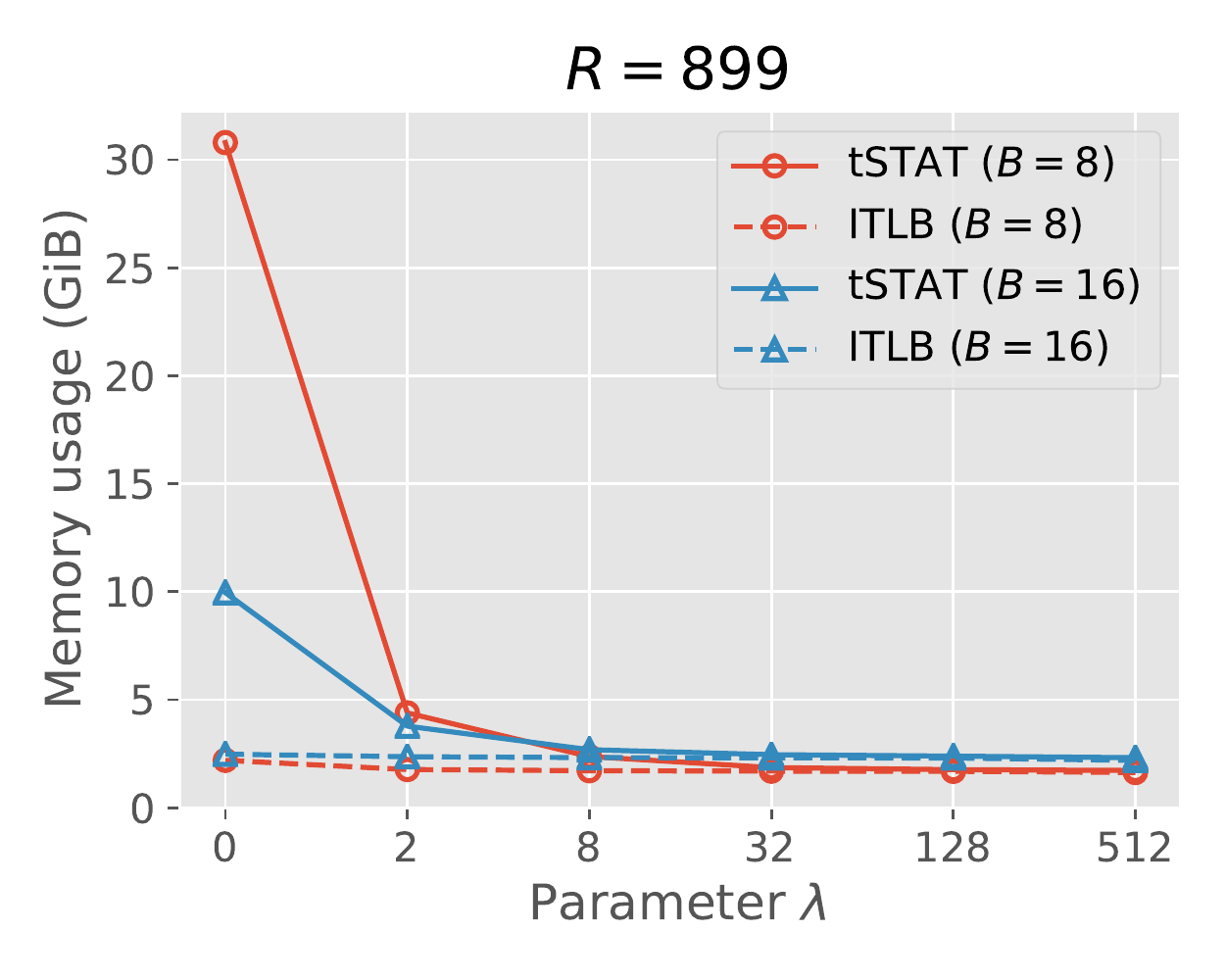}
\includegraphics[width=\ChartWidthApp]{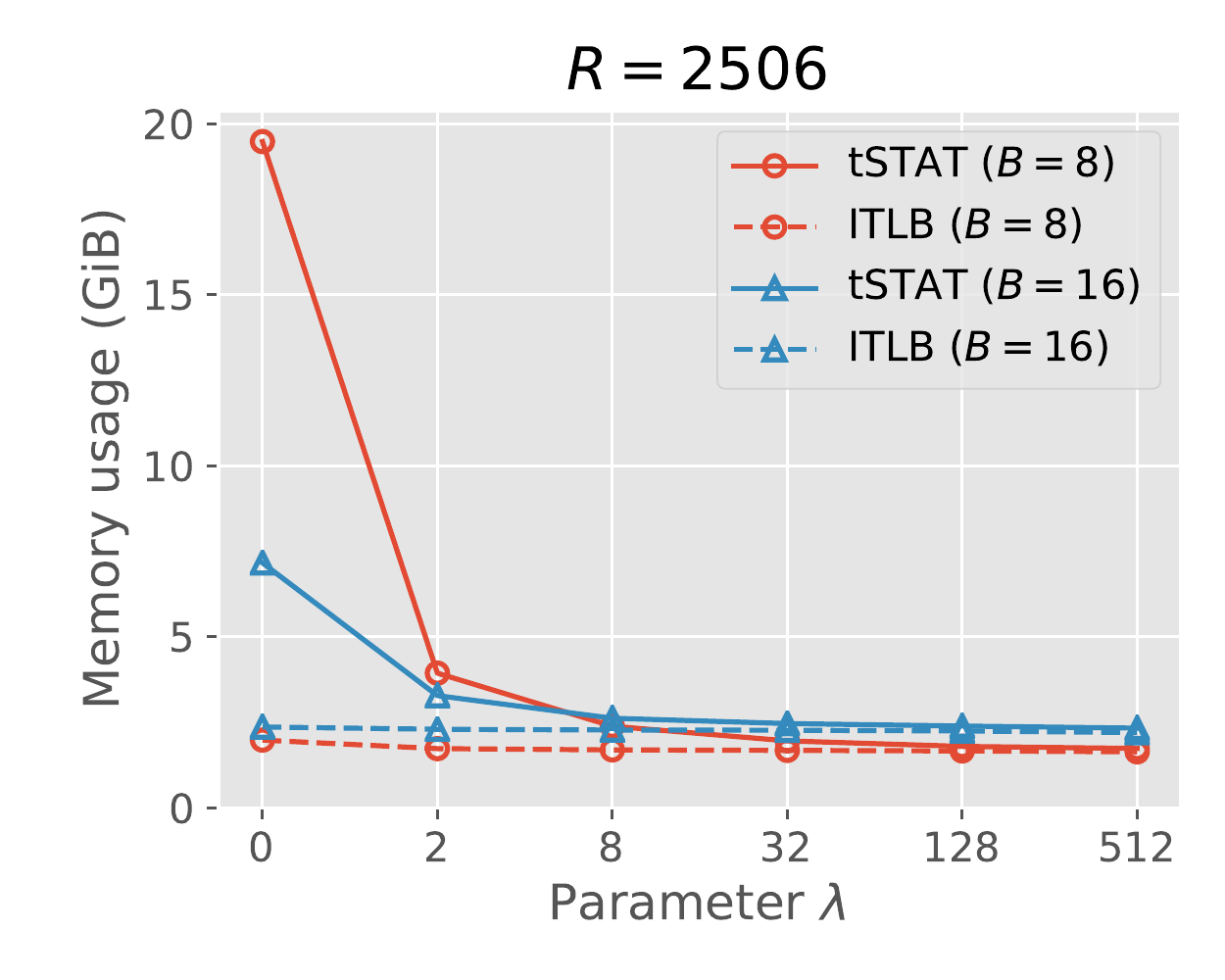}
\includegraphics[width=\ChartWidthApp]{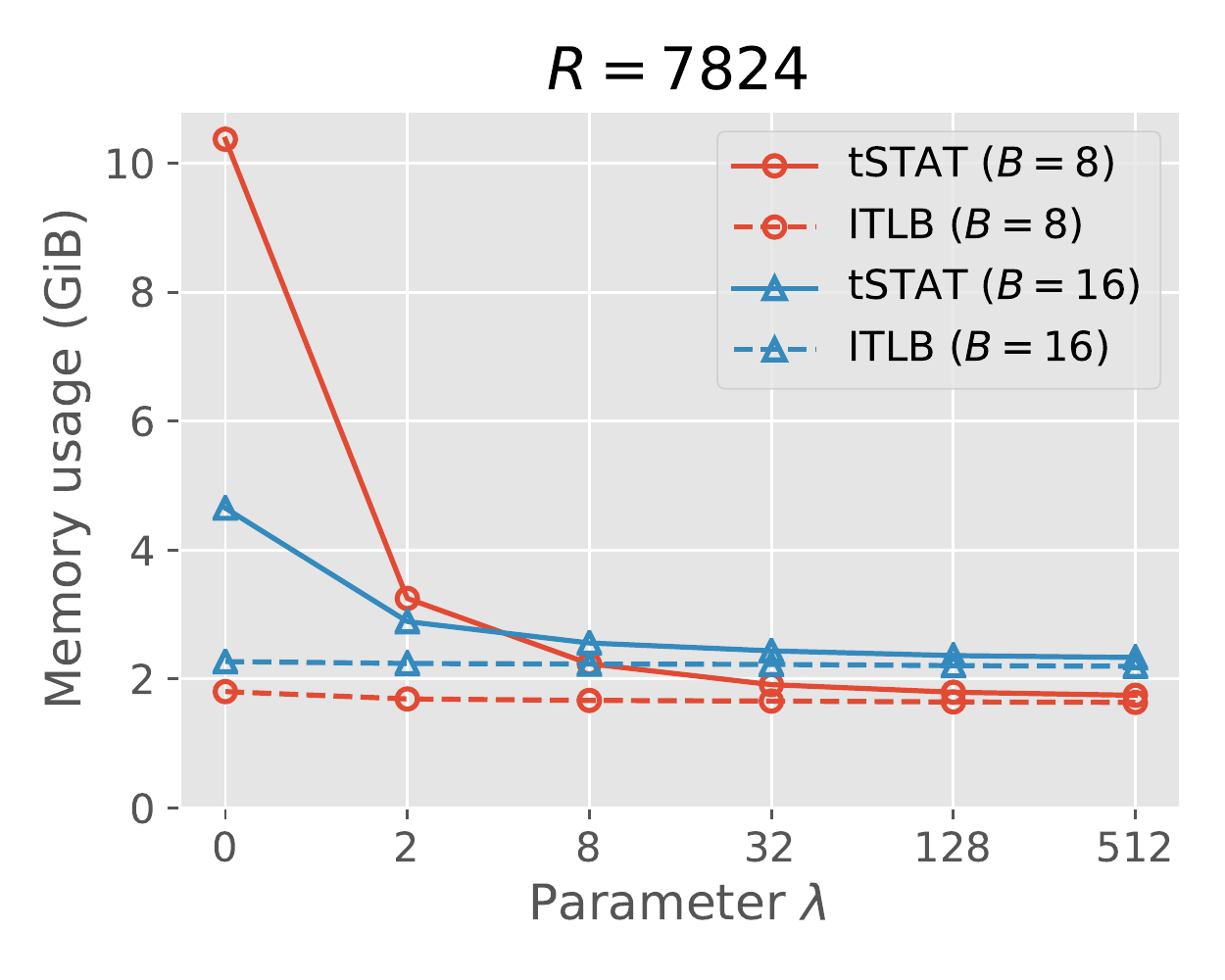}
}\\
\subfloat[Number of candidates $|\Cand|$]{
\begin{tabular}{c}
\includegraphics[width=\ChartWidthApp]{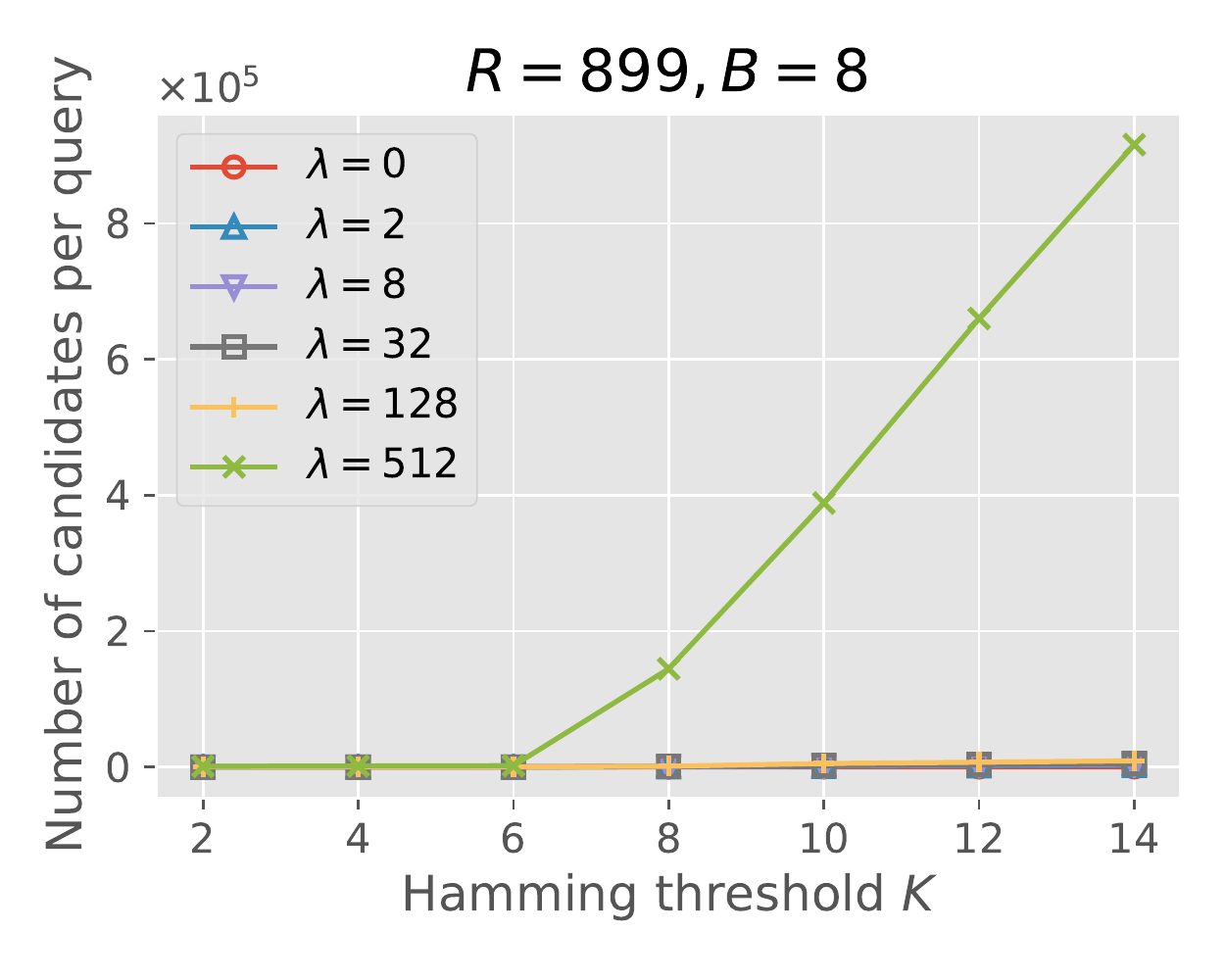}
\includegraphics[width=\ChartWidthApp]{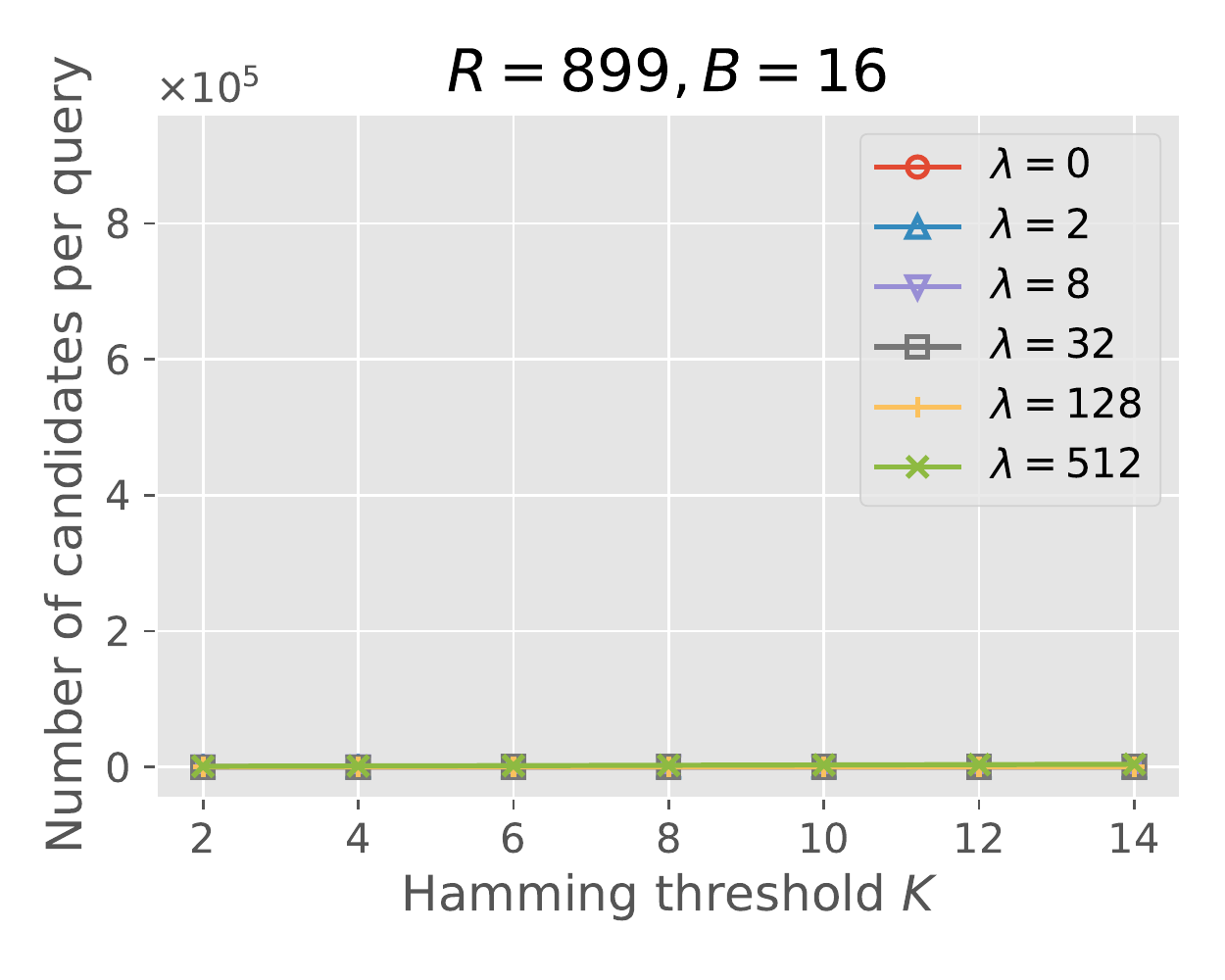}\\
\includegraphics[width=\ChartWidthApp]{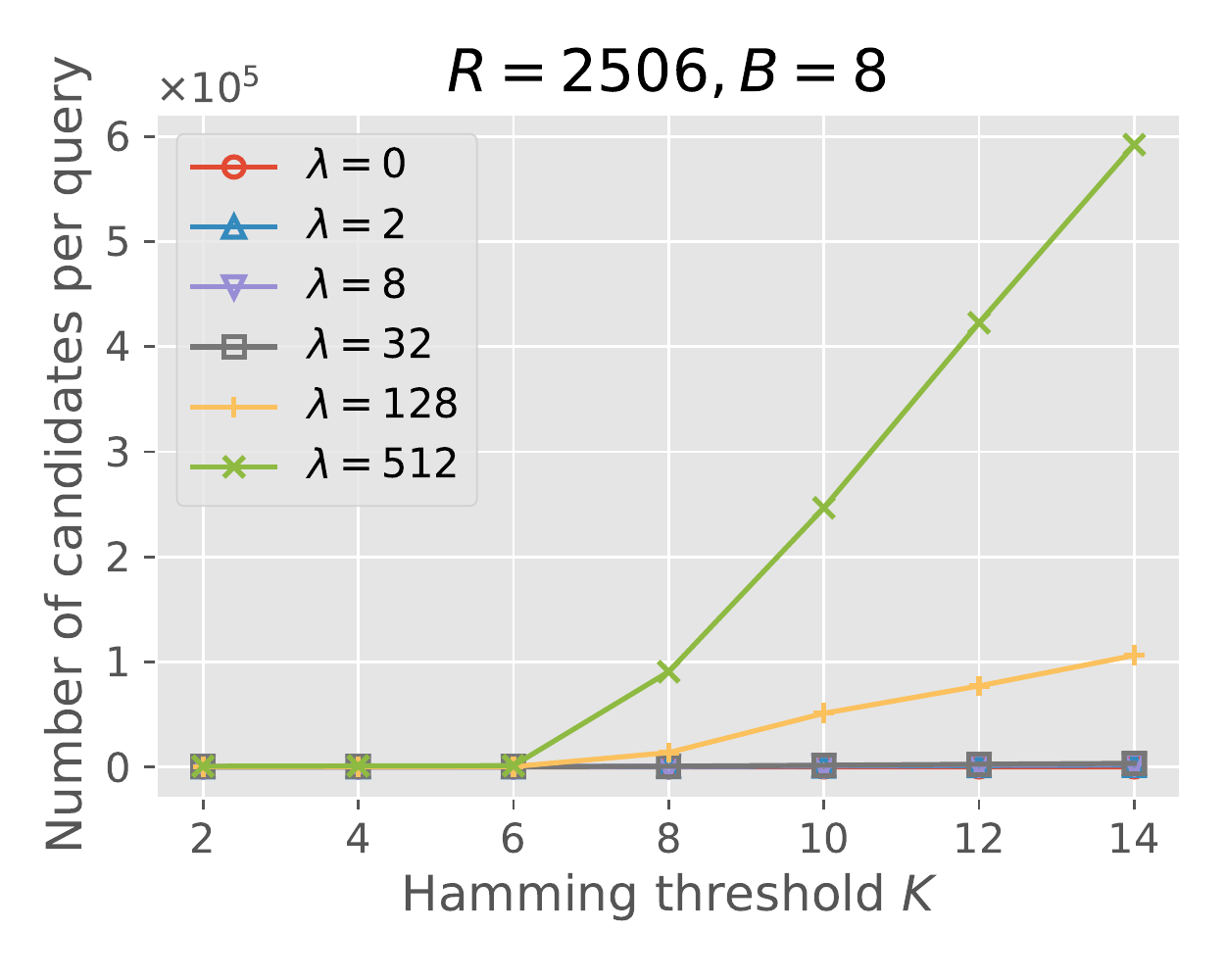}
\includegraphics[width=\ChartWidthApp]{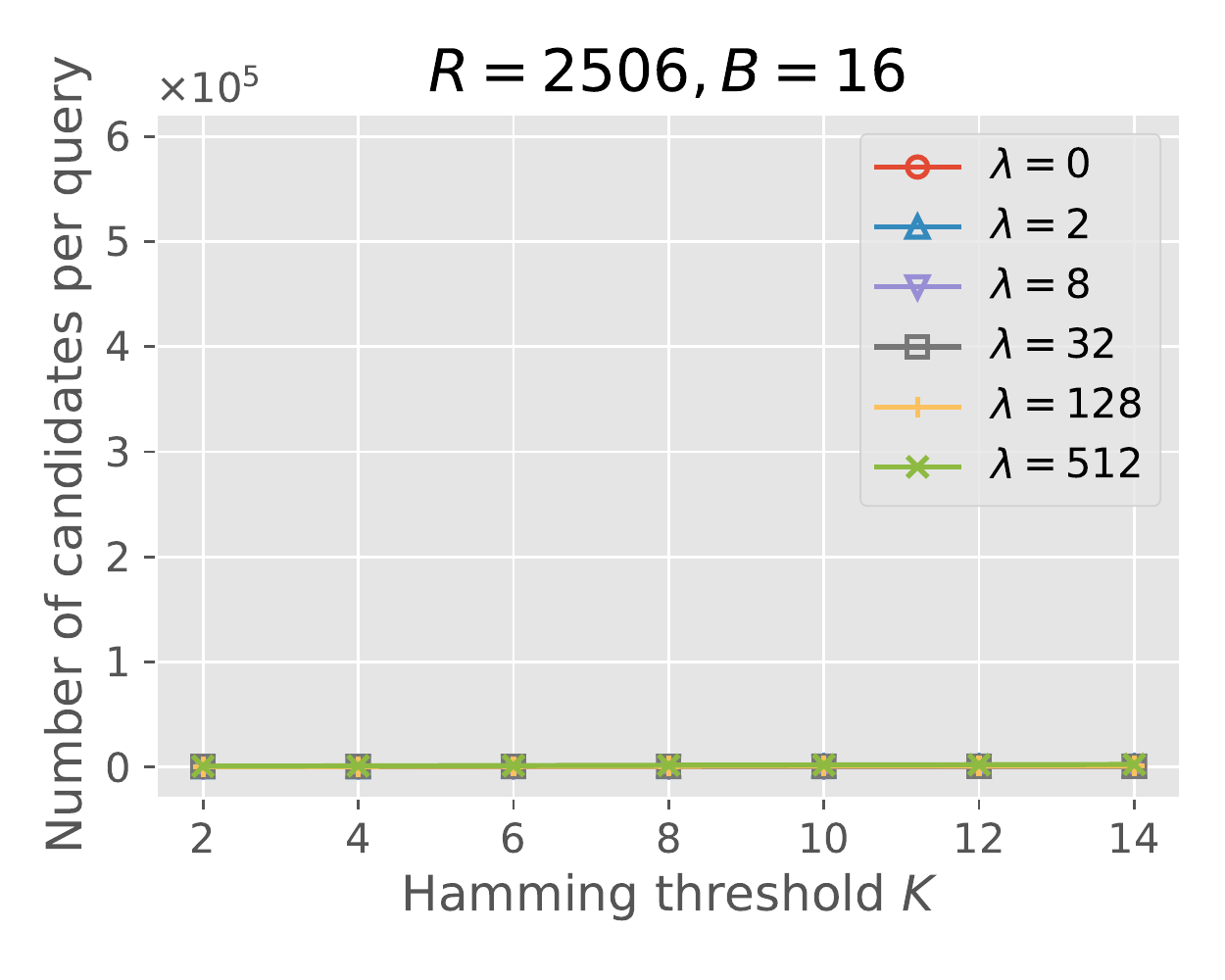}\\
\includegraphics[width=\ChartWidthApp]{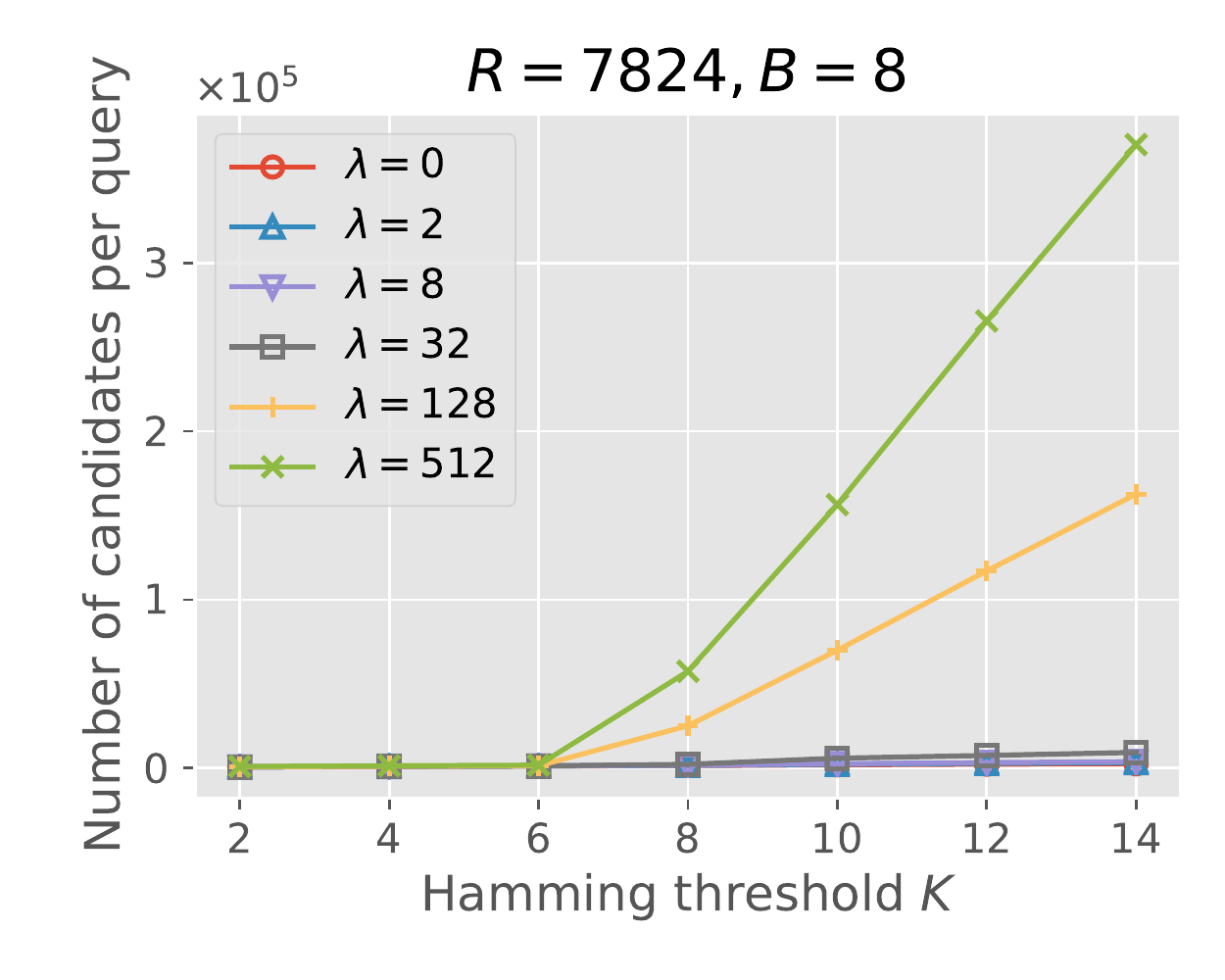}
\includegraphics[width=\ChartWidthApp]{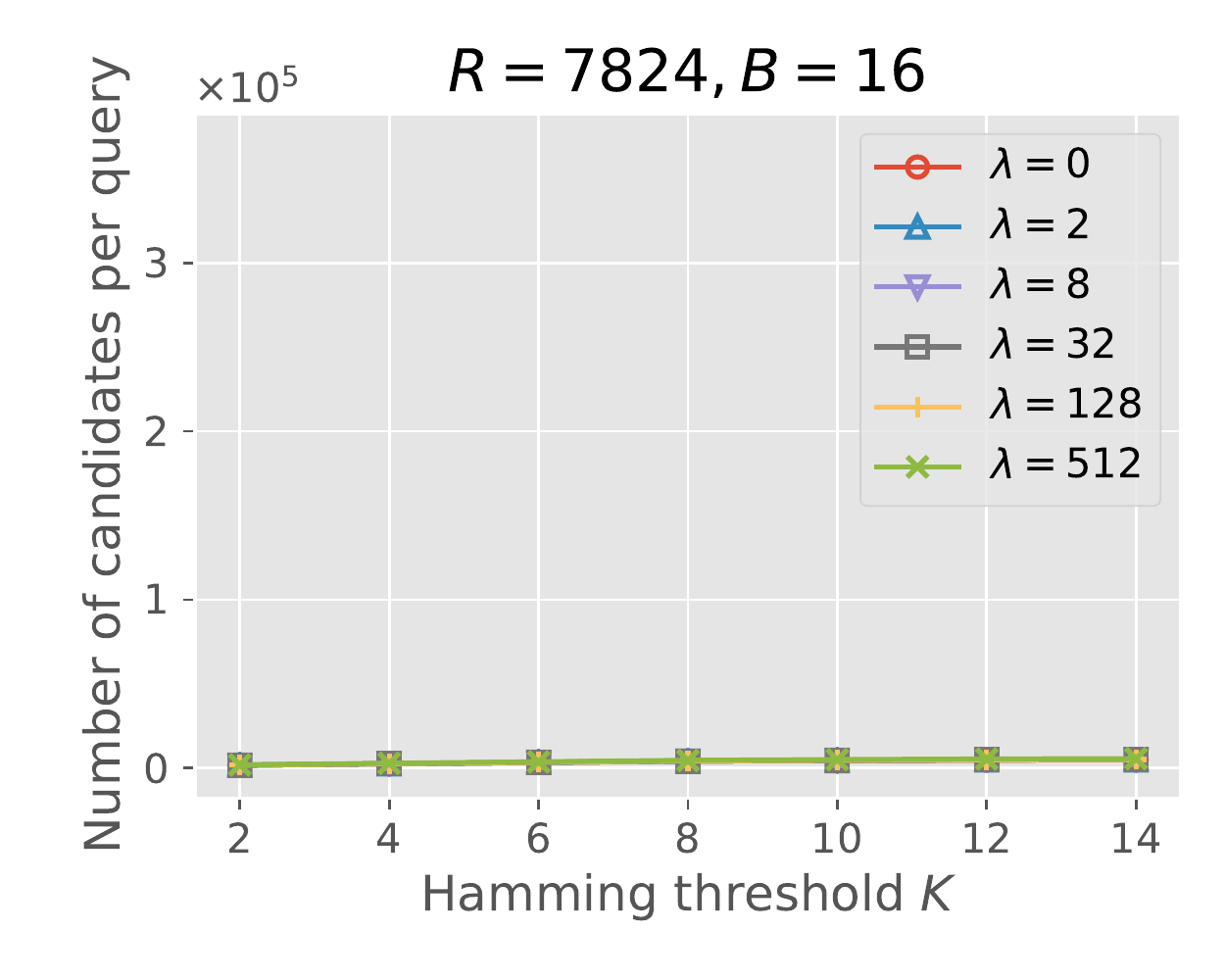}
\end{tabular}
}
\subfloat[Search time in milliseconds (ms) per query]{
\begin{tabular}{c}
\includegraphics[width=\ChartWidthApp]{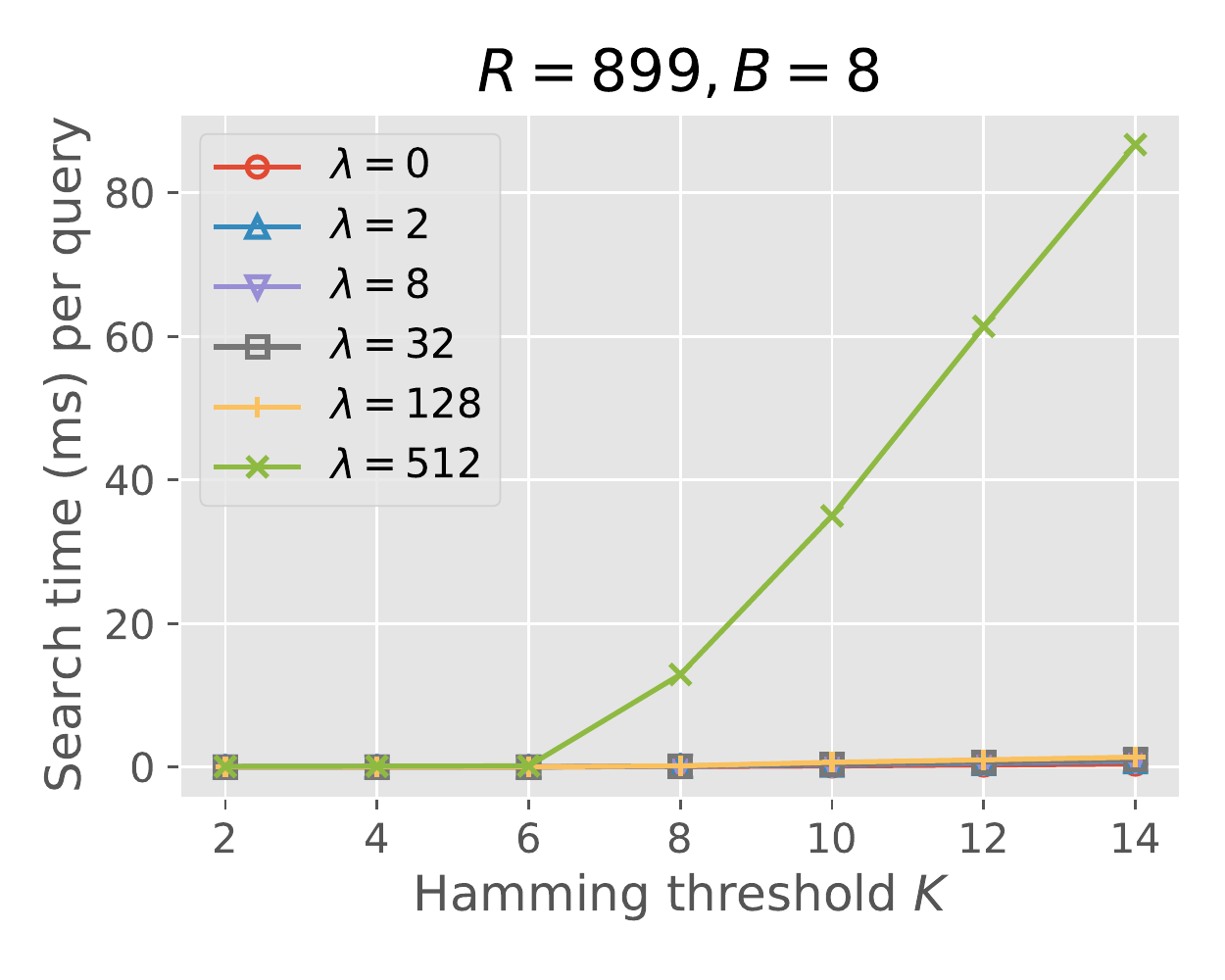}
\includegraphics[width=\ChartWidthApp]{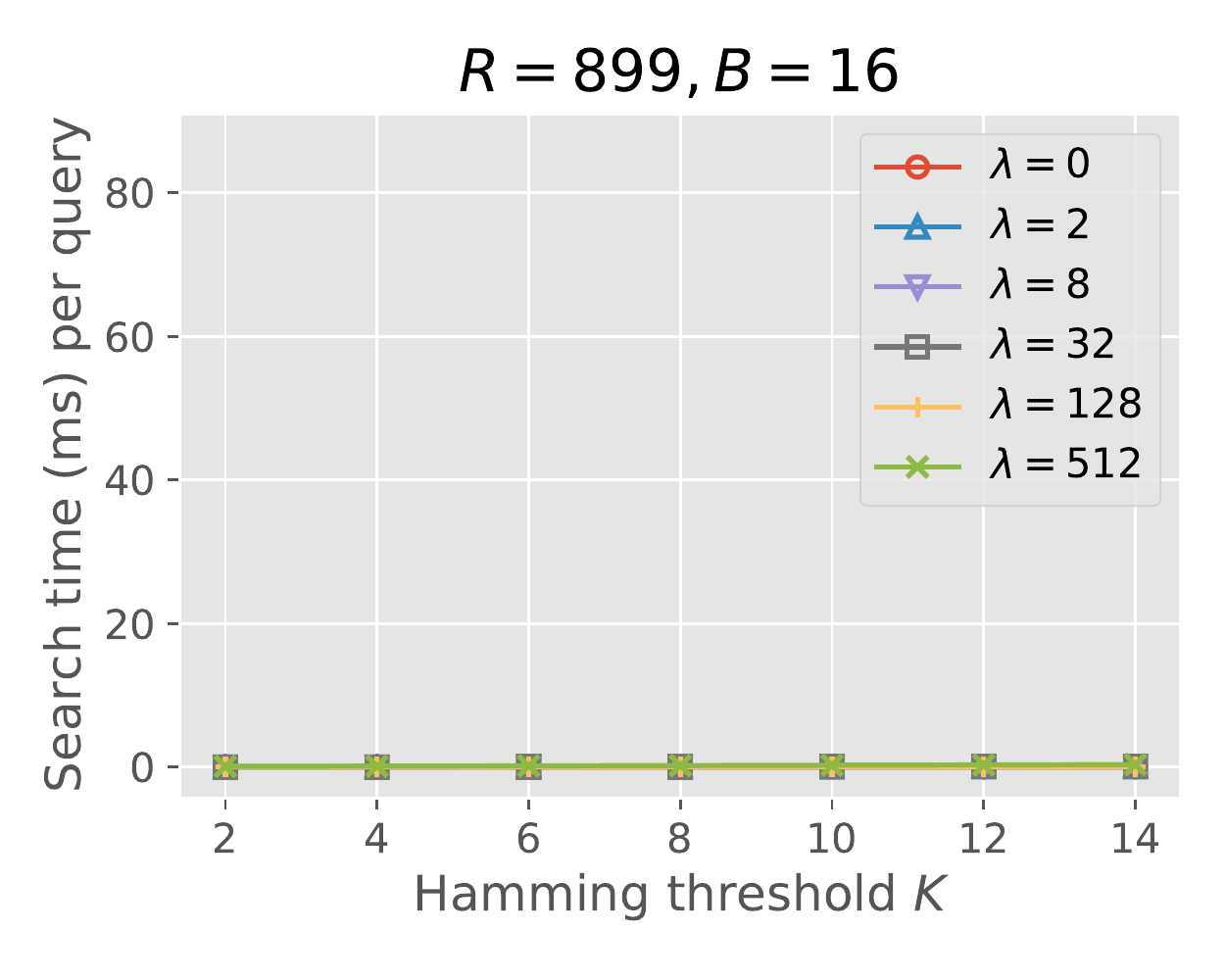}\\
\includegraphics[width=\ChartWidthApp]{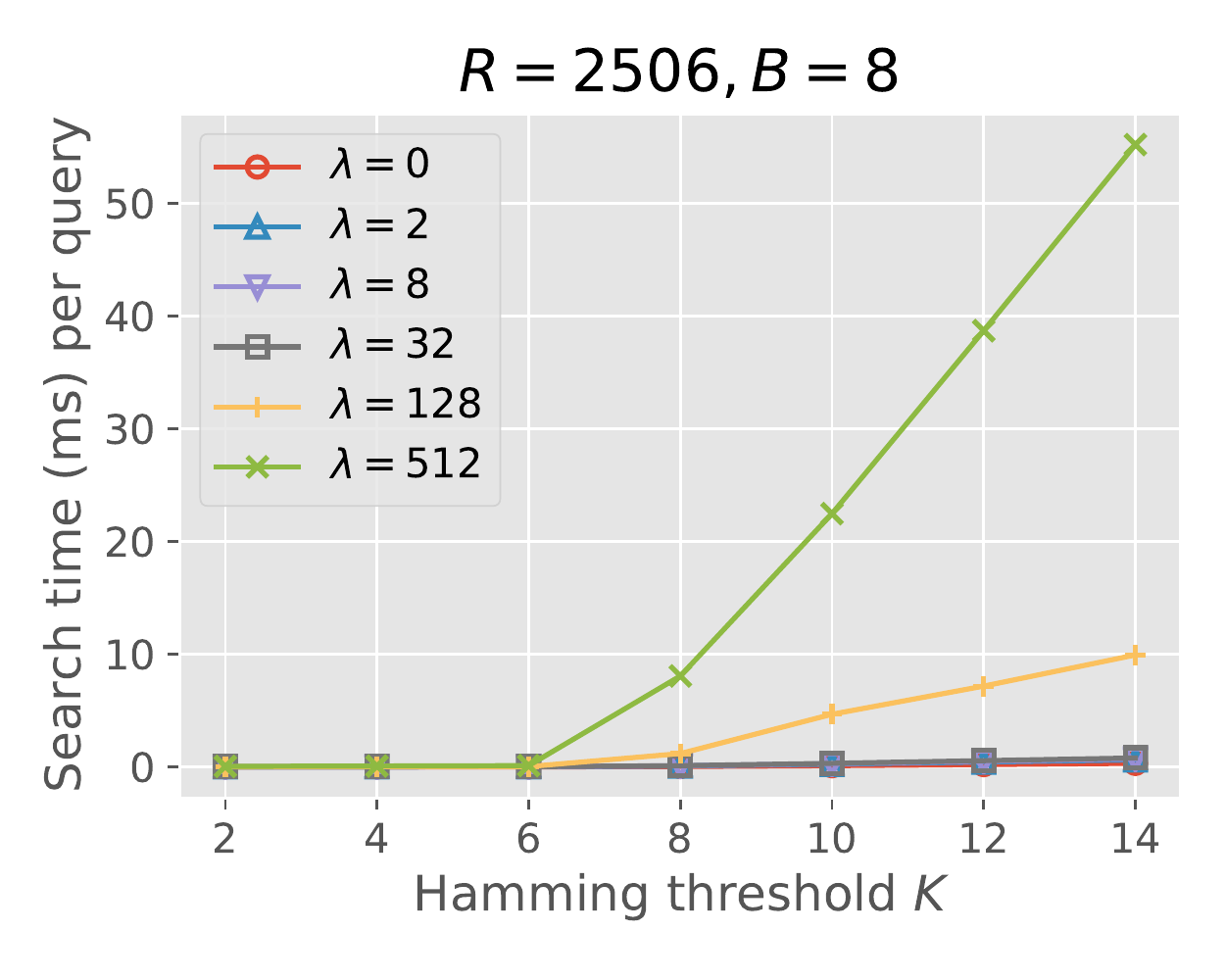}
\includegraphics[width=\ChartWidthApp]{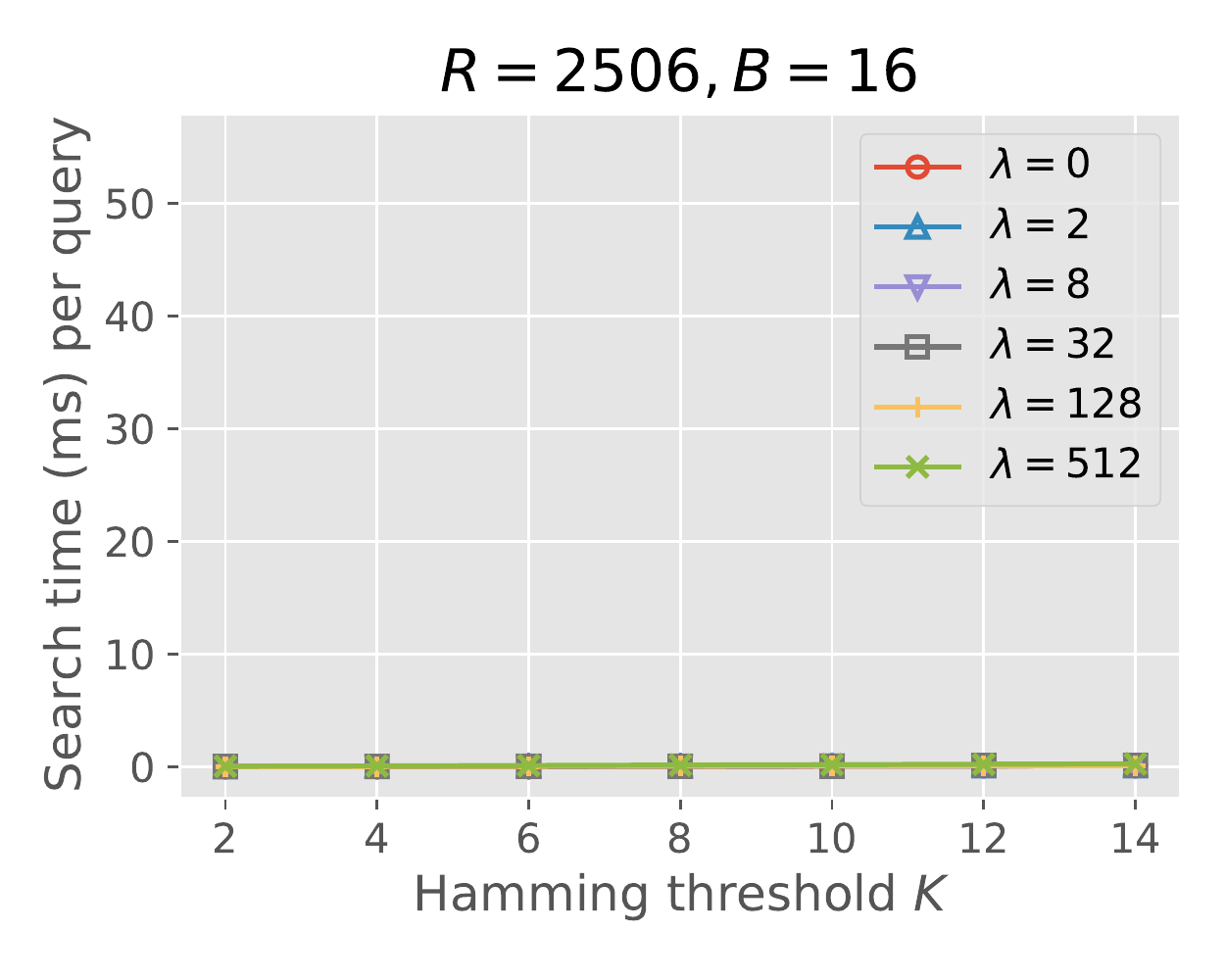}\\
\includegraphics[width=\ChartWidthApp]{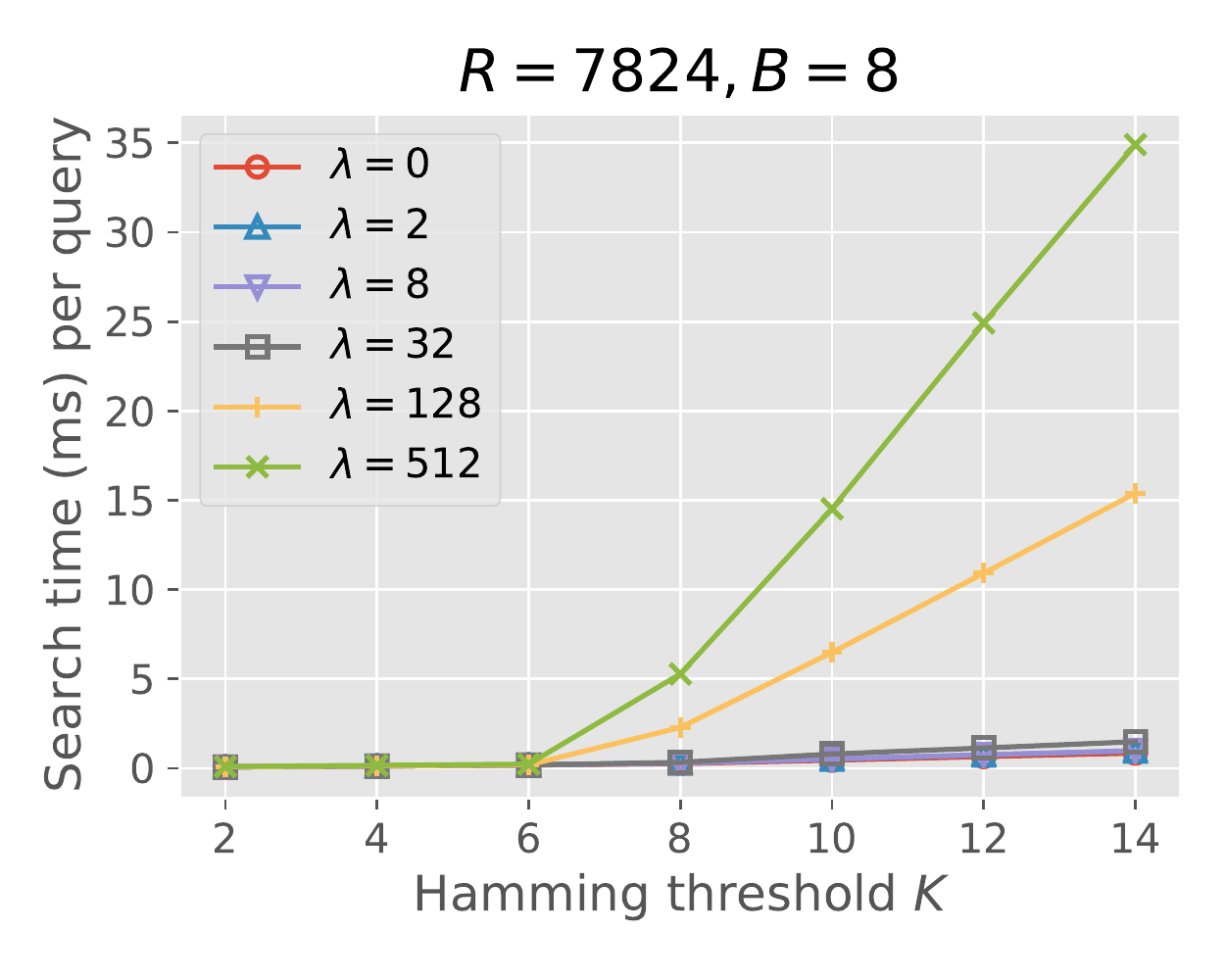}
\includegraphics[width=\ChartWidthApp]{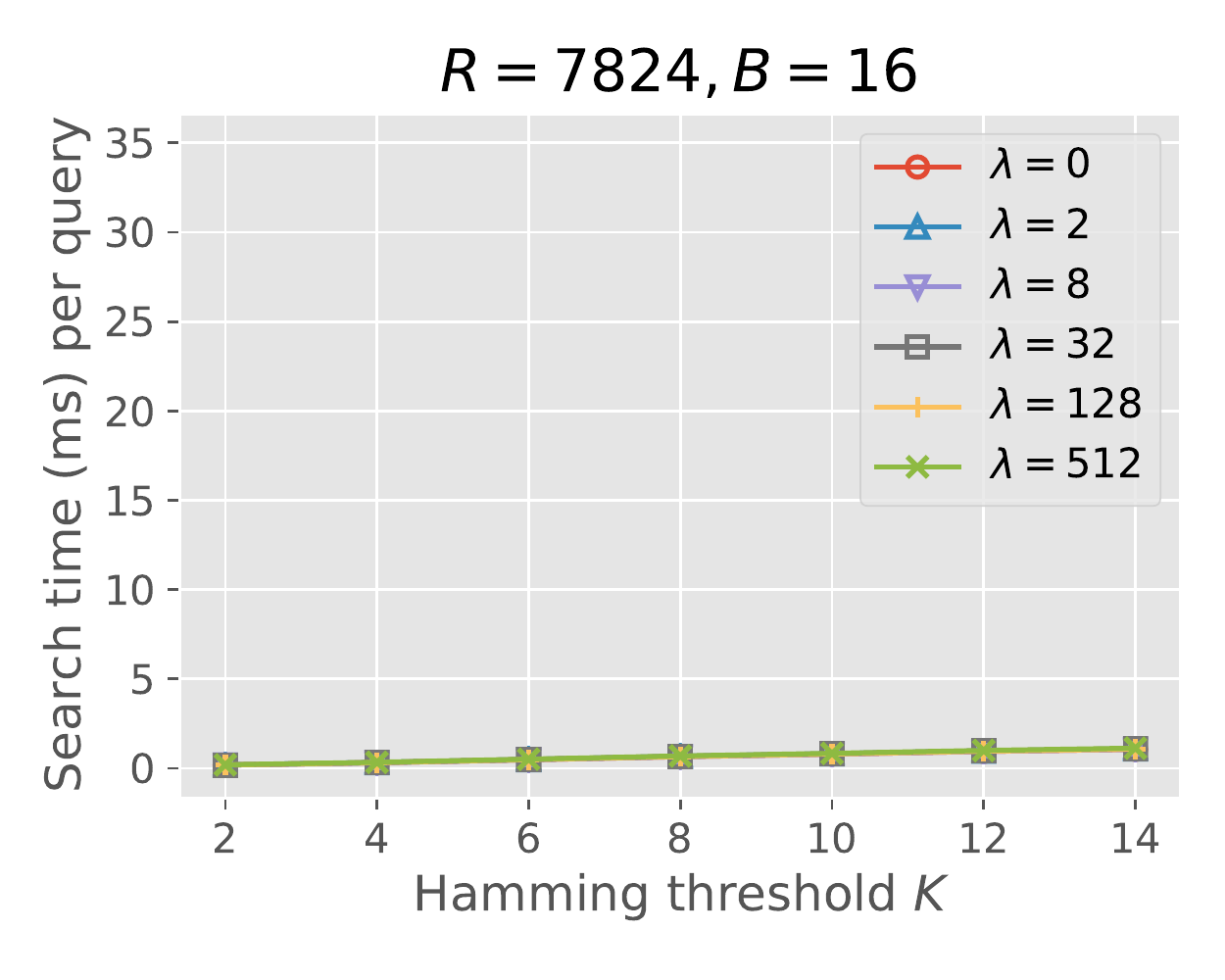}
\end{tabular}
}
\caption{Results of node reduction on OSM.}
\label{charts:app:reduce:OSM}
\end{figure*}

\clearpage

\newcommand{\ChartWidthAppB}{48mm}

\begin{figure*}[tb]
    \centering
    \setlength{\tabcolsep}{0mm}
    \subfloat[Taxi]{
        \begin{tabular}{c}
        \includegraphics[width=\ChartWidthAppB]{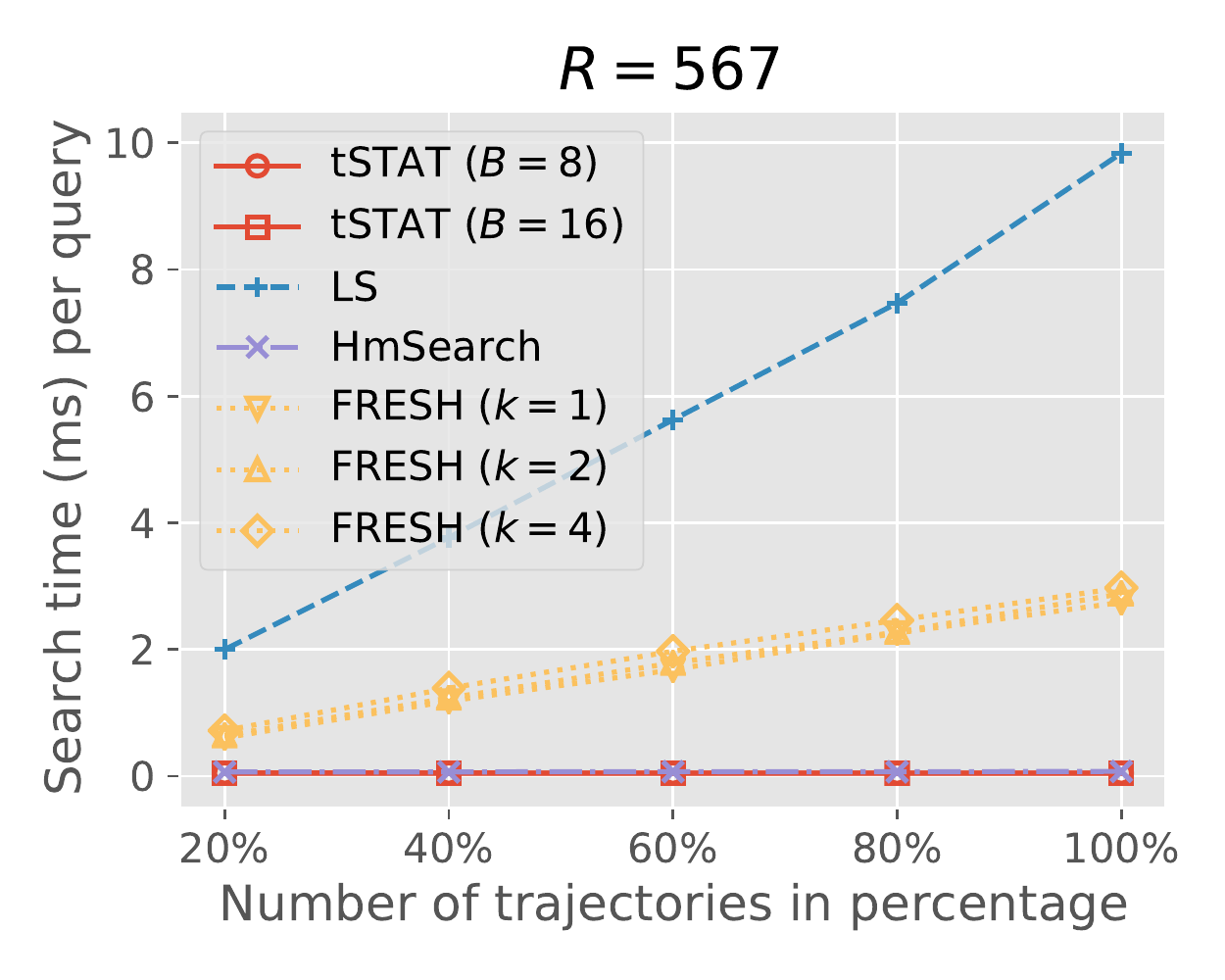}\\
        \includegraphics[width=\ChartWidthAppB]{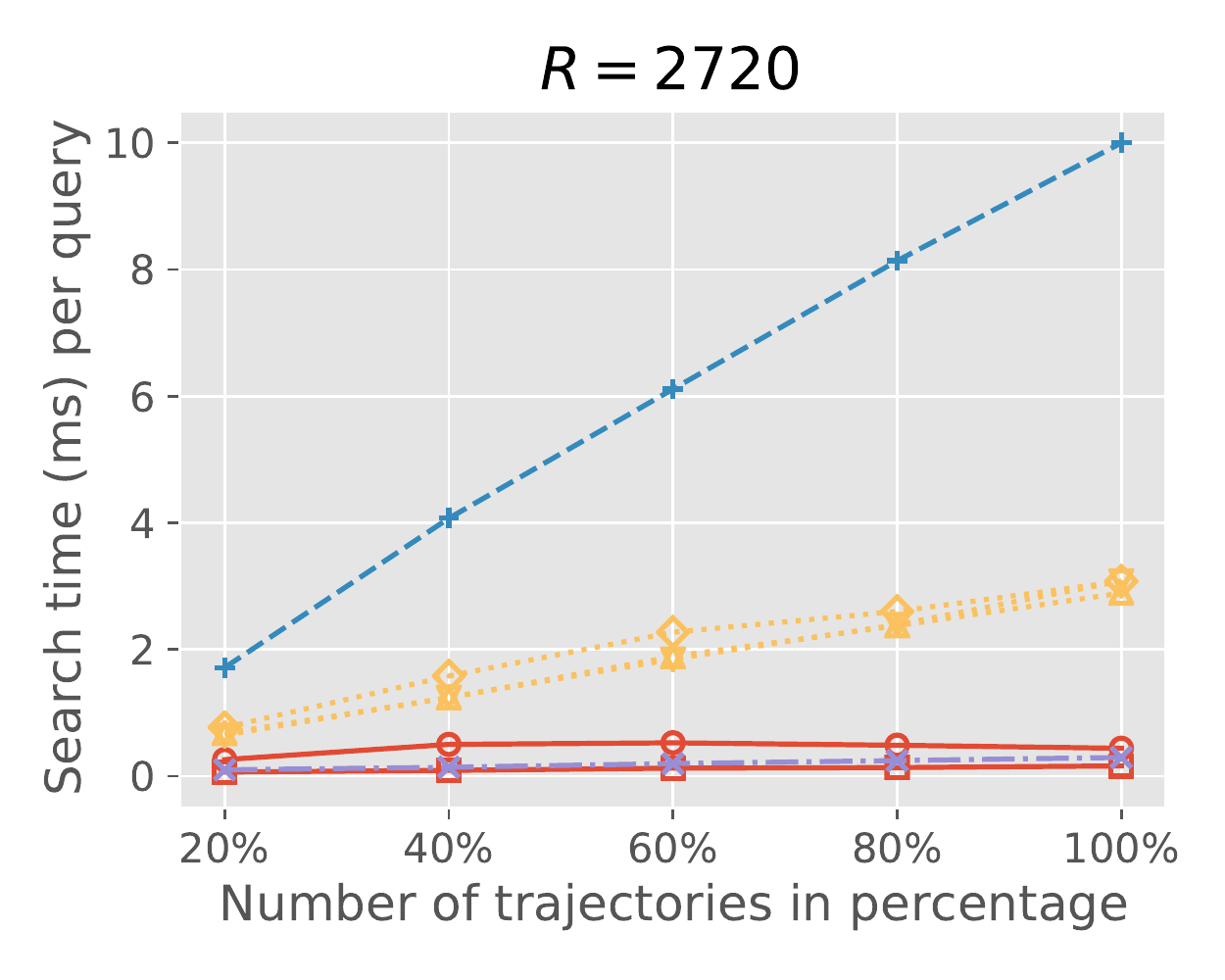}\\
        \includegraphics[width=\ChartWidthAppB]{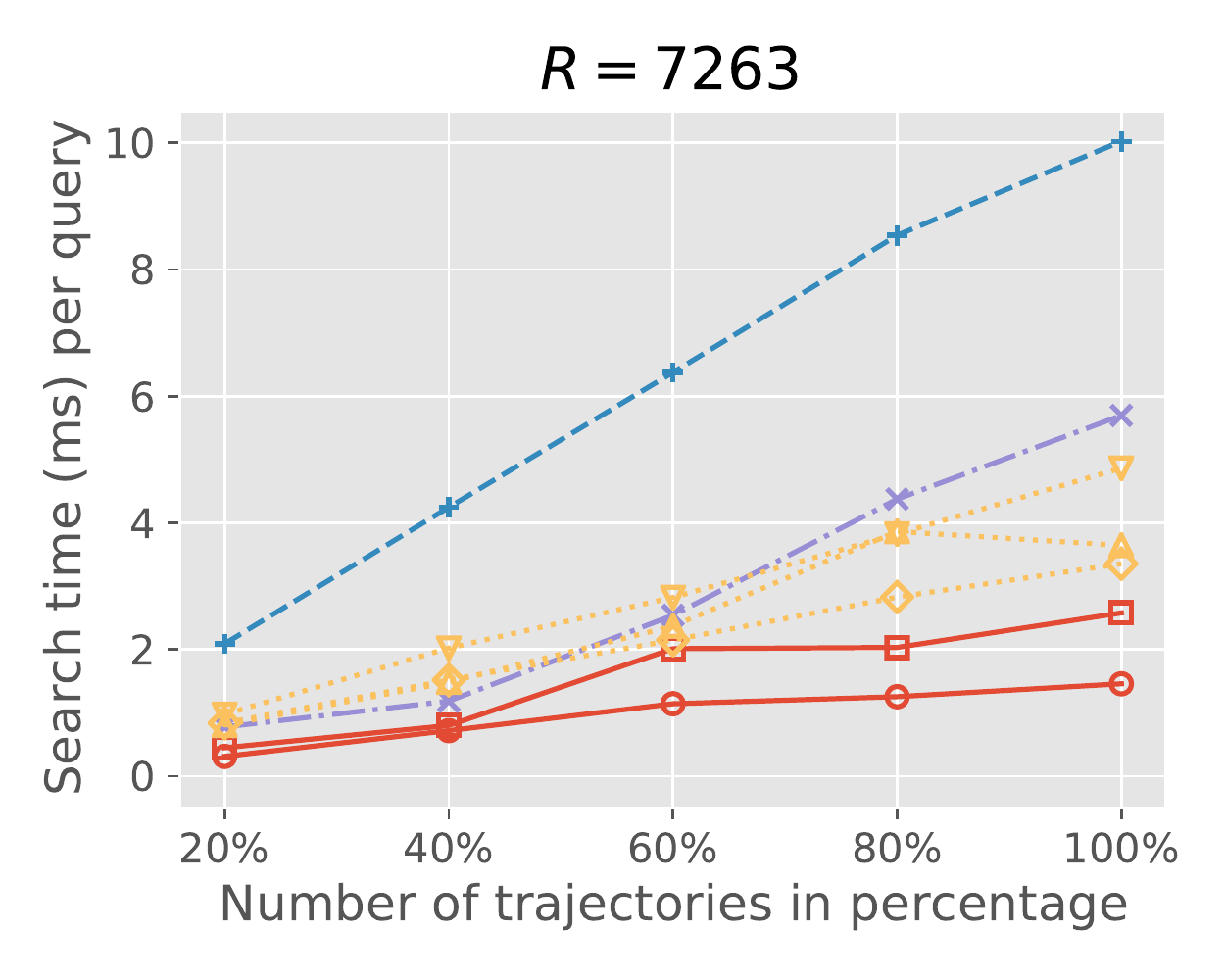}
        \end{tabular}
        \label{charts:app:search_scale:Taxi}
    }
    \subfloat[NBA]{
        \begin{tabular}{c}
        \includegraphics[width=\ChartWidthAppB]{charts/scale_time-NBA-0_9-0_1455R-64L.pdf}\\
        \includegraphics[width=\ChartWidthAppB]{charts/scale_time-NBA-0_9-0_2601R-64L.pdf}\\
        \includegraphics[width=\ChartWidthAppB]{charts/scale_time-NBA-0_9-0_4519R-64L.pdf}
        \end{tabular}
        \label{charts:app:search_scale:NBA}
    }
    \subfloat[OSM]{
        \begin{tabular}{c}
        \includegraphics[width=\ChartWidthAppB]{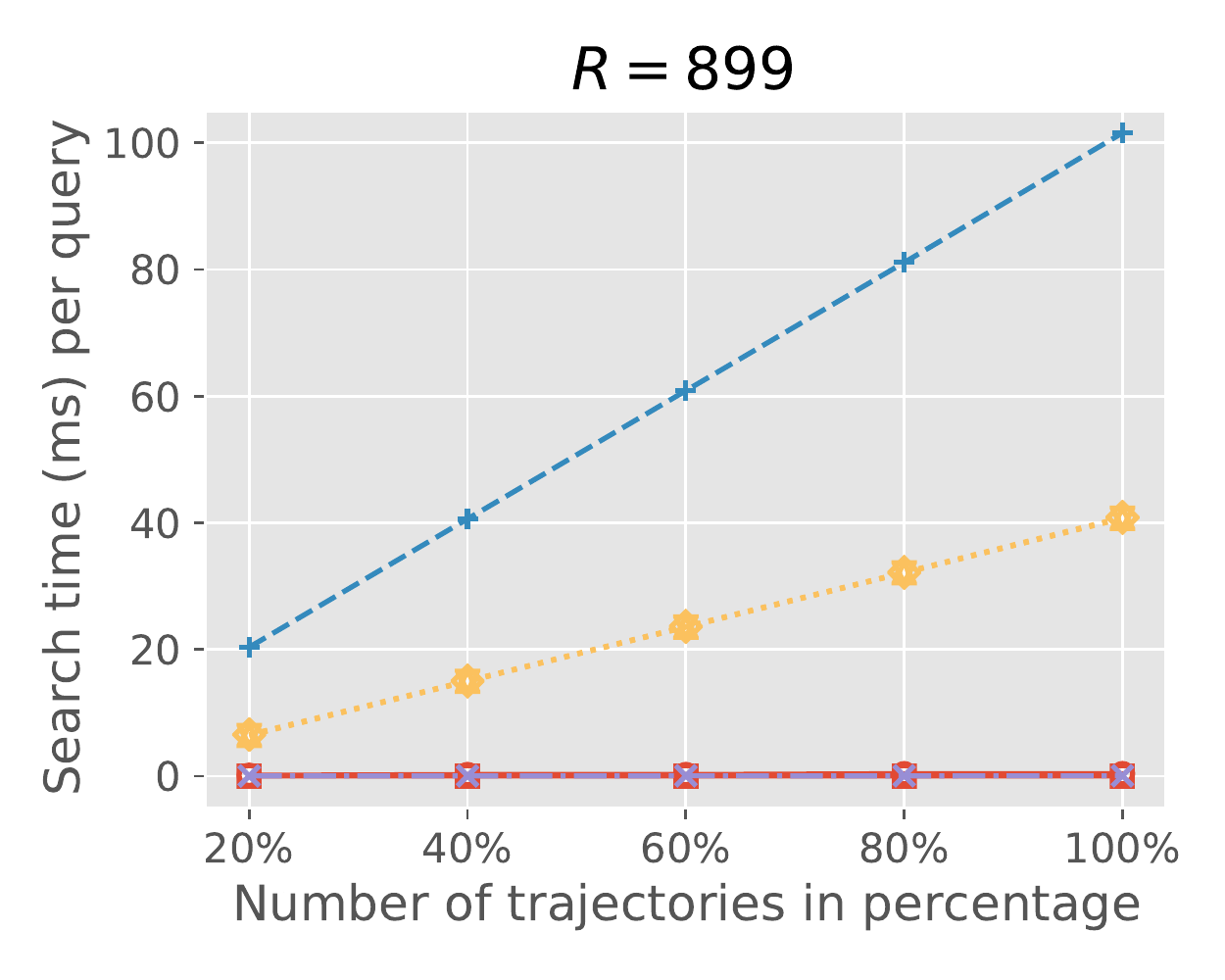}\\
        \includegraphics[width=\ChartWidthAppB]{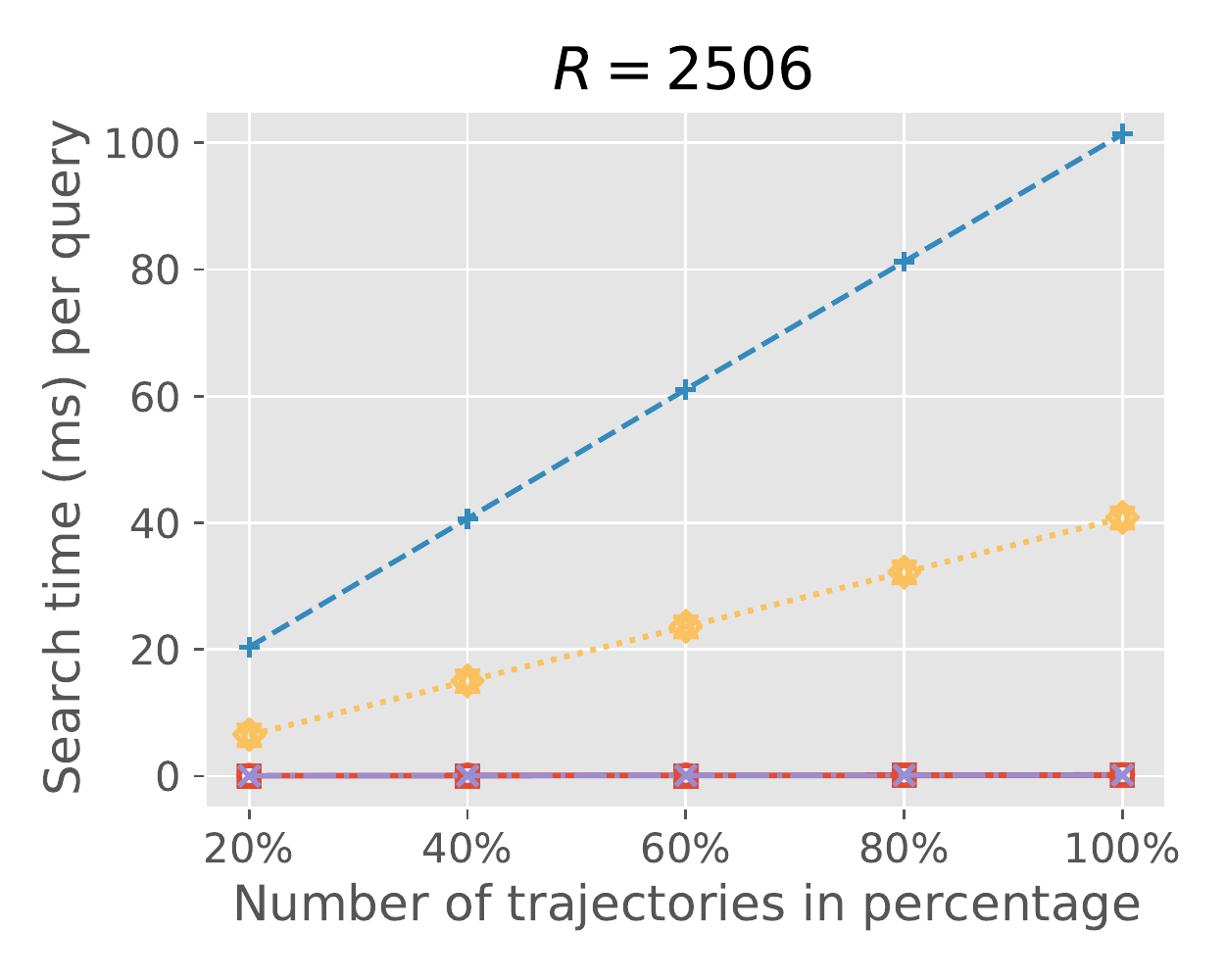}\\
        \includegraphics[width=\ChartWidthAppB]{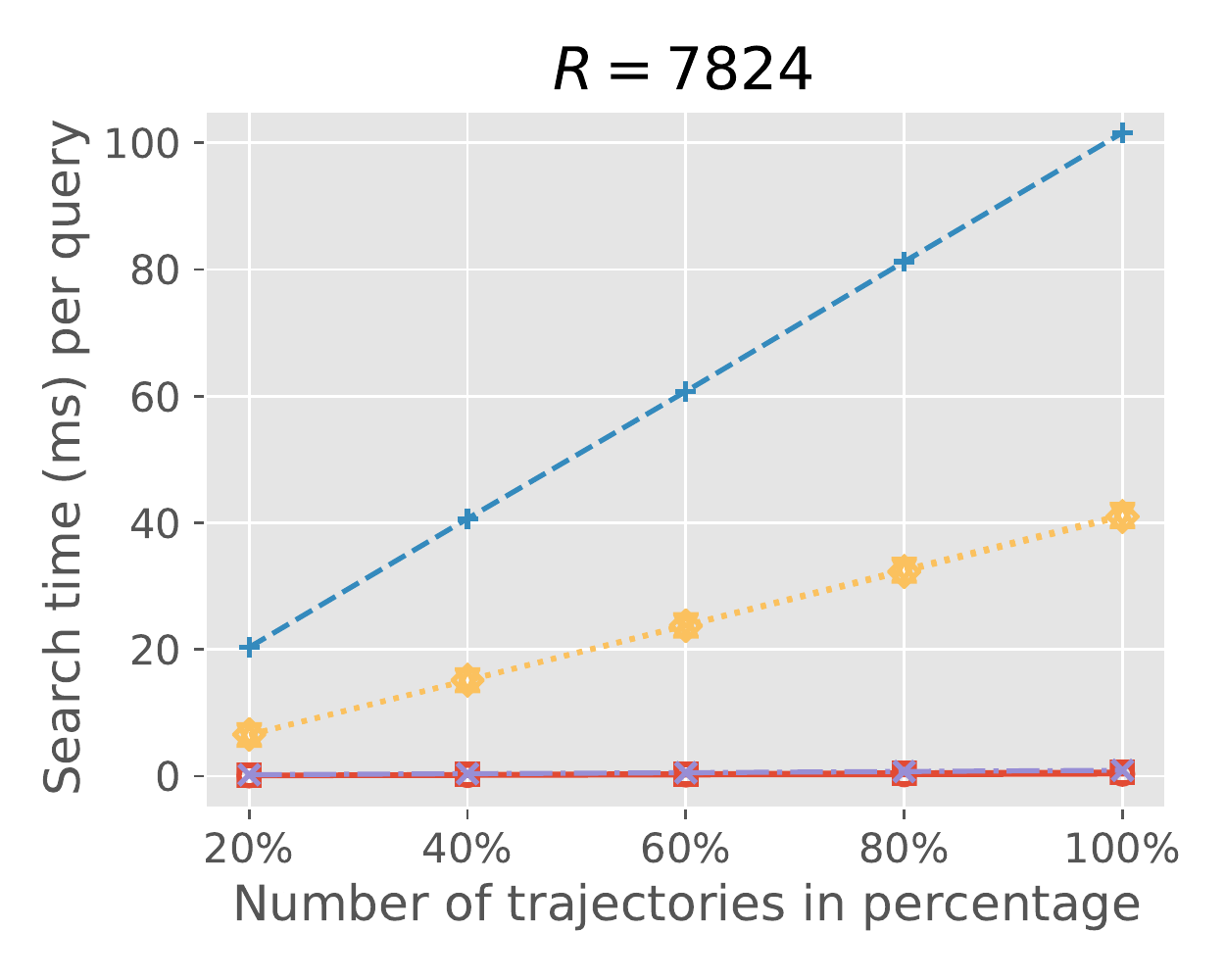}
        \end{tabular}
        \label{charts:app:search_scale:OSM}
    }
    \caption{
    Results of average search times per query in ms for varying the number of trajectories.}
    \label{charts:app:results:search_scale}
\end{figure*}

\begin{figure*}[tb]
    \centering
    \setlength{\tabcolsep}{0mm}
    \subfloat[Taxi]{
        \begin{tabular}{c}
        \includegraphics[width=\ChartWidthAppB]{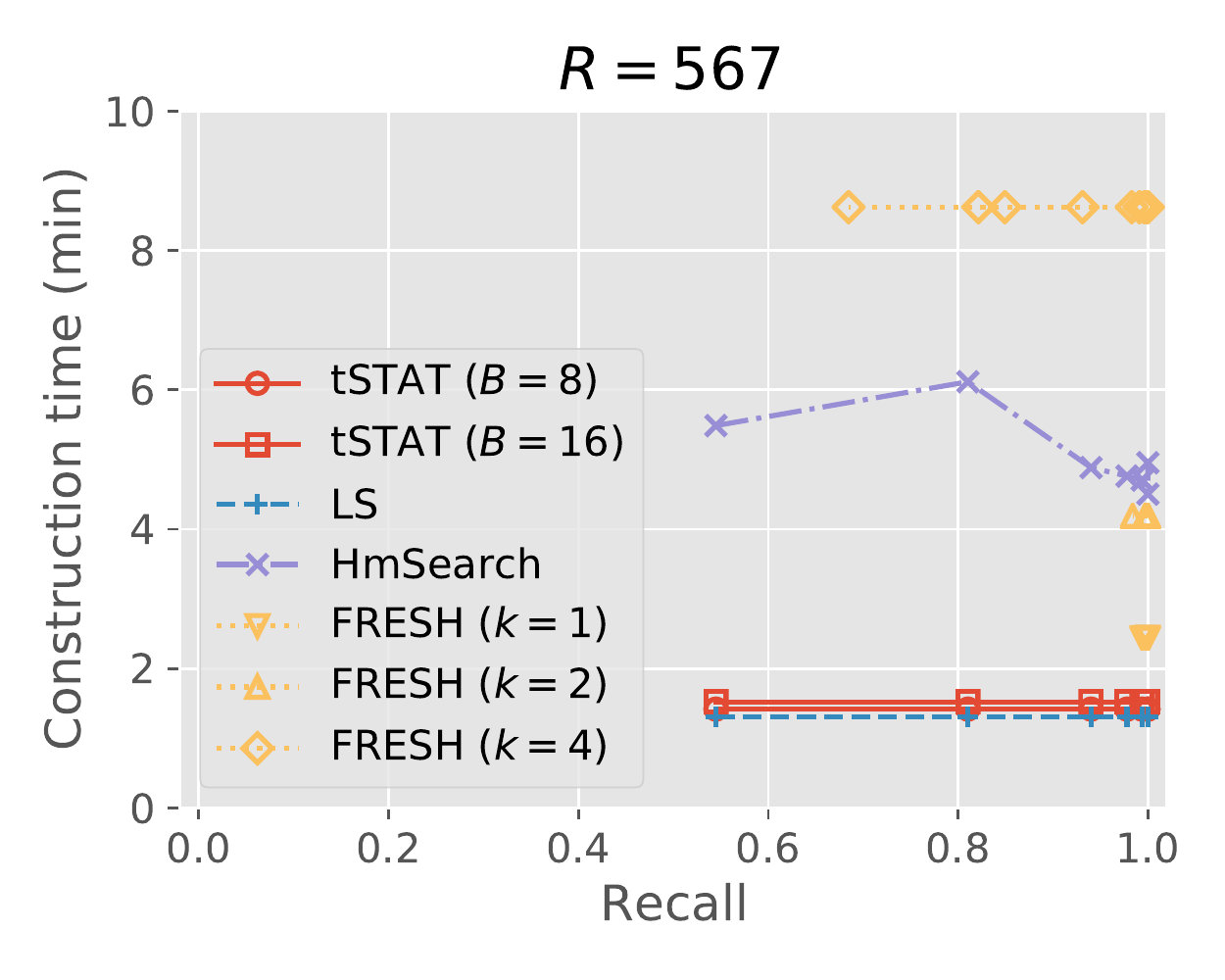}
        \end{tabular}
        \label{charts:app:constr:Taxi}
    }
    \subfloat[NBA]{
        \begin{tabular}{c}
        \includegraphics[width=\ChartWidthAppB]{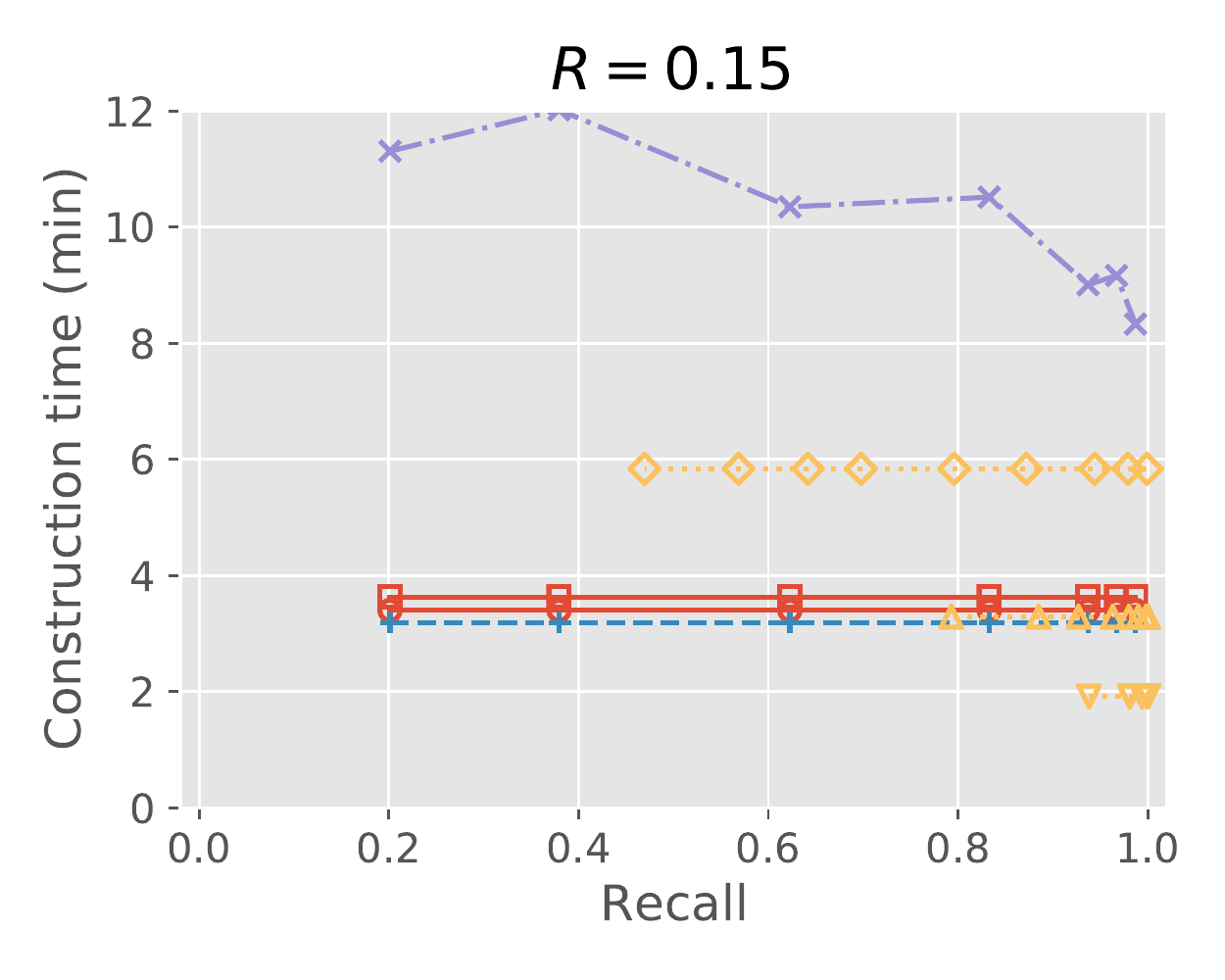}
        \end{tabular}
        \label{charts:app:constr:OSM}
    }
    \subfloat[OSM]{
        \begin{tabular}{c}
        \includegraphics[width=\ChartWidthAppB]{charts/recall_vs_constr-us-west-latest_base-us-west-latest_query-899_0R-64L.pdf}\\
        \end{tabular}
        \label{charts:app:constr:OSM}
    }
    \caption{Results for construction times in minutes for varying recalls.}
    \label{charts:app:results:constr}
\end{figure*}

\end{document}